\documentclass[a4paper,11pt,reqno]{amsart} 
\usepackage{amsmath}
\usepackage{amsfonts}
\usepackage[T1]{fontenc}  
\usepackage{lmodern}
\usepackage{microtype}
\usepackage[utf8]{inputenc}
\usepackage[mathscr]{euscript}
\usepackage{verbatim}
\usepackage{amssymb,latexsym}
\usepackage{amsthm}
\usepackage{color}
\usepackage{graphicx}
\usepackage[hidelinks]{hyperref}
\usepackage[font=scriptsize]{caption}
\usepackage{subcaption}
\usepackage{bbm}
\usepackage{cite}
\usepackage{amsmath}
\usepackage{eurosym}

\usepackage[lmargin=3 cm,rmargin=3 cm,tmargin=3.5cm,bmargin=2.5cm,paper=a4paper]{geometry}
\DeclareMathOperator{\curl}{curl}

\DeclareMathOperator{\supp}{supp}

\DeclareMathOperator{\dom}{\mathrm{Dom}}
\newtheorem{thm}{Theorem}[section]

\newtheorem{lem}[thm]{Lemma}
\newtheorem{cor}[thm]{Corollary}

\newtheorem{theorem}[thm]{Theorem}
\newtheorem{assumption}[thm]{Assumption}

\newtheorem{lemma}[thm]{Lemma}
\newtheorem{proposition}[thm]{Proposition}

\theoremstyle{remark}
\newtheorem{rem}[thm]{Remark}

\newcommand{\R}{\mathbb{R}}
\newcommand{\Fb}{\mathbf{F}}

\newcommand{\clb}{\color{blue}}
\newcommand{\N}{\mathbb{N}}

\newcommand{\Om}{\Omega}

\newcommand{\tO}{\tilde{\mathcal O}}
\numberwithin{equation}{section}
\title{Semi-classical eigenvalue estimates under magnetic steps}
\author[W. Assaad]{Wafaa Assaad}
\address{Lebanese International  University,  Faculty of Arts and Sciences,  Beirut,  Lebanon}
\email{wafaa\_assaad@hotmail.com}
\author[B. Helffer]{Bernard Helffer}
\address{Nantes Universit\'e, Laboratoire Jean Leray, Nantes, France}
\email{Bernard.Helffer@univ-nantes.fr}
\author[A. Kachmar]{Ayman Kachmar}
\address{Lebanese University, Department of Mathematics, 1700 Nabatiye,  Lebanon}
\address{Center for Advanced Mathematical Sciences (CAMS), American University of Beirut}
\email{akachmar@ul.edu.lb}
\date{\today}
\begin{document}
\maketitle
\begin{abstract} 
 We establish accurate eigenvalue asymptotics and, as a by-product,  sharp estimates of the splitting between two consecutive eigenvalues, for the Dirichlet magnetic Laplacian with a non-uniform magnetic field having a jump discontinuity along a smooth curve. The asymptotics hold in the semiclassical limit which also corresponds to a large magnetic field limit, and is valid under a geometric assumption on the curvature of the discontinuity curve.
\end{abstract}
\section{Introduction}\label{sec:int}

The paper studies a semiclassical Schr\"odinger operator with a step magnetic field and Dirichlet boundary conditions, in a smooth bounded domain. The aim is to give accurate estimates of the lower eigenvalues in the semiclassical limit.

\begin{figure}
\includegraphics[scale=1]{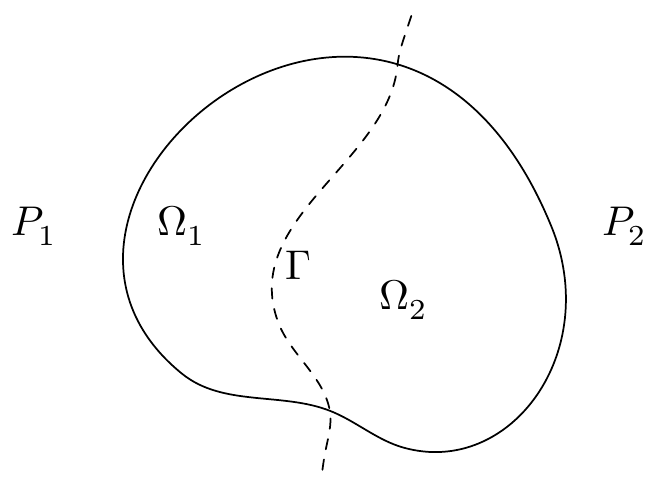}
	\caption{The curve $\Gamma$ transversally cuts $\partial\Om$ at two points and splits $\Om$ into two regions, $\Om_1$ and  $\Om_2$.}\label{fig1}
\end{figure}

Let $\Om$ be an open,  bounded, and simply connected subset of $\R^2$ with smooth  $C^1$ boundary.   We consider a simple smooth curve $\Gamma\subset\R^2$ that splits $\R^2$ into two disjoint unbounded open sets, $P_{1}$ and $P_{2}$, and such  that $\Gamma$ is a semi-straight line when $|x|$ tends to $+\infty$. We assume that $\Gamma$ decomposes $\Om$ into two sets $\Om_1$ and $\Om_2$ as follows (see~Figure~\ref{fig1}):
\begin{enumerate}
\item  $\Gamma$ intersects  transversally $\partial \Om$ at two distinct points. 
\item 	$\Omega_1:=\Omega\cap P_{1}\not=\emptyset$ and  $\Omega_2:=\Omega\cap P_{2}\not=\emptyset$.
\end{enumerate} 
Let $h>0$ and  $\Fb=(F_1,F_2) \in  H^1_\mathrm{loc}(\R^2)$ be a magnetic potential whose associated magnetic field is 
\begin{equation}\label{eq:F}
	\curl \Fb=a_1\mathbbm 1_{P_1}+a_2\mathbbm 1_{P_2},\quad \mathbf{a}:=(a_1,a_2)\in \R^2,\ a_1\neq a_2.
	\end{equation}
When restricted to $\Om$, the vector field $\Fb$ satisfies
\begin{equation}\label{eq:curl-F}
\curl \Fb=a_1\mathbbm 1_{\Omega_1}+a_2\mathbbm 1_{\Omega_2},\quad \mathbf{a}:=(a_1,a_2)\in \R^2,\ a_1\neq a_2 \mbox{ and } \Fb \in L^4(\Omega)\,.
\end{equation}
Note that the curve $\Gamma$ separates the two regions $\Om_1$ and $\Om_2$ which are assigned with different values of the magnetic field. For this reason,  we refer to $\Gamma$ as the \emph{magnetic edge}.
We consider the  quadratic form on $H_0^1(\Omega)$ 
\begin{equation}\label{eq:qf-MLD}
u \mapsto \mathcal Q_h(u)=\int_\Omega|(h\nabla-i\Fb)u|^2\,dx\,.
\end{equation}
This quadratic form is closed on the form domain $H^1_0(\Omega)$. By the Friedrichs extension procedure, we can associate its  Dirichlet realization in $\Omega$
  \begin{equation}\label{eq:P}
\mathcal P_{h}:=-(h\nabla-i\Fb)^2=-\sum_{j=1}^2 (h \partial_{x_j} - i F_j)^2,\end{equation}
 whose domain is
\begin{equation}\label{eq:DomP}
\dom(\mathcal P_{h})=\{u\in L^2(\Om)~:~(h\nabla-i\Fb)^j u \in L^2(\Om),\, j\in\{1,2\},u|_{\partial \Om}=0 \}. \end{equation}

The operator $\mathcal P_{h}$ is self-adjoint, has  compact resolvent, and its spectrum is an increasing sequence, $(\lambda_n(h))_{n\in \N}$, of real eigenvalues listed with multiplicities. 

In this contribution, we aim at giving the asymptotic expansion of the low-lying eigenvalues of $\mathcal P_{h}$, in the semiclassical limit, i.e. when $h$ tends to $0$.

Schr\"odinger operators with a discontinuous magnetic field, like $\mathcal P_h$, appear in many models in nanophysics such as in quantum transport while studying the transport properties of a bidimensional electron gas~\cite{reijniers2000snake,peeters1993quantum}.   In that context,   the magnetic edge is \emph{straight} and  bound states  interestingly  feature currents flowing along the magnetic edge. 

The present contribution addresses another appealing question on the influence of the shape of the magnetic edge on the energy of the bound states. We give an affirmative answer by providing sharp semiclassical eigenvalue asymptotics  under a single `well'   hypothesis on the curvature of the magnetic edge  (cf. Assumption~\ref{kmax} and Theorem~\ref{thm:main} below). Loosely speaking, our hypothesis says  that we perform a local deformation of the magnetic edge so that its curvature has a unique non-degenerate maximum.

  Another important occurrence of  magnetic  Laplace operators is  in the Ginzburg--Landau model of superconductivity~ \cite{saint1963onset}. 
In bounded domains, the spectral properties of these operators can describe interesting physical situations. In the context of superconductivity, an accurate information about the lowest eigenvalues  is important for giving a precise description of the concentration of superconductivity in a type-II superconductor. Moreover, it improves the estimates of the third critical field, $H_{C_3}$, that marks the onset of\break superconductivity in the domain. We refer the reader to~\cite{assaad2020band,Assaad3} for discontinuous field cases, and  to~\cite{fournais2006accurate,helffer2003upper,lu2000gauge,lu1999eigenvalue,lu1999estimates,bonnaillie2007superconductivity,bonnaillie2006asymptotics, bernoff1998onset,tilley1990superfluidity}   for a further discussion in smooth fields cases. In the present paper, the Dirichlet realization of $\mathcal P_h$ in the bounded domain $\Om$ can physically correspond to a superconductor which is set in the normal (non superconducting) state at its boundary.

Using symmetry and scaling arguments, one can reduce the problem to the study of cases of $\mathbf a=(a_1,a_2)$, where $a_1=1$ and $a_2=a\in[-1,1)$. Moreover, we will soon make a more restrictive choice of cases of $\mathbf a$ (see~\eqref{eq:a} below). Towards justifying the upcoming choice of $\mathbf a$,  we introduce  the effective operator $\mathfrak h_a[\xi]$ with a discontinuous field, defined on $\R$ and parametrized by $\xi\in\R$:

\begin{equation}\label{eq:beta00}
\mathfrak h_a[\xi]=-\frac{d^2}{d\tau^2}+\big(\xi+b_a(\tau) \tau\big)^2\,,
\end{equation}
where 
\begin{equation}\label{eq:beta01}
b_a(\tau) =\mathbf{1}_{\R_+}(\tau)+a\mathbf{1}_{\R_-}(\tau)\,.
\end{equation}
This operator arises from the approximation by  the case where $\Omega=\R^2$ and $\Gamma=\{x_2=0\},$ $\tau$ corresponding to the variable $x_2$ and $\xi$ being the dual variable of $x_1$.
The known spectral properties of $\mathfrak h_a[\xi]$,  obtained earlier in \cite{hislop2016band, Assaad2019, assaad2020band},  are recalled in Subsection~\ref{sec:step1}. Here, we only present some features of this operator that are useful to this introduction.  The bottom of the spectrum of $\mathfrak h_a[\xi]$, denoted by $\mu_a(\xi)$, is a simple eigenvalue for $a\not=0$, usually called \emph{band function} in the literature. Minimizing the band function leads us to introduce
\begin{equation}\label{eq:beta0}
\beta_a=\inf_{\xi\in \R}\mu_a(\xi)\,.
\end{equation}
We list the following properties of $\beta_a$, depending on the values of $a$:
\paragraph{\emph {Case $a=-1$}} ~\\
In the case where $\Omega=\R^2$ and $\Gamma=\{x_2=0\}$, this case is called the `symmetric trapping magnetic steps', and is well-understood in the literature (see~e.g.~\cite{hislop2016band}). In this case, the study of $\mathfrak h_a[\xi]$ can be reduced to that of the de Gennes operator (a harmonic oscillator on the half-axis with
	Neumann condition at the origin). We refer the reader to~\cite{fournais2010spectral} and the references therein for the spectral properties of this operator. Here, 
	\begin{equation}\label{eq:teta1}
\Theta_0:=\beta_{-1}\approxeq 0.59 
	\end{equation}
	is attained by $\mu_{-1}(\cdot)$ at a unique and non-degenerate minimum $\xi_0=-\sqrt{\Theta_0}$. Moreover, $\beta_{-1}=\mu_{-1}(\xi_0)$ is a simple eigenvalue of  $\mathfrak h_{-1}[\xi_0]$.
\paragraph{\emph {Case $-1<a<0$}}~\\
 This case is called the `asymmetric trapping magnetic steps', and is studied in many works (see~\cite{assaad2020band,Assaad2019,hislop2016band}).  We have
	$|a|\Theta_0<\beta_a<\min(|a|,\Theta_0)$ and
	$\beta_a$ is attained by  $\mu_{a}(\cdot)$ at a unique $\zeta_a<0$ \cite{assaad2020band}
\begin{equation}\label{eq:mub}
\mu_{a}(\zeta_a)=\beta_a\,.
\end{equation}
 Moreover,  the minimum is non-degenerate, i.e. $\mu_a''(\zeta_a)>0\,$. 
\paragraph{\emph {Case $a=0$}}~\\
 This corresponds to the `magnetic wall' case studied for instance in~\cite{reijniers2000snake,hislop2016band}.  We refer to~\cite[Section~2]{hislop2016band} for this case. \\
  For $\xi\leq0$, we have $$\sigma (h_a[\xi])=\sigma _{ess}(h_a[\xi])=[\xi^2,+\infty)\,,$$
  where $\sigma $ and $\sigma _{ess}$ respectively denote the spectrum and essential spectrum.\\
    For $\xi>0$, $$\sigma _{ess}(h_a[\xi])=[\xi^2,+\infty)$$  and $h_a[\xi]$ may have positive eigenvalues $\lambda<\xi^2$. 
    Consequently, $\beta_0=\mu_0(0)=\inf \sigma _{ess}\mathfrak h_0[0]=0$, and $\beta_0$ is not an eigenvalue of $\mathfrak h_a[\xi]$, for all $\xi\in\R$.

\paragraph{\emph {Case $0<a<1$}}~\\ This corresponds with the 'non-trapping magnetic steps' case (see~\cite{Assaad2019,hislop2015edge,iwatsuka1985examples}).  Here, $\beta_a=a$ and $\mu_a(\cdot)$ doesn't achieve a minimum; the infimum is attained at $+\infty$. \\

A key-ingredient in establishing the asymptotics of the  eigenvalues $\lambda_n(h)$ is that  $\beta_a$
is an eigenvalue of $\mathfrak h_a[\xi],$ for some $\xi\in\R$. We will use the corresponding eigenfunction in constructing quasi-modes of the operator $\mathcal P_h$. The above discussion  shows that $\beta_a$ is an eigenvalue only when $a\in[-1,0)$. The case $a=-1$ is excluded from our study, despite the fact that $\beta_{-1}$ is an eigenvalue of $\mathfrak h_{-1}[\xi_0]$.  Except if  $\Gamma$ is  an axis of symmetry of $\Omega$ as in \cite{hislop2016band}, the situation is more difficult and the curvature will play a more  important role. We hope to treat this case in a future work. This explains our choice to work under the following assumption on $\mathbf a$ (thus on the magnetic field $\curl\mathbf F$) throughout the paper:
\begin{equation}\label{eq:a}
\mathbf a=(1,a),\quad \mathrm{with}\ -1<a<0\,.
\end{equation}
Under assumption~\eqref{eq:a}, we introduce two spectral invariants: 
\begin{equation}\label{eq:main-ct}
c_2(a)=\frac 12 \mu''_a(\zeta_a)>0\quad \mbox{and}\quad M_3(a)=\frac 13\Big(\frac 1a-1\Big)\zeta_a\phi_a(0)\phi_a'(0)<0\,,
\end{equation}
where $\mu_a$ and $\zeta_a$ are  introduced  in~\eqref{eq:beta0} and~\eqref{eq:mub}, and $\phi_a$ is the positive $L^2$-normalized eigenfunction of $\mathfrak h_a[\zeta_a]$ corresponding to $\beta_a$.

 Furthermore, we work under the following assumption:
\begin{assumption}\label{kmax}~\\
The curvature $\Gamma\ni s\mapsto k(s)$ at the magnetic edge has a unique maximum 
\[k(s)<k(s_0)=:k_{\max},\ \mbox{for}\ s\neq s_0.\]
This maximum is attained in $\Gamma\cap\Omega$ and is non-degenerate
\[ k_2:=k''(s_0)<0.\]
\end{assumption}

The goal of this paper is to prove the following theorem:
\begin{theorem}\label{thm:main}
Let $n\in\N^*$ and $\mathbf a=(1,a)$ with $-1<a<0$\,.  Under Assumption~\ref{kmax},  the $n$'th eigenvalue $\lambda_n(h)$ of $\mathcal P_{h}$,  defined in~\eqref{eq:P}, satisfies as $h\rightarrow 0$, 
\[\lambda_n(h)= h\beta_a+h^\frac 32k_{max} M_3(a)+h^\frac 74(2n-1)\sqrt{\frac{{ k_2}M_3(a)c_2(a)}{2}}+\mathcal O(h^{\frac{15}8}),\]
where $\beta_a$,  $c_2(a)$ and $M_3(a)$ are the spectral quantities introduced in~\eqref{eq:beta0} and~\eqref{eq:main-ct}.
\end{theorem}
\begin{rem} This theorem extends \cite[Theorem~4.5]{assaad2020band}) where the first two terms in the expansion of the first eigenvalue  were determined with a remainder in $\mathcal O(h^{\frac 53})$\,.  The proof of Theorem~\ref{thm:main} partially relies on decay estimates of the  eigenfunctions with the right scale (see Sec.~\ref{sec:local} and \cite{assaad2020band}).  In fact,  away from the edge  $\Gamma$,  the  eigenfunctions decay exponentially at the scale $h^{-1/2}$  of the distance  to $\Gamma$,   while,   along $\Gamma$,  they decay exponentially with a scale  of $h^{-1/8}$ of the tangential distance  on $\Gamma$ to the point with maximum curvature. 
\end{rem}
\paragraph{\emph{Comparison with earlier situations}}
It is useful to compare the asymptotics of $\lambda_n(h)$ in Theorem~\ref{thm:main} with those obtained in the literature, for regular domains submitted to uniform magnetic fields. In bounded planar domains with smooth boundary, subject to unit magnetic fields and when the \emph{Neumann} boundary condition is imposed, the low-lying eigenvalues of the linear operator, analogous to $\mathcal P_h$, admit the following asymptotics as $h$ tends to $0$ (see e.g.~\cite{fournais2006accurate})
\[\lambda_n(h)= h\Theta_0-h^\frac 32\tilde k_{max} C_1+h^\frac 74C_1\Theta_0^{\frac 14}(2n-1)\sqrt{\frac 32 \tilde k_2}+\mathcal O(h^{\frac{15}8}),\]
where $\Theta_0$ is as in~\eqref{eq:teta1}, $C_1>0$ is some spectral value,  and  $\tilde k_{max}$ and $\tilde k_2$ are positive constants introduced in what follows. In this uniform field/Neumann condition situation, the eigenstates localize near the boundary of the domain. More precisely, they localize near the point $\tilde s$ with maximum curvature $ k(\tilde s)$ of this boundary, assuming the uniqueness and non-degeneracy of this point. We define $\tilde k_{max}=k(\tilde s)$ and $\tilde k_2=-k''(\tilde s)>0$. In~\cite{fournais2006accurate}, the foregoing localization of eigenstates restricted the study  to the boundary, involving 
a family of 1D effective operators which act in the normal direction to the boundary. These are  the de Gennes operators
\[\mathfrak h^{N}[\xi]=-\frac{d^2}{d\tau^2}+(\xi+\tau)^2,\]
defined on  $\R_+$ with Neumann boundary condition at $\tau=0$, and parametrized by $\xi\in \R$. We recover the value $\Theta_0$ as an \emph{effective energy} associated to $(\mathfrak h^{N}[\xi])_\xi$
\[\Theta_0=\inf_{\xi\in\R}\mu^{N}(\xi),\]
where $\mu^{N}(\xi)$ is the bottom of the spectrum $\sigma ( \mathfrak h^{N}[\xi])$ of $\mathfrak h^{N}[\xi]$, for $\xi\in\R$.

Back to our discontinuous field case with \emph{Dirichlet} boundary condition, we prove that our eigenstates are localized near the magnetic edge $\Gamma$,  and more particularly,  near the point with maximum curvature of this edge (see Section~\ref{sec:local}). Analogously to the aforementioned uniform field/Neumann condition situation, our study near $\Gamma$ involves the family of 1D effective operators $(\mathfrak h_a[\xi])_{\xi\in \mathbb R}$ which act in the normal direction to the edge $\Gamma$, along with the associated effective energy $\beta_a$.

At this stage, it is natural to discuss our problem when the Dirichlet boundary conditions are replaced by  Neumann boundary ones. In this situation, one can prove the concentration of the eigenstates of the operator $\mathcal P_h$ near the points of intersection between the edge $\Gamma$ and the boundary $\partial \Om$. This was shown in~\cite{Assaad3} at least for the lowest eigenstate (see Theorem~6.1 in this reference). In such  settings, a geometric condition is usually imposed related to the angles formed at the intersection $\Gamma\cap\partial \Om$ (see~\cite[Assumption~1.3 and Remark~1.4]{Assaad3}). The localization of the eigenstates near $\Gamma\cap\partial\Om$ will involve effective models that are genuinely 2D, i.e. they can not be fibered to 1D operators (see~\cite[Section~3]{Assaad3}). Studying this case may show similarity features with the case of piece-wise smooth bounded domains with corners submitted to uniform magnetic fields, treated  in~\cite{bonnaillie2006asymptotics} (see also~\cite{bonnaillie2007computations, bonnaillie2007superconductivity,bonnaillie2005fundamental,bonnaillie2003analyse} for studies on corner domains). Such similarities were first revealed in~\cite{Assaad3} (see Subsection~1.3 in this reference).  More precisely, one expects the result in the discontinuous field/Neumann condition situation to be similar to that in~\cite[Theorem~7.1]{bonnaillie2006asymptotics}.  Such a result is worth to be established in a future work.\\

Theorem~\ref{thm:main} permits to deduce the splitting between the ground-state energy (lowest eigenvalue) and the energy of the first excited state of $\mathcal P_h$.   More precisely, introducing the spectral gap
\[\Delta(h):=\lambda_2(h)-\lambda_1(h),\]
we get by Theorem~\ref{thm:main}:
\begin{cor}
	Under the conditions in Theorem~\ref{thm:main}, we have as $h\rightarrow 0$
	\[\Delta(h)=h^\frac 74\sqrt{2k_2M_3(a)c_2(a)}+\mathcal O(h^{\frac{15}8}).\]
\end{cor}
 Apart from its own interest,   estimating   the foregoing spectral gap has potential applications  in   non-linear bifurcation problems, for instance,    in the context of the  
Ginzburg-Landau model of superconductivity (cf. \cite[Sec.~13.5.1]{fournais2010spectral}).

\begin{rem}  Altering the regularity/geometry  of the  edge $\Gamma $ may lead to radical changes in Theorem~\ref{thm:main}. 
\begin{itemize}
\item  If  $\Gamma$ is  a piecewise smooth  curve (a broken edge) then we have to analyze a new model in the full plane (reminiscent of  a model in \cite{Assaad3}).   We expect  analogies with  domains with corners in a uniform magnetic field  \cite{bonnaillie2003analyse}.
 \item  If we relax Assumption~\ref{kmax} by allowing the curvature $k$ to  have two symmetric maxima,   then a tunnel effect may occur and the splitting in Theorem~\ref{thm:main} becomes of exponential order.   This is recently analyzed in \cite{FHK} based on  the analysis of this paper  and the recent work \cite{BHR21}.
 \item If the curvature along $\Gamma$ or a part  of  $\Gamma$  is constant,  then  we expect that the magnitude of the splitting in Theorem~\ref{thm:main} will change too,  probably  leading to  multiple eigenvalues.  It would be desirable to get  accurate estimates in this setting.  We expect analogies with disc domains in a non-uniform magnetic field \cite{FP}. 
 \end{itemize}
\end{rem}

\subsubsection*{Heuristics  of the proofs}

 Our proof of Theorem~\ref{thm:main} is purely variational. The derivation of the eigenvalue upper bound is rather standard. It is obtained  by computing the energy of a well chosen  trial state, $v^{\rm  app}_{h,n}$, constructed by expressing the operator  in a Frenet  frame near  the point of maximum curvature and doing  WKB like expansions (for the operator and the trial state). 

Proving the eigenvalue lower bound is  more involved. The idea is to project the actual bound state, $v_{h,n}$, on the trial state $v_{h,n}^{\rm app}$, and to prove that  this provides us with a well chosen trial state  for a 1D effective operator, $H_{a}^{\rm harm}=-c_2(a)\partial_\sigma^2-\frac{k_2M_3(a)}{2}\sigma^2$. To validate this method, we need sharp estimates of the  tangential derivative  of the  actual bound state, which we derive via a simple, but lengthy and quite technical method  involving Agmon estimates and other implementations from  1D model operators. At this stage, one advantage of our approach seems  its applicability  with weaker regularity assumptions on the magnetic edge or the magnetic field, which could be useful in other situations as well, like the study of the 3D problem in \cite{HM3d}. 
 
 \subsubsection*{Outline of the paper}
 
 The paper is organized as follows. Sections \ref{sec:fo} and \ref{sec:step2} contain the necessary material on the  model 1D problems, for flat and curved magnetic edges, respectively. Section~\ref{sec:up} is devoted to the eigenvalue upper bounds matching with the asymptotics of Theorem~\ref{thm:main}. Here, we give the  construction of the aforementioned trial state  $v_{h,n}^{\rm app}$.
 
 In Sections~\ref{sec:loc-fct} and  \ref{sec:local},  we estimate the tangential derivative of the actual bound states, after being truncated  and properly expressed in rescaled variables. The tangential derivative estimate of the $L^2$ norm will  follow straightforwardly from the main result of Section~\ref{sec:loc-fct}. However, a higher regularity    estimate will  require additional  work in  Section \ref{sec:local}.
 
In Section~\ref{sec:lb}, using the actual bound states, we  construct  trial  states for the effective 1D operator, and eventually  prove the eigenvalue lower bounds of Theorem~\ref{thm:main}. Finally, we  give two appendices, Appendix~\ref{sec:frenet} on the Frenet coordinates near the magnetic edge, and Appendix~\ref{sec:app-r} on the control of a remainder term that we meet in Section~\ref{sec:lb}. 

\section{Fiber operators}\label{sec:fo}

\subsection{Band functions}\label{sec:step1}

Let  $a \in [-1,0)$. We first introduce some constants whose definition involves the following family of  fiber  operators in $L^2(\R)$
\begin{equation}\label{eq:ha}
\mathfrak h_a[\xi]=-\frac{d^2}{d\tau^2}+V_a(\xi,\tau),
\end{equation}
where $\xi\in\R$ is a parameter, 
\begin{equation}\label{eq:potential}
V_a(\xi,\tau)=\big(\xi+b_a(\tau)\tau\big)^2\,,\, b_a(\tau)=\mathbf{1}_{\R_+}(\tau)+a\mathbf{1}_{\R_-}(\tau),
\end{equation}
and the domain of $\mathfrak h_a[\xi]$ is given by~:
$${\rm Dom}\big(\mathfrak h_a[\xi]\big)= B^2(\mathbb R)\,.$$
Here the space $B^n(I)$  is  defined for a positive integer $n$ and an open interval $I\subset \R$ as follows
\begin{equation} \label{eq:B_n}
B^n(I)=\{u \in L^2(I):\,\tau^i
\frac{d^ju}{d\tau^j} \in L^2(I),\,\forall i, j \in \N\ \mathrm{s.t.}\ i+j\leq n \}.
\end{equation} 
The operator $\mathfrak h_a[\xi]$ is  essentially  self-adjoint  and has compact resolvent.  Actually, it can also be defined as the Friedrichs realization starting from the closed quadratic form 
\begin{equation} \label{eq:quad}
u \mapsto q_a[\xi](u)=\int_{ \R} \big(|u'(\tau)|^2+V_a(\xi,\tau)|u(\tau )|^2 \big)\,d\tau
\end{equation}
defined on $B^1(\R)$. \\
For $(a,\xi) \in [-1,0)\times \mathbb R$,  the ground-state energy (bottom of the spectrum)  $\mu_a(\xi)$  of $\mathfrak h_a[\xi]$  can be characterized by
\begin{equation}\label{mu_a_1}
\mu_a(\xi)=\inf_{u\in B^1(\R),u\neq0} \frac{q_a[\xi](u)}{\|u\|^2_{L^2(\R)}}\,,
\end{equation}
and $\xi \mapsto \mu_a(\xi)$ will be called the band function.\\
We then introduce the \emph{step constant} at $a$ by
\begin{equation}\label{eq:beta}
\beta_a:=\inf_{\xi \in \R} \mu_a(\xi)\,.
\end{equation}
For $a=-1$, it is easy to identify by symmetrization $\mu_{-1} (\xi)$ with the ground-state energy of the Neumann realization of $-\frac{d^2}{d\tau^2}+ (\tau +\xi)^2$ in $\mathbb R_+$ and therefore
\begin{equation}\label{eq:deG}
\beta_{-1} = \Theta_0\,,
\end{equation}
where $\Theta_0$ is the celebrated de Gennes constant. \\
 By the general theory for the Schr\"odinger operator, $\mu_a(\xi)$ is, for each $\xi \in \mathbb R$, a simple eigenvalue, that we associate with a unique \emph{positive} $L^2$-normalized eigenfunction denoted by $\varphi_{a,\xi}$, i.e. satisfying 
\begin{equation}\label{eq:phi-a-p}
\varphi_{a,\xi}>0,~(\mathfrak h_a[\xi]-\mu_a(\xi))\varphi_{a,\xi}=0~\mbox{ and }~\int_{\R}|\varphi_{a,\xi}(\tau)|^2\,d\tau=1. 
\end{equation}
By Kato's theory, the band function $\mu_a$ is an analytic function on $\mathbb R$. Its  derivative was computed  in \cite{hislop2015edge} (see also~\cite[Prop. A.4]{Assaad2019}):
\begin{equation}\label{eq:mu_deriv}
\mu_a'(\xi)=\Big(1-\frac1a\Big)\Big ({\varphi'_{a,\xi}(0)}^2+\big(\mu_a(\xi)-\xi^2\big){\varphi_{a,\xi}(0)}^2\Big),
\end{equation}
which results from the following Feynman-Hellmann formula (see~\cite[Eq.~(A.9)]{Assaad2019}, and also~\cite{bolley1993application,dauge1993eigenvalues})
\begin{equation}\label{eq:FH}
\mu_a'(\xi)= 2\int_\R (\xi+b_a(\tau) \tau)|\varphi_{a,\xi}(\tau)|^2\,d\tau\,.
\end{equation} 
\subsection{Properties of band functions/states}\label{sec:p-bnd-fc}
For $a\in(-1,0)$, the following results were recently established in~\cite{assaad2020band,Assaad2019,hislop2016band}. 
\begin{enumerate}
\item $|a|\Theta_0<\beta_a<\min(|a|,\Theta_0)$\,.
\item There exists a unique  $\zeta_a\in\R$ such that
	$\beta_a=\mu_a(\zeta_a)$\,.
\item	$\zeta_a<0$, $\mu_a''(\zeta_a)>0 $ and the ground state  $\phi_a:=\varphi_{a,\zeta_a}$ satisfies
 \[ \phi'_{a}(0)<0 \mbox{ and  } \zeta_a=-\sqrt{\beta_a+(\phi^{'2}_{a}(0)/\phi_{a}^2(0))}\,.\]
\end{enumerate}
 In particular, using \eqref{eq:FH}  for $\xi=\zeta_a$, we observe that the functions $\phi_a$ and $(\zeta_a+b_a(\tau) \tau)\phi_{a}$  are orthogonal
\begin{equation}\label{eq:FH-orth}
\int_\R (\zeta_a+b_a(\tau) \tau)|\phi_{a}(\tau)|^2\,d\tau=0\,.
\end{equation} 
Moreover, the ground-state $\phi_a$ satisfies the following decay estimates 
\begin{proposition}\label{prop:decayfi}
	Let $a\in[-1,0)$. For any $\gamma>0$, there exists a positive constant $C_\gamma$  such that 
	\[\int_\R e^{\gamma|\tau|}(|\phi_a(\tau)|^2+|\phi'_a(\tau)|^2)\,d\tau\leq C_\gamma\,.\]
\end{proposition}
Consequently, for all $n \in \N^*$ there exists $C_n>0$ such that
\begin{equation}\label{eq:finmom}
\int_{\R}|\tau|^n|\phi_a(\tau) |^2\,d\tau\leq C_n\,.
\end{equation}
The proof is classical by using Agmon's approach for proving decay estimates.  We omit it and refer the reader to~\cite[Theorem~7.2.2]{fournais2010spectral}  or to the proof of Lemma~\ref{lem:res} below.
\subsection{Moments}
Later in the paper, we will encounter the following \emph{moments}
\begin{equation}\label{eq:moments}
M_n(a)=\int_{-\infty}^{+\infty}\frac 1{b_a(\tau)}(\zeta_a+b_a(\tau)\tau)^n|\phi_a(\tau)|^2\,d\tau\,, \end{equation}
which are finite according to \eqref{eq:finmom}.\\
For $n\in\{1,2,3\}$, they  were computed in~\cite{assaad2020band} and we have:
	\begin{align}
	M_1(a)&=0\,,\label{eq:m1}\\
	M_2(a)&=-\frac 12 \beta_a\int_{-\infty}^{+\infty}\frac 1{b_a(\tau)}|\phi_a(\tau)|^2\,d\tau+\frac 14\Big(\frac 1a-1\Big)\zeta_a\phi_a(0)\phi_a'(0)\,, \label{eq:m2}\\
	M_3(a)&=\frac 13\Big(\frac 1a-1\Big)\zeta_a\phi_a(0)\phi_a'(0)\,.\label{eq:m3}
	\end{align}
\begin{rem}\label{rem:moment}~
From the properties of the band function recalled in Subsection~\ref{sec:p-bnd-fc}, we get that $M_3(a)$ is negative for $-1<a<0$, and vanishes for $a=-1$.
\end{rem}
\begin{rem}\label{prop:mom}
The next identities follow in a straightforward manner from the foregoing  formulae of the moments:
\[\int_{-\infty}^{+\infty}\tau(\zeta_a+b_a(\tau) \tau)|\phi_a(\tau)|^2\,d\tau=M_2(a),\]
\[\int_{-\infty}^{+\infty}\tau(\zeta_a+b_a(\tau) \tau)^2|\phi_a(\tau)|^2\,d\tau=M_3(a) -\zeta_aM_2(a),\]
\[\int_{-\infty}^{+\infty}b_a(\tau) \tau^2(\zeta_a+b_a(\tau) \tau)|\phi_a(\tau)|^2\,d\tau=M_3(a)-2\zeta_aM_2(a),\]
\[\int_{-\infty}^{+\infty}\tau|\phi_a(\tau)|^2\,d\tau=-\zeta_a\int_{-\infty}^{+\infty}\frac 1{b_a(\tau) }|\phi_a(\tau)|^2\,d\tau,\]
\[\int_{-\infty}^{+\infty}\tau|\phi'_a(\tau)|^2\,d\tau=\beta_a\zeta_a\int_{-\infty}^{+\infty}\frac 1{b_a(\tau) }|\phi_a(\tau)|^2\,d\tau+2M_3(a) -2\zeta_aM_2(a).\]
\end{rem} 
We will also encounter the moment:
\begin{equation}\label{eq:I2}  
I_2(a):=\int_{\R}(\zeta_a+b_a(\tau) \tau)\phi_a \mathfrak R_a[(\zeta_a+b_a(\tau) \tau)\phi_a]\,d\tau,
\end{equation}
involving the resolvent $\mathfrak R_a$, which is an operator  defined on $L^2(\R)$ by means of the following lemma:
\begin{lemma}\label{lem:res}
If  $u \in L^2(\R)$ is orthogonal to  $ \phi_a$, we define $(\mathfrak h_a[\zeta_a]-\beta_a)^{-1}u$ in  $L^2(\R)$ as the  unique solution $v$  orthogonal to  $ \phi_a$ to
	\[(\mathfrak h_a[\zeta_a]-\beta_a)v=u\,.\]
	We introduce the regularized resolvent $\mathfrak R_a$ in $\mathcal L(L^2(\R))$ by
	\begin{equation}\label{eq:R}
	\mathfrak R_a(u)=
	\begin{cases}
	0&\mathrm{if}~u\parallel\phi_a\\
	(\mathfrak h_a[\zeta_a]-\beta_a)^{-1}u&\mathrm{if}~u\perp \phi_a
	\end{cases}
	\end{equation}
	(extended by linearity). Then, for any $\gamma\geq0$, $\mathfrak R_a$  and $\frac{d}{d\tau}  \circ \mathfrak R_a$ are two bounded operators on  $L^2(\R,\exp(\gamma |\tau|)\,d\tau)$.
\end{lemma}
\begin{proof}
We follow Agmon's approach. 
Consider $v\in {\rm Dom}(\mathfrak h_a[\zeta_a])$ and $u \in L^2(\R,\exp(\gamma |\tau|)\,d\tau)$ such that
\[ (\mathfrak h_a[\zeta_a]-\beta_a)v=u\,.\]
For all $\gamma>0$ and $N> 1$, consider the continuous function on $\R$
\[\Phi_{\gamma,N}(\tau)=\min(\gamma|\tau|,N)\,.\]
Observe that $\Phi_{\gamma,N}\in H^1_{\rm loc}(\R)$ and 
 \[|\Phi_{\gamma,N}'(\tau)|=\begin{cases}
\gamma& {\rm if~}\gamma|\tau|< N\\
0&{\rm if~}\gamma |\tau|>N
\end{cases}\,.\]
Integration by parts yields
\begin{align*}
 \langle u,e^{2\Phi_{\gamma,N}}v\rangle &=\langle (\mathfrak h_a[\zeta_a]-\beta_a)v,e^{2\Phi_{\gamma,N}}v\rangle\\
 &=\Big\|\left(e^{\Phi_{\gamma,N}}v\right)'\Big\|^2+\int_{\R}\Big((\zeta_a+b\tau)^2-\beta_a\Big)|e^{\Phi_{\gamma,N}}v|^2d\tau-\|\Phi_{\gamma,N}'e^{\Phi_{\epsilon,N}}v\|^2\\
 &\geq\Big\|\left(e^{\Phi_{\gamma,N}}v\right)'\Big\|^2+\int_{\R}\Big((\zeta_a+b\tau)^2-\beta_a-\gamma^2\Big)|e^{\Phi_{\gamma,N}}v|^2d\tau\,.
\end{align*}
Choose $A_\gamma>1$ so that, for  $|\tau| \geq A_\gamma$, we have $(\zeta_a+b\tau)^2-\beta_a-\gamma^2\geq 1$; consequently, for $N\geq \gamma A_\gamma$,
\[
 \langle u,e^{2\Phi_{\gamma,N}}v\rangle\geq 
\Big\|\left(e^{\Phi_{\gamma,N}}v\right)'\Big\|^2+\int_{\{|\tau|\geq A_\gamma\}}
|e^{\Phi_{\gamma,N}}v|^2d\tau
-(\beta_a +\gamma^2)e^{2\gamma A_\gamma}\|v\|^2\,.
\]
Using the Cauchy-Schwarz inequality, we get further
\[
  \| e^{\Phi_{\gamma,N}} u\| \,\|e^{\Phi_{\gamma,N}}v\| \geq
\Big\|\left(e^{\Phi_{\gamma,N}}v\right)'\Big\|^2+\int_{\{|\tau|\geq A_\gamma\}}
|e^{\Phi_{\gamma,N}}v|^2d\tau
-(\beta_a +\gamma^2)e^{2\gamma A_\gamma}\|v\|^2\,.\]
Rearranging the terms in \eqref{eq:dec-reg-res} and using Cauchy's inequality 
\[\| e^{\Phi_{\gamma,N}} u\| \,\|e^{\Phi_{\gamma,N}}v\|\leq 2\| e^{\Phi_{\gamma,N}} u\|^2 + \frac12\|e^{\Phi_{\gamma,N}}v\|^2\,,\] we get
\[\Big\|\left(e^{\Phi_{\gamma,N}}v\right)'\Big\|^2+\frac12 \int_{\{|\tau|\geq A_\gamma\}}
|e^{\Phi_{\gamma,N}}v|^2d\tau\leq(\beta_a +\gamma^2+1)e^{2\gamma A_\gamma}\|v\|^2 +
2\| e^{\Phi_{\gamma,N}} u\|^2\,. \]
We end up with the following estimate 
\[
\int 
|e^{\Phi_{\gamma,N}}v'|^2d\tau +  \int 
|e^{\Phi_{\gamma,N}}v|^2d\tau\leq C_\gamma \left( \|v\|^2 +
\| e^{\Phi_{\gamma}} u\|^2\right)\,, \]
where we note that the right hand side is independent of $N$. \\
Since $\Phi_{\gamma,N}$ is non negative and  monotone increasing with respect to $N$, we get by monotone convergence that $e^{\Phi_{\gamma}}v$ and $e^{\Phi_{\gamma}}v'$ belong to $L^2(\R)$ and satisfy
\begin{equation}\label{eq:dec-reg-res}
\int 
|e^{\Phi_{\gamma}}v'|^2d\tau +  \int 
|e^{\Phi_{\gamma}}v|^2d\tau\leq C_\gamma \left( \|v\|^2 +
\| e^{\Phi_{\gamma}} u\|^2\right)\,, \end{equation}
where
\[\Phi_\gamma(\tau)= \lim_{N\to+\infty}\Phi_{\gamma,N}(\tau)=\gamma|\tau|\,.\]
To finish the proof, we note that, since the regularized resolvent is bounded and $\Phi_\gamma \geq 0$,
\[ \|v\|^2=\|\mathfrak R_au\|^2\leq \|\mathfrak R_a\|^2\|u\|^2\leq  \|\mathfrak R_a\|^2\|e^{\Phi_\gamma}u\|^2\,.\]
\end{proof}   
\begin{proposition}\label{prop:I2}
For any $a\in (-1,0)$, it holds
\begin{equation}\label{eq:I2a}
 \mu''_a(\zeta_a)=2\big(1-4I_2(a)\big)>0.
\end{equation}
\end{proposition} 
\begin{proof}
	First we notice that $(\zeta_a+b_a(\tau) \tau )\phi_a$ is orthogonal to  $\phi_a$ in $L^2(\R)$ (see \eqref{eq:FH}).  Thus $\mathfrak R_a[(\zeta_a+b_a(\tau) \tau )\phi_a]$ is well defined as $(\mathfrak h_a[\zeta_a]-\beta_a)^{-1}(\zeta_a+b_a(\tau) \tau )\phi_a$. Let $z\in \R$, and $E_a(z)$ be the lowest eigenvalue of the operator $H_a(z)$, defined  on 
		 $L^2(\R)$ as follows
	\[H_a(z):=\mathfrak h_a[\zeta_a+z]=-\frac{d^2}{d\tau^2}+(\zeta_a+z+b_a(\tau) \tau )^2.\]
We adopt the same proof of~\cite[Proposition~A.3]{fournais2006accurate} (replacing $P_0$ by $H_a(0)-\beta_a$ there) to get the identity in \eqref{eq:I2a}.  Finally, by \cite{assaad2020band}, $\mu''(\zeta_a)>0$.
	\end{proof}

\section{1D model involving the curvature}\label{sec:step2}

We consider a new family of fiber operators which are obtained by adding  to the fiber operators in Section~\ref{sec:fo}  new  terms that will be related to the geometry of the magnetic edge. This family was introduced  earlier in \cite{assaad2020band} and their definition  is reminiscent of the weighted operators introduced in  the context of the Neumann  Laplacian with a uniform magnetic field \cite{HM}.

We introduce the following parameters  
\[a\in(-1,0),~\delta\in(0,\frac1{12}),~M>0,~h_0>0\mbox{ and } \kappa  \in [-M,M] \,,\]  that satisfy 
\[M h_0^{\frac12-\delta}<\frac13\,,\]
and will be fixed throughout this section.\\
Consider   on $(-h^{-\delta}, h^{-\delta})$,  the positive function $a_{\kappa,h} (\tau)=(1-\kappa  h^\frac12 \tau)$, the Hilbert space
$L^2\big( (-h^{-\delta},h^{-\delta});a_{\kappa,h} \,d\tau\big)$ with the inner product
\[\langle u,v\rangle=\int_{-h^{-\delta}}^{h^{-\delta}} u(\tau)\overline{v(\tau)}\,(1-\kappa  h^\frac 12\tau)\,d\tau, \]
and for $\xi\in\R$, the following  operator  
\begin{multline}\label{eq:H-beta}
\mathcal H_{a,\xi,\kappa ,h}=-\frac {d^2}{d\tau^2}+(b_a(\tau) \tau +\xi)^2+\kappa  h^\frac 12(1- \kappa h^\frac 12\tau)^{-1}\partial_\tau+2\kappa   h^\frac 12 \tau\left(b_a(\tau) \tau +\xi-\kappa  h^\frac 12 b_a(\tau) \frac {\tau^2}2\right)^2\\-\kappa  h^\frac 12 b_a(\tau) \tau ^2 (b_a(\tau) \tau +\xi)+\kappa ^2 hb_a(\tau) ^2\frac {\tau^4}4,
\end{multline}
where $b_a $ is the function in~\eqref{eq:potential} and 
\begin{equation}\label{eq:dom-1dw}
\dom(\mathcal H_{a,\xi,\kappa ,h})=\{u\in H^2(-h^{-\delta},h^{-\delta})~:~u(\pm h^{-\delta})=0\}.
\end{equation}
The operator $\mathcal H_{a,\xi,\kappa ,h}$ is a self-adjoint operator   in $L^2\big( (-h^{-\delta},h^{-\delta});a_{\kappa,h} \,d\tau\big)$  with compact resolvent. We denote by $\big(\lambda_n(\mathcal H_{a,\xi,\kappa ,h})\big)_{n\geq 1}$ its sequence of min-max eigenvalues. The first eigenvalue can be expressed as follows
\begin{equation}\label{eq:ev-fo-w}
\lambda_1(\mathcal H_{a,\xi,\kappa ,h})= \inf\{q_{a,\xi,\kappa ,h}(u) ~:~u\in H^1_0\big( -h^{-\delta},h^{-\delta}\big) \mbox{ and } \| u\|_{L^2\big( (-h^{-\delta},h^{-\delta});a_{\kappa,h} d\tau\big)}=1\}\,,
\end{equation}
where
\begin{equation}\label{eq:qf-fo-w}
q_{a,\xi,\kappa ,h}(u)=\int_{-h^{-\delta}}^{h^{-\delta}}\Big(|u'(\tau)|^2+(1+2\kappa  h^\frac 12\tau)\Big(b_a(\tau) \tau +\xi-\kappa  h^\frac 12 b_a(\tau) \frac {\tau^2}2\Big)^2u^2(\tau)\Big)(1-\kappa  h^\frac 12\tau)\,d\tau.\end{equation}
 By Cauchy's inequality, we write for any $\varepsilon\in(0,1)$,
\[\Big(b_a(\tau) \tau +\xi-\kappa  h^\frac 12  b_a(\tau) \frac {\tau^2}2\Big)^2\geq (1-\varepsilon)(b_a(\tau) \tau +\xi)^2
-\varepsilon^{-1}\kappa^2  h b_a(\tau)^2 \frac {\tau^4}4\,.\]
Noticing that $h\tau^4\leq h^{1-4\delta}$ for $\tau\in(h^{-\delta},h^{\delta})$ and optimizing with respect to $\varepsilon$, we choose $\varepsilon=h^{\frac12-2\delta}$ and get
\begin{equation}\label{eq:c1}
\Big(b_a(\tau) \tau +\xi-\kappa  h^\frac 12 b_a(\tau) \frac {\tau^2}2\Big)^2\geq (1-h^{\frac12-2\delta})(b_a(\tau) \tau +\xi)^2
-\kappa^2  b_a(\tau)^2 h^{\frac12-2\delta}\,.\end{equation}
We plug ~\eqref{eq:c1} in~\eqref{eq:qf-fo-w} to get, for some $C_0 >0$, 
\begin{equation}\label{eq:qf-fo-w1}
q_{a,\xi,\kappa ,h}(u)\geq (1- C_0h^{\frac12-2\delta}) q_a[\xi](u)-C_0h^{\frac12-2\delta}\|u\|^2_{L^2(-h^{-\delta},h^{-\delta})}\,,
\end{equation}
 where $q_a[\xi]$ is the quadratic form  in \eqref{eq:quad}. The min-max principle ensures that
\begin{equation}\label{eq:qf-fo-w2}
q_a[\xi](u)\geq \beta_a \|u\|^2_{L^2(-h^{-\delta},h^{-\delta})},\quad \mbox{for all}\ u\in H^1_0\big( -h^{-\delta},h^{-\delta}\big)\,.
\end{equation}
Since $\beta_a>0$,~\eqref{eq:qf-fo-w1} and~\eqref{eq:qf-fo-w2} imply 
\begin{equation}\label{eq:min-max-1dw}
q_{a,\xi,\kappa ,h}(u)\geq (1- Ch^{\frac12-2\delta}) q_a[\xi](u)\,,
\end{equation}
with $C=(1+\beta_a^{-1})C_0$. From \eqref{eq:min-max-1dw} and the min-max principle we deduce the lower bounds in Lemma~\ref{lem:lb-wop}  below (see \cite[Subsection~4.2]{assaad2020band} for details).
\begin{lem}\label{lem:lb-wop}
 Given $a\in(-1,0)$, there exist positive  constants $\varepsilon_0(a),\varepsilon_1(a),\varepsilon_2(a),c_0(a), h_0(a), C_0(a)$   such that, for all $h\in (0,h_0(a))$, 
\begin{itemize}
\item For $|\xi-\zeta_a|\geq \varepsilon_0(a)$, we have
\[\lambda_1(\mathcal H_{a,\xi,\kappa ,h}) \geq \beta_a+c_0(a)\,.\]
\item For
$\varepsilon_2(a) h^{\frac14-\delta} \leq |\xi-\zeta_a| \leq \varepsilon_0(a)$, we have
\[\lambda_1(\mathcal H_{a,\xi,\kappa ,h}) \geq \beta_a +\varepsilon_1(a)(\xi-\zeta_a)^2\,.\]
\item For $|\xi-\zeta_a|\leq \varepsilon_2(a)h^{\frac14-\delta}$, we have
\[ \lambda_1(\mathcal H_{a,\xi,\kappa ,h}) \geq \beta_a +c_2(a)|\xi-\zeta_a|^2+\kappa M_3(a)h^{1/2} - C_0(a) \max(h^{1/2}|\xi-\zeta_a|,|\xi-\zeta_a|^3,h) \]
where
\begin{equation}\label{eq:c2(a)}
c_2(a)=\frac12\mu_a''(\zeta_a)>0\,. 
\end{equation}
\end{itemize}
\end{lem}
 We can now state the following:
\begin{proposition}\label{prop:1dw}
 There exists $
\hat c_0(a)>0$ and for all $\varepsilon\in(0,1)$, there exist $C_\epsilon, h_\epsilon>0$ such that, for all $h\in(0,h_\epsilon )$ and $\xi\in\R$, the following inequality holds
\[ \lambda_1(\mathcal H_{a,\xi,\kappa ,h})\geq \beta_a+\hat c_0(a)\min\Big( (\xi-\zeta_a)^2,\varepsilon \Big)+\kappa M_3(a)h^{1/2} - C_\epsilon  h \,.\]
\end{proposition}
\begin{proof}
In the third item of Lemma \ref{lem:lb-wop}, we estimate the remainder term
\[\max(h^{1/2}|\xi-\zeta_a|,|\xi-\zeta_a|^3,h)\leq (\eta^{-1}+1)h+\eta|\xi-\zeta_a|^2+|\xi-\zeta_a|^3\]
for all $\eta\in(0,1)$.  Choosing $\eta=\frac{c_2(a)}{4C_0(a)}$, where $C_0(a)$ is the constant in Lemma~\ref{lem:lb-wop}, we deduce from Lemma~\ref{lem:lb-wop} the  lower bound for the eigenvalue $\lambda_1(\mathcal H_{a,\xi,\kappa ,h})$, with
\[\hat c_0(a)=\frac12\min\Big( \varepsilon_1(a) ,\frac{c_0(a)}{\varepsilon_0(a)^2},c_0(a) \Big)\,.\]
\end{proof}

\section{Upper bound}\label{sec:up}
We establish an upper bound of the $n$'th  eigenvalue $\lambda_n(h)$ of $\mathcal P_h$ which was   defined in~\eqref{eq:P}.  This will involve the spectral value $\beta_a$ introduced   in \eqref{eq:beta}, the moment $M_3(a)<0$ introduced  in \eqref{eq:m3}, and $c_2(a)>0$  the value defined in~\eqref{eq:c2(a)}.  In this section, we consider  two parameters  $\eta\in(0,1/8)$ and $\delta\in (0,1/2)$.
 
\begin{theorem}\label{thm:up}
Let $n\in\N^*$ and $\mathbf a=(1,a)$ with $-1<a<0$.  Under Assumption~\ref{kmax},  there exist $h_0>0$ and $C_0>0$ such that for all $h\in(0,h_0)$, the $n$'th eigenvalue $\lambda_n(h)$ of the operator $\mathcal P_{h}$ defined in~\eqref{eq:P} satisfies
\begin{equation}\label{eq:0}
\lambda_n(h)\leq h\beta_a+h^\frac 32k_{max}M_3(a)+h^\frac 74(2n-1)\sqrt{\frac{k_2M_3(a)c_2(a)}{2}} +C_0 \, h^{\frac{15}8},\end{equation}
where $c_2(a)$ and $M_3(a)$ are introduced in \eqref{eq:main-ct}.
\end{theorem}

\begin{proof} The approach is similar to the one used in the literature in establishing  upper bounds for the low-lying eigenvalues of operators defined on smooth bounded domains, like Schr\"odinger operators with uniform magnetic fields (and Neumann boundary conditions) or the Laplacian (with Robin boundary conditions). For instance, one can see~\cite{bernoff1998onset, fournais2006accurate, helffer2017eigenvalues}.
 The proof relies on the construction of quasi-modes localized near the point of maximal curvature on $\Gamma$.
	
	 Let $h\in(0,1)$. Working near $\Gamma$, we start by  expressing the operator $\mathcal P_{h}$  in the adapted $(s,t)$-coordinates there (see Appendix~\ref{sec:frenet}):
\begin{equation}\label{eq:op-Ph-st}
\tilde{\mathcal P}_{h}=-\mathfrak a^{-1}(h\partial_s-i\tilde{F}_{1})\mathfrak a^{-1}(h\partial_s-i\tilde{F}_{1})-\mathfrak a^{-1}(h\partial_t-i\tilde{F}_{2})\mathfrak a(h\partial_t-i\tilde{F}_{2}),\end{equation}
Recall that we assume that the maximum is attained for $s=0$, hence $k_{max}=k(0)$, and having Lemma~\ref{lem:Anew2}, we perform a global change of gauge $\omega$ such that  the magnetic potential $\Fb$ satisfies in $\Omega$ near the edge $\Gamma$, when expressed in the $(s,t)$ coordinates
	\begin{equation}\label{eq:gauge}\tilde{\Fb}(s,t)=
	\begin{pmatrix}
	- b_a(t)\big(t-\frac {t^2}2 k(s)\big)\\0
	\end{pmatrix}, \end{equation}
where $t\mapsto b_a(t)$ is defined by 
 \begin{equation*}\label{eq:b}
			b_a(t)=\mathbbm 1_{\R_+}(t)+a\mathbbm 1_{\R_-}(t),\quad t\in\R.
			\end{equation*}
Performing  the following change of variables: \[\sigma=h^{-1/8}s \mbox{  and } \tau=h^{-1/2}t\,,\] the operator $\tilde{\mathcal P}_{h}$  becomes in the $(\sigma,\tau)$-coordinates 
\begin{equation}\label{eq:op-sigma-tau}
\check{\mathcal P}_{h}=- \check {\mathfrak a}^{-1}(h^{\frac 78} \partial_\sigma+ih^{\frac 12}b_a(\tau) \tau \check  {\mathfrak a}_2)\check {\mathfrak a}^{-1}(h^{\frac 78} \partial_\sigma+ih^{\frac 12}b_a(\tau) \tau\check {\mathfrak a}_2)-h\check {\mathfrak a}^{-1}\partial_\tau\check{\mathfrak a}\partial_\tau,
\end{equation}
	with 
\begin{equation}\label{eq:a,a2} 
\check {{\mathfrak a}}(\sigma,\tau;h)=1-h^\frac 12\tau k(h^\frac 18\sigma)\quad{\rm and}\quad\check {\mathfrak a}_2(\sigma,\tau;h)=1-h^\frac 12\tau k(h^\frac 18\sigma)/2\,.
\end{equation}
It is convenient to introduce the following operator 
	\begin{equation}\label{eq:P1}
	\mathcal P_{h}^{\rm new}=e^{-i\sigma \zeta_a/h^{\frac 38}}h^{-1}\check{\mathcal P}_{h}e^{i\sigma \zeta_a/h^{\frac 38}}-\beta_a,\end{equation}
	where $\zeta_a$ is introduced in Subsection~\ref{sec:p-bnd-fc} 
	and we  get
	\begin{multline*}
	\mathcal P_{h}^{\rm new}= 
	 -\check{\mathfrak a}^{-1}\partial_\tau\check {\mathfrak a}\partial_\tau-\beta_a\\ \qquad 
	 - \check{\mathfrak a}^{-1}\Big(h^{\frac 38} 
	\partial_\sigma+i(\zeta_a+b_a(\tau) \tau )-ib_a(\tau) \tau (1-\check{ \mathfrak a}_2)\Big){\check{\mathfrak a}}^{-1}\Big(h^{\frac 38} \partial_\sigma+i(\zeta_a+b_a(\tau) \tau ) 
	 -ib_a(\tau) \tau (1-\check {\mathfrak a}_2)\Big)\,.
	\end{multline*}
	Using  the boundedness and the smoothness of $k$, and the fact that $k'(0)=0$ and $k''(0)<0$, we write 
	\begin{align*}
	\check{\mathfrak a}(\sigma,\tau;h)&=1-h^\frac 12 \tau k(0)-h^\frac 34 \tau\sigma^2\frac {k''(0)}{2}+h^{\frac {7}8} e_{1,h}(\sigma,\tau),\\
	\check{\mathfrak a}_2(\sigma,\tau;h)&=1-h^\frac 12 \tau \frac {k(0)}2-h^\frac 34 \tau\sigma^2\frac {k''(0)}{4}+h^{\frac {7}8} e_{2,h}(\sigma,\tau),\\
	\check{\mathfrak a}^{-1}(\sigma,\tau;h)&=1+h^\frac 12 \tau k(0)+h^\frac 34 \tau\sigma^2\frac {k''(0)}{2}+ h^{\frac {7}8} e_{3,h}(\sigma,\tau),\\
	\check{\mathfrak a}^{-2}(\sigma,\tau;h)&=1+2h^\frac 12 \tau k(0)+h^\frac 34 \tau\sigma^2 k''(0)+h^{\frac {7}8} e_{4,h}(\sigma,\tau),
	\end{align*}
where $(e_{i,h})_{i=1,\cdots,4}$ are functions of $\sigma$ and $\tau$   having the property that there exist $C$ and $h_0$ such that\footnote{The following conditions on the length scales of $\tau$ and $\sigma$ (namely that  $\sigma\in (-h^{-\delta},h^{-\delta})$ and $\tau\in (-h^{-\rho},h^{-\rho})$), as well as~\eqref{eq:e} and~\eqref{eq:w2} below are set for a later use in the paper.}, for $h\in (0,h_0)$,  $\sigma\in (-h^{-\eta},h^{-\eta})$ and $\tau\in (-h^{-\rho},h^{-\rho})$ we have, 
\begin{equation}\label{eq:e}
|e_{1,h}(\sigma,\tau)|+|e_{2,h}(\sigma,\tau)|\leq C|\tau\sigma^3|\,,\quad |e_{3,h}(\tau,\sigma)|+|e_{4,h}(\tau,\sigma)|\leq C\big(\sigma^6 +\tau^4+1\big)\,,
\end{equation}	
and 
\begin{equation}\label{eq:w2}
\sum_{i=1}^4\Big(\sum_{j=1}^2\big(|\partial_\tau^j e_{i,h}(\sigma,\tau)|+|\partial_\sigma^j e_{i,h}(\sigma,\tau)|\big)+ |\partial^2_{\sigma\tau} e_{i,h}(\sigma,\tau)|\Big)\leq
 C\big( |\sigma|^{5}+|\tau|^{3}+1\big)\,.
\end{equation}
Hence,
\begin{equation}\label{eq:P2}
\mathcal P_{h}^{\rm new}=P_0+h^{\frac 38}P_1+h^{\frac 12}P_2+h^{\frac 34}P_3+h^{\frac 78}Q_h\,,\end{equation}
where 
\begin{equation}\label{eq:LOT-Ph}
\begin{aligned}
P_0&=-\partial_\tau^2+(\zeta_a+b_a(\tau) \tau )^2-\beta_a,\\
P_1&=-2i(\zeta_a+b_a(\tau) \tau )\partial_\sigma,\\
P_2&=k(0)[2\tau(\zeta_a+b_a(\tau) \tau )^2-b_a(\tau) \tau ^2(\zeta_a+b_a(\tau) \tau )]+k(0)\partial_\tau,\\
P_3&=-\partial^2_{\sigma}+\frac {k''(0)}2\sigma^2[2\tau(\zeta_a+b_a(\tau) \tau )^2-b_a(\tau) \tau ^2(\zeta_a+b_a(\tau) \tau )]+\frac {k''(0)}2\sigma^2\partial_\tau,\end{aligned}
\end{equation}
 and
\begin{equation}\label{eq:Q(h)}
Q_h={\mathcal E}_{1,h}(\sigma,\tau)\partial^2_{\sigma}+{\mathcal E}_{2,h}(\sigma,\tau)\partial_{\sigma}+{\mathcal E}_{3,h}(\sigma,\tau)\partial_{\tau}+{\mathcal E}_{4,h}(\sigma,\tau)\,.
\end{equation}
Here the terms $({\mathcal E}_{i,h})_{i=1,\cdots,4}$ are functions in $\sigma$ and $\tau$ having the property, that there exist $C$ and $h_0$ such that, for $h\in (0,h_0)$,  $\sigma\in (-h^{-\eta},h^{-\eta})$ and $\tau\in (-h^{-\rho},h^{-\rho})$ we have 
\begin{equation}\label{eq:wafaafinal} 
|\mathcal E_{i,h}(\sigma,\tau)|+|\partial_\sigma\mathcal E_{i,h}(\sigma,\tau)|+|\partial_\tau\mathcal E_{i,h}\sigma,\tau)| \leq C\, (|\sigma|^6+|\tau|^6 +1)\,.
\end{equation}
 In what follows, we will construct, for each  $n\in\N^*$, a trial function  $\phi_n \in \dom \mathcal P_{h}^{\rm new}$ satisfying the following
\begin{multline}\label{eq:1}
\Big\|\mathcal P_{h}^{\rm new}\phi_n-\Big(h^\frac 12k_{max}M_3(a)+h^\frac 34(2n-1)\sqrt{\frac{k_2M_3(a)c_2(a)}{2}}\Big)\phi_n\Big\|_{L^2(\R^2,h^{\frac 58} \tilde {\mathfrak a}\, d\sigma d\tau)}\\=\mathcal O(h^{\frac{7}8})\|\phi_n\|_{L^2(\R^2,h^{\frac 58} \tilde {\mathfrak a}\, d\sigma d\tau)},
\end{multline}
(recall $k_2=k''(0)$).

The result in~\eqref{eq:1}, once established,
will imply  by the spectral theorem the existence of an eigenvalue $\lambda_n^{\rm new}(h)$ of $\mathcal P_{h}^{\rm new}$ such that 
\begin{equation}\label{eq:2}
\lambda_n^{\rm new}(h)=h^\frac 12k_{max}M_3(a)+h^\frac 34(2n-1)\sqrt{\frac{k_2M_3(a)c_2(a)}{2}}+\mathcal O(h^{\frac{7}8}).
\end{equation}
 Furthermore, by  the definition of $\mathcal P_{h}^{\rm new}$ in~\eqref{eq:P1} we have:
\[\sigma (\mathcal P_{h})=h\, \sigma (\mathcal P_{h}^{\rm new}).\]
Thus,~\eqref{eq:2} will yield the result in~\eqref{eq:0}. Hence, the discussion above shows that establishing~\eqref{eq:1} is sufficient to complete the proof of the theorem.\\

 We construct the trial functions in the form 
\begin{equation}\label{eq:phin}
\phi_h(\sigma,\tau)=h^{-5/16}\chi(h^\eta\sigma)\chi(h^\rho\tau)g (\sigma,\tau)\,,\end{equation}
where  $\chi$ is a smooth cut-off function supported in $(-1,1)$ and $g= g[h]$ will be determined in $L^2(\mathbb R^2)$ with rapid decay at infinity.
First we set
\begin{equation}\label{eq:g}
g[h]=g_0+h^{\frac 38}g_1+h^{\frac 12}g_2+h^{\frac 34}g_3,\end{equation}
with $g_i\in L^2(\R^2)$ for $i=0,\cdots,3$, and 
\begin{equation}\label{eq:mu}
\mu=\mu(h)=\mu_0+h^{\frac 38}\mu_1+h^{\frac 12}\mu_2+h^{\frac 34}\mu_3
\end{equation}
with $\mu_i\in\R$ for $i=0,\cdots,3$. We will search for  $\mu$ and $g$ satisfying on $\mathbb R^2$
\begin{equation}\label{eq:3} 
(\mathcal P_{h}^{\rm new}-\mu)g =\mathcal O(h^{\frac{7}8})\,.
\end{equation}
More precisely,
using the expansion of $\mathcal P_{h}^{\rm new}$ in~\eqref{eq:P2}, we will search for $\mu_i$ and $g_i$ satisfying the following system of equations:
\begin{equation*}
\begin{cases}
(e_0): (P_0-\mu_0)g_0=0\,,\\
(e_1): (P_0-\mu_0)g_1+(P_1-\mu_1)g_0=0\,,\\
(e_2): (P_0-\mu_0)g_2+(P_2-\mu_2)g_0=0\,,\\
(e_3): (P_0-\mu_0)g_3+(P_1-\mu_1)g_1+(P_3-\mu_3)g_0=0\,.
\end{cases}
\end{equation*}

Let $u_0=\phi_a$ be the positive normalized eigenfunction of the operator $\mathfrak h_a[\zeta_a]$ (in~\eqref{eq:ha}) corresponding to the lowest eigenvalue $\beta_a$.

Obviously, the pair   
\begin{equation}\label{eq:g0}
(\mu_0,g_0)=(0,u_0f)\end{equation} is a solution of $(e_0)$, for any $f\in \mathcal S(\R_\sigma)$. 

We implement this choice of $(\mu_0,g_0)$ in $(e_1)$ and write
\[P_0g_1=-(P_1-\mu_1)g_0=[2i(\zeta_a+b_a(\tau) \tau )\partial_\sigma+\mu_1]u_0 f.\]
Noticing that $(\zeta_a+b_a(\tau) \tau )u_0$ is orthogonal to $ u_0$ in $L^2(\R)$,  $\mathfrak R_a[(\zeta_a+b_a(\tau) \tau )u_0]$ is well defined  with  $\mathfrak R_a$ in~\eqref{eq:R} (see \eqref{eq:FH-orth} and Remark~\ref{rem:moment}), and the pair   
\begin{equation}\label{eq:g1}(\mu_1,g_1)=\big(0,2i\mathfrak R_a[(\zeta_a+b_a(\tau) \tau )u_0]\partial_\sigma f\big)
\end{equation} 
is a solution of $(e_1)$.

Similarly,
\[P_0 g_2=-(P_2-\mu_2)g_0=\big[-k_{max}\big(2\tau(\zeta_a+b_a(\tau) \tau )^2-b_a(\tau) \tau ^2(\zeta_a+b_a(\tau) \tau )\big)+\mu_2\big]u_0f-k_{max}f\partial_\tau u_0.\]
From Remark~\ref{prop:mom}, we observe that $[2\tau(\zeta_a+b_a(\tau) \tau )^2-b_a(\tau) \tau ^2(\zeta_a+b_a(\tau) \tau )-M_3(a)]u_0$ is orthogonal to $ u_0$ in $L^2(\R)$. 
Moreover, the normalization of  $u_0$ in $L^2(\R)$ yields that $\partial_\tau u_0 \perp u_0$. Hence, the pair 
\begin{multline}\label{eq:g2}
(\mu_2,g_2)=\\
\quad \left(k_{max}M_3(a),-k_{max}\mathfrak R_a\Big( [2\tau(\zeta_a+b_a(\tau) \tau )^2-b_a(\tau) \tau ^2(\zeta_a+b_a(\tau) \tau )-M_3(a)]u_0+\partial_\tau u_0\Big)f\right)
\end{multline}
is a solution of Equation $(e_2)$. \\
Finally, we consider Equation $(e_3)$:
\[P_0g_3=-P_1g_1-(P_3-\mu_3)g_0.\]
We will search for $\mu_3$ and $f$ satisfying  
\begin{equation}\label{eq:orth}
\big(P_1 g_1(\sigma,\cdot)+(P_3-\mu_3)g_0(\sigma,\cdot)\big)\perp u_0(\cdot),
\end{equation}for every fixed $\sigma$. This orthogonality result will allow us to choose
\begin{equation}\label{eq:g3}
g_3(\sigma,\cdot)=-\mathfrak R_a[P_1g_1(\sigma,\cdot)+(P_3-\mu_3)g_0(\sigma,\cdot)]\end{equation}
in order to satisfy $(e_3)$.
To that end, the aforementioned choice of $g_0$, $g_1$ and $g_2$ gives for any fixed $\sigma$
\begin{align}
&\langle P_1 g_1(\sigma,\cdot)+(P_3-\mu_3)g_0(\sigma,\cdot)\big),u_0(\cdot)\rangle_{L^2(\R)} \nonumber\\
&\qquad =4\partial_\sigma^2f(\sigma) \int_\R (\zeta_a+b_a(\tau) \tau )u_0\mathfrak R_a[(\zeta_a+b_a(\tau) \tau )u_0]\,d\tau+\frac {k_2}2\sigma^2 f(\sigma) \int_{\R}u_0\partial_\tau u_0\,d\tau \nonumber\\
&\qquad \quad+\int_\R\Big(-\partial^2_{\sigma}f(\sigma)+\frac {k_2}2\sigma^2f(\sigma)[2\tau(\zeta_a+b_a(\tau) \tau )^2-b_a(\tau) \tau ^2(\zeta_a+b_a(\tau) \tau )]-\mu_3f(\sigma)\Big)u_0^2\,d\tau \nonumber\\
&\qquad =-(1-4I_2(a))\partial_\sigma^2f(\sigma)+\frac {k_2 M_3(a)}2\sigma^2 f(\sigma)-\mu_3f(\sigma) \qquad (\mathrm{using}\ \|u_0\|_{L^2(\R)}=1)\nonumber\\
&\qquad =-c_2(a)\partial_\sigma^2f(\sigma)+\frac {k_2 M_3(a)}2\sigma^2 f(\sigma)-\mu_3f(\sigma),\label{eq:mu3}
\end{align}
where $I_2(a)$ is  introduced  in  \eqref{eq:I2} and \eqref{eq:I2a}, and $c_2(a)$ is introduced in \eqref{eq:main-ct}. \\
We consider the harmonic oscillator on $\mathbb R$
\begin{equation}\label{eq:harm}
H_a^\mathrm{harm}:=-c_2(a)\frac{d^2}{d\sigma^2}+\frac 12 k_2M_3(a)\sigma^2\,. \end{equation}
 For each $n\in\N^*$,  let $f_n\in \mathcal S(\R)$ be the n$^{th}$ normalized eigenfunction of $H_a^\mathrm{harm}$ 
 corresponding to the eigenvalue $(2n-1)\sqrt{\frac{k_2M_3(a)c_2(a)}2}$. The choice 
 \begin{equation}\label{eq:f}
 f=f_n\qquad \mathrm{and} \qquad\mu_3=(2n-1)\sqrt{\frac{k_2M_3(a)c_2(a)}2}
 \end{equation} makes the expression in~\eqref{eq:mu3} equal to zero, hence realizing the orthogonality result in~\eqref{eq:orth}.
	
We can now gather the above results.  For each $n\in\N^*$, we choose $\mu$ in~\eqref{eq:mu} and $g=g_{(n)}$ in~\eqref{eq:g} such that $\mu_i$, $g_i$ and $f$ are as in~\eqref{eq:g0}--\eqref{eq:g2},~\eqref{eq:g3} and~\eqref{eq:f}. 

For $h$ sufficiently small, using the properties of $Q_h$ in \eqref{eq:Q(h)}  and~\eqref{eq:wafaafinal},  the fact that $f\in \mathcal S(\R)$, the decay properties of $\phi_a$ in Proposition~\ref{prop:decayfi} and those of the resolvent  $\mathfrak R_a$ in~\eqref{eq:R}, the foregoing choice of $g$ and $\mu$ implies~\eqref{eq:3}.

Now, we consider the trial function (see~\eqref{eq:phin}) associated with $g_{(n)}$. 
 Using again the decay properties of $u_0$ and $f$, and  Lemma~\ref{lem:res} for getting the same properties for the $g_j$, one can neglect the effect of the cut-off functions in the computation  while concluding from~\eqref{eq:3} the desired result
in~\eqref{eq:1}.
We omit further details of the computation, and refer the reader to~\cite[Sections 2\&3]{fournais2006accurate}.
\end{proof}
 \begin{rem}
 	The formal construction of the pairs $(\mu_i,g_i)_{i=0,\cdots,3}$ in the proof of Theorem~\ref{thm:up} can be pushed to any order,  assuming that the curve $\Gamma$ is $C^\infty$ smooth. Using the same approach we can construct pairs  $(\mu_i,g_i)_{i\in\N^*}$  for defining quasimodes yielding an accurate upper bound of the eigenvalue $\lambda_n(h)$, which is an infinite expansion of powers of $h^{\frac 18}$. This upper bound will agree with the one  in  Theorem~\ref{thm:up} up to the order $h^{\frac 74}$ (see~\cite{bernoff1998onset, fournais2006accurate, helffer2017eigenvalues}).
 \end{rem}
 
 \begin{rem}\label{rem:eff-qm}
 In the derivation of the lower bound in Section~\ref{sec:lb}, the operator $H_a^{\rm harm}$ introduced in \eqref{eq:harm} plays the role of an effective operator in the tangential variable. In light of  \eqref{eq:g}, \eqref{eq:g0}, \eqref{eq:g1}, \eqref{eq:g2} and \eqref{eq:f}, the quasi-mode 
 \[ v_{h,n}^{\rm app}=\phi_a(\tau)f_n(\sigma)+2ih^{3/8}\mathfrak R_a\big((\zeta_a+b_a(\tau)\tau)\phi_a(\tau)\big) \partial_\sigma f_n(\sigma)+h^{1/2}g_2(\sigma,\tau)\,,\]
 is  a candidate for the profile of an actual eigenfunction of the operator $\mathcal P_h$,  after rescaling and a gauge transformation.
 \end{rem}

\section{Functions localized near the magnetic edge}\label{sec:loc-fct}

In this section, we consider functions  satisfying the energy bound\,\footnote{This is coherent   with  \eqref{eq:0} if we consider the function a normalized bound state.} in \eqref{eq:hyp-g-qf},   which are  consequently  localized near the maximum of the curvature of the magnetic edge  $\Gamma$. We will be able to estimate the tangential derivative  of such functions.

As we shall see in Subsection \ref{sec:loc-fct*}, bound states and their first order tangential derivatives are examples of the functions we discuss in this section.

\subsection{Localization hypotheses}\label{sec:loc-fct*}
We fix   $t_0>0$ so  that the Frenet  coordinates recalled in Appendix~\ref{sec:frenet} are valid in $\{d(x,\Gamma)<t_0\}$. We recall our  assumption that the curvature of $\Gamma$ attains its maximum at a  unique point defined by the tangential coordinate  $s=0$. 

Let $\theta\in(0,\frac38)$ be a fixed constant. Consider a family of functions   $(g_h)_{h\in(0,h_0]}$ in $ H^1(\Omega)$ for which there exist positive constants  $C_1$, $C_2$ such that for $h\in (0,h_0]$, 
\begin{equation}\label{eq:hyp-g-qf}
\mathcal Q_h(g_{h}) \leq \big(h\beta_a+h^{3/2}M_3(a)k_{\max}+C_1 h^{7/4}\big)\|g_h\|_{L^2(\Omega)}^2+C_2  h^{\frac52-\theta}\,,
\end{equation}
where $\mathcal Q_h$ is the quadratic form introduced in \eqref{eq:qf-MLD}.\\
Suppose also  that there exist   constants $\alpha,C>0$ and a family $(r_h)_{h\in(0,h_0]}\subset\R_+$ such that
\begin{equation}
\limsup_{h\to0_+}r_h<+\infty\,,
\end{equation}
and the following two estimates hold,
\begin{equation}\label{eq:hyp-g-dec}
\int_\Omega\big( |g_h|^2+h^{-1}|(h\nabla-i\Fb)g_h|^2\big)\exp \left(\alpha h^{-1/2}d(x,\Gamma)\right)dx\leq Cr_h\,,
\end{equation}
and
\begin{equation}\label{eq:hyp-g-dec*}
\int_{d(x,\Gamma)\leq t_0} \big(|g_h(x)|^2+h^{-1}|(h\nabla-i\Fb)g_h|^2\big)\exp\left(\alpha h^{-1/8}|s(x)|\right)dx\leq Cr_h\,.
\end{equation}
We can derive  from the decay estimates in \eqref{eq:hyp-g-dec} and \eqref{eq:hyp-g-dec*}  four estimates.\\
 The two first estimates  follow  from the  inequality $e^{z}\geq \frac{z^N}{N!}$ for $z\geq 0$ and read:\\
 For $N\geq 1$, there exist $C_N,h_N>0$ such that, for all $h\in(0,h_N]$, we have 
\begin{equation}\label{eq:dec-g-bnd1}
 A_N(g_h):=\int_\Omega \big(d(x,\Gamma)\big)^N\big(|g_{h}(x)|^2+h^{-1}|(h\nabla-i\Fb)g_h (x) |^2\big)\,dx\leq C_N h^{N/2}r_h\,,
\end{equation} 
and for $\rho\in(0,1/2)$, there exist $C_{N,\rho},h_{N,\rho}>0$ such that, for all $h\in(0,h_{N,\rho}]$,
\begin{equation}\label{eq:dec-g-bnd1*}
 B_N(g_h):= \int_{d(x,\Gamma)\leq h^{\rho}} |s(x)|^N\big(|g_{h}(x)|^2+h^{-1}|(h\nabla-i\Fb)g_h (x) |^2\big)\,dx\leq C_N 
h^{N/8}r_h\,.
\end{equation}  
 
  The two last  estimates imply  that, for a fixed $\rho\in(0,\frac12)$,  and $N\geq 1$, there exist  $C_{N,\rho},h_{N,\rho}>0$ such that, for all $h\in(0,h_{N,\rho}]$, we have, 
\begin{equation}\label{eq:dec-g-bnd2}
\int_{d(x,\Gamma)\geq h^{\rho}} \big(|g_h(x)|^2+h^{-1}|(h\nabla-i\Fb)g_{h}(x) |^2\big)\,dx\leq C_{N,\rho}\, h^{N}r_h\,,
\end{equation}
and for $\eta\in(0,1/8)$, there exist $C_{N,\rho,\eta},h_{N,\rho,\eta}>0$ such that, for all $h\in(0,h_{N,\rho,\eta}]$, we have  
\begin{equation}\label{eq:dec-g-bnd2*}
\int_{\substack{d(x,\Gamma)\leq h^{\rho}\\ |s(x)|\geq h^\eta}} \big(|g_h(x)|^2+h^{-1}|(h\nabla-i\Fb)g_h|^2\big)\,dx\leq C_{N,\rho,\eta}\, h^{N}r_h\,.
\end{equation}

In fact,  \eqref{eq:dec-g-bnd2} and \eqref{eq:dec-g-bnd2*} follow in a straightforward manner  from \eqref{eq:hyp-g-dec} and \eqref{eq:hyp-g-dec*} after noticing that
\[ \int_{d(x,\Gamma)\geq h^{\rho}} \big(|g_h(x)|^2+h^{-1}|(h\nabla-i\Fb)g_{h}|^2\big)\,dx\leq Cr_h \exp (-\alpha h^{\rho -\frac 12} ) \,,\]
and
\[ \int_{\substack{d(x,\Gamma)\leq h^{\rho}\\ |s(x)|\geq h^\eta}}\big(|g_h(x)|^2+h^{-1}|(h\nabla-i\Fb)g_{h}|^2\big)\,dx\leq C r_h\exp (-\alpha h^{\eta -\frac 18} )\,.\]

\subsection{Rescaled functions and tangential estimates}

Let  $\delta\in(0,\frac1{12})$ and  $\eta\in(0,\frac18)$ be two fixed constants. 
Consider the function $w_h$ defined as follows
\begin{equation}\label{eq:g-tranc}
 w_h(\sigma,\tau)=h^{5/16}\chi(h^{\eta}\sigma)\chi(h^{\delta}\tau)\tilde g_h\big(h^{1/8}\sigma,h^{1/2}\tau\big)\,,
\end{equation}
where $\tilde g_h$ is the function assigned to $g_h$ by the Frenet coordinates as in \eqref{eq:u-(s,t)}  namely
\[
\tilde g_h (s,t) = g_h (x) \,, \]
and   $\chi\in C_c^\infty(\R)$, ${\supp}\,\chi\subset[-1,1]$,  $0\leq \chi\leq 1$  and $\chi=1$ on $[-1/2,1/2]$. \\
Note that, due to our conditions on $\delta$ and $\eta$,  $  w_h$  can be seen as a function on $\R^2$, and its $L^2$-norm can be estimated by using \eqref{eq:A_tild2} and \eqref{eq:dec-g-bnd1} as follows
\begin{equation}\label{eq:g-norm}
\|w_h\|^2_{L^2(\R^2)}=\big(1+\mathcal O(h^{1/2})\big)\|g_h\|^2_{L^2(\Om)}\,.
\end{equation}
Under our hypotheses on the function $g_h$ (particularly \eqref{eq:hyp-g-qf} for $\theta\in(0,\frac38)$ and \eqref{eq:hyp-g-dec}-\eqref{eq:hyp-g-dec*}), we can estimate the tangential derivative of the function $w_h$.

\begin{proposition}\label{prop:g-tang-freq} 
For all $\theta\in(0,\frac38)$,  there exist constants $C_\theta,h_\theta>0$ such that, if $h\in(0,h_\theta]$, and
$g_h$ satisfies $(5.1)_\theta$, \eqref{eq:hyp-g-dec} and \eqref{eq:hyp-g-dec*}, then the function $w_h$ introduced in \eqref{eq:g-tranc} satisfies the following estimate
\begin{equation}\label{eq:5.9a}
 \|(h^{3/8}\partial_\sigma-i\zeta_a)  w_{h} \|_{L^2(\R^2)}\leq Ch^{\frac38-\frac\theta2}\Big( \|w_h\|_{L^2(\R^2)}+\sqrt{r_h}+h^{\frac38-\frac{3\theta}4}\Big)\,.  
 \end{equation}
\end{proposition}
\begin{proof}
The proof is split into four steps. \medskip

\indent{\it Step~1.}\medskip

We   localize the integrals defining the $L^2$-norm and the quadratic form of $g_{h}$ to
 the neighborhood, $\mathcal N_h=\{x\in\Om~:~d(x,\Gamma)\leq h^{
 \frac12-\delta},~ |s(x)|\leq h^\eta\}$, of the point of maximal curvature, $s=0$. In fact, by the decay estimates in \eqref{eq:dec-g-bnd2} and \eqref{eq:dec-g-bnd2*},   
 \[
 \|g_h\|_{L^2(\Omega)}^2=\int_{\mathcal N_h}|g_h(x)|^2dx+\mathcal O(h^\infty)\quad{\rm and}\quad  \mathcal Q_h(g_h)=\int_{\mathcal N_h}|(h\nabla-i\Fb)g_h|^2dx+\mathcal O(h^\infty)\,.
 \]
We  refine the localization of these integrals by using the decay estimates in \eqref{eq:dec-g-bnd1} and \eqref{eq:dec-g-bnd1*},   the change of variable formulas in \eqref{eq:A_tild2} and the following expansions
\[k(s)=\kappa+\mathcal O(s^2)\,,\quad \mathfrak a(s,t)=1-t\kappa+\mathcal O(s^2t)\,,\quad \mathfrak a^{-2}=1+2t\kappa+\mathcal  O(s^2t)\,,\]
where we set $\kappa =k_{\max}$. More precisely, 
\begin{equation*}
\|g_h\|_{L^2(\Omega)}^2=\int_{\R}\int_{-h^{\frac12-\delta}}^{h^{\frac12-\delta}}|\tilde g_h|^2(1-t\kappa) dsdt+\int_{\R}\int_{-h^{\frac12-\delta}}^{h^{\frac12-\delta}}\mathcal O(s^2t) |\tilde g_h|^2  dsdt+\mathcal O(h^\infty)\,.
\end{equation*}
To estimate the second term in the right hand side we use the Cauchy-Schwarz inequality  to obtain
\[
\int_{\R}\int_{-h^{\frac12-\delta}}^{h^{\frac12-\delta}}s^2|t|\, |\tilde g_h|^2  dsdt \leq \left(\int_{\R}\int_{-h^{\frac12-\delta}}^{h^{\frac12-\delta}}t^2 |\tilde g_h|^2  dsdt \right)^{\frac 12} 
\left(\int_{\R}\int_{-h^{\frac12-\delta}}^{h^{\frac12-\delta}}s^4 |\tilde g_h|^2  dsdt\right)^{\frac 12} \,.
\]
Hence by \eqref{eq:dec-g-bnd1} (with $N=2$) and  \eqref{eq:dec-g-bnd1*} (with $N=4$) we get
\[
\int_{\R}\int_{-h^{\frac12-\delta}}^{h^{\frac12-\delta}}s^2|t|\, |\tilde g_h(s,t)|^2  dsdt = \mathcal O(h^{3/4})r_h\,.
\]
Implementing the above, we have 
\begin{equation} 
\|g_h\|_{L^2(\Omega)}^2 \leq \int_{\R}\int_{-h^{-\delta}}^{h^{-\delta}}|w_h|^2(1-h^{1/2}\tau\kappa)d\sigma d\tau
  +\mathcal O(h^{3/4 }) r_h  +\mathcal O(h^\infty)\,,
\end{equation}
and
\begin{multline}\label{eq:Qh-gh}
\mathcal Q_h(g_h)=\int_{\R}\int_{-h^{\frac12-\delta}}^{h^{\frac12-\delta}}\left(|h\partial_t \tilde g_h|^2+(1+2\kappa t )\Big|\Big(h\partial_s+i b_a(t)\Big(t  - \frac {\kappa t^2}2\Big)\Big)\tilde g_h\Big|^2\right)(1-\kappa t)\, dsdt\\
+\mathcal O(h^\infty)+\mathcal O(R_h)\,,
\end{multline}
where 
\begin{align*}
R_h&= \int_{\R^2} s^2|t|\left(|h\partial_t \tilde g_h|^2+\Big|\Big(h\partial_s+i b_a(t)\Big(t  - \frac {k(s) t^2}2\Big)\Big)\tilde g_h\Big|^2\right)\, dsdt\\
&\quad+ \int_{\R^2}s^4t^4 |\tilde g_h|^2 dsdt+\left(  \int_{\R^2}s^4t^4 |\tilde g_h|^2 dsdt\right)^{1/2}\|(h\nabla-i\Fb)g_h\|_{L^2(\Omega)}\,.
\end{align*}
Proceeding as above for the treatment of $\int_{\R^2}s^4t^4 |\tilde g_h|^2 dsdt$,  we infer from \eqref{eq:hyp-g-qf}, \eqref{eq:dec-g-bnd1} and \eqref{eq:dec-g-bnd1*} that
 \begin{align*}
R_h&\leq C \Big( \big(A_2(g_h)  B_4(g_h)\big)^{1/2}h+\big(A_8(g_h)B_8(g_h)\big)^{1/2}+ \big(A_8(g_h)B_8(g_h)\big)^{1/4}h^{1/2}\Big) \\
&= \mathcal O(h^{7/4}r_h)\,.
\end{align*}
Now, coming back to  \eqref{eq:hyp-g-qf}, we get  after performing a change of variable and dividing by $h$  
that\,\footnote{Replacing the cut-off functions in \eqref{eq:g-tranc} by $1$ in the integrals produces $\mathcal O(h^\infty)$ errors by \eqref{eq:dec-g-bnd2} and \eqref{eq:dec-g-bnd2*}.}
\begin{multline}\label{eq:qf-u-tranc}
\int_{\R}\int_{-h^{-\delta}}^{h^{-\delta}}\left(|\partial_\tau w_h|^2+(1+2\kappa  h^\frac 12\tau)\Big|\Big(h^{3/8}\partial_\sigma+i \Big(b_a(\tau) \tau  -\kappa  h^\frac 12 b_a(\tau) \frac {\tau^2}2\Big)\Big)w_h\Big|^2\right)(1-\kappa  h^\frac 12\tau)\,d\sigma d\tau\\
\leq \big(\beta_a+h^{1/2}M_3(a)\kappa +\mathcal O(h^{3/4})\big)m_h+\mathcal O(h^{3/4}r_h)+\mathcal O(h^{\frac32-\theta})\,,
\end{multline}
where
\begin{equation}\label{eq:norm-u-w}
 m_h:= \int_{\R}\int_{-h^{-\delta}}^{h^{-\delta}}|w_h|^2(1-\kappa  h^\frac 12\tau)\,d\sigma d\tau=\big(1+o(1)\big)\|w_h\|_{L^2(\R^2)}^2\,.
\end{equation}
 In the sequel, we set
\begin{equation}\label{eq:def-Mh}
M_h=m_h+r_h\,.
\end{equation}

 Next we perform a Fourier transform with respect to $\sigma$  and denote the transform of $w_{h}$  by 
 \[\hat w_{h}(\xi,t)=\frac1{\sqrt{2\pi}}\int_\R w_h(\sigma,t)e^{-i\sigma \xi}d\sigma\,.\]
It results then immediately from 
 \eqref{eq:qf-u-tranc} and \eqref{eq:norm-u-w}  
 the following, 
\begin{multline}\label{eq:qf-u-tranc*}
\int_{\R}\int_{-h^{-\delta}}^{h^{-\delta}}\left(|\partial_\tau \hat w_h|^2+(1+2\kappa  h^\frac 12\tau)\Big|\Big(h^{3/8}\xi+b_a(\tau) \tau  -\kappa  h^\frac 12 b_a(\tau) \frac {\tau^2}2\Big)\hat w_h\Big|^2\right)(1-\kappa  h^\frac 12\tau)\,d\xi d\tau\\
\leq  \big(\beta_a+h^{1/2}M_3(a)\kappa\big)m_h +\mathcal O(h^{3/4}M_h)+\mathcal O(h^{\frac32-\theta})\,,
\end{multline}
and $m_h$ introduced in \eqref{eq:norm-u-w} now satisfies
\begin{equation}\label{eq:norm-u-w*}
m_h=  \int_{\R}\int_{-h^{-\delta}}^{h^{-\delta}}|\hat w_h|^2(1-\kappa  h^\frac 12\tau)\,d\xi d\tau\,.
\end{equation}

{\it Step~2.}

We introduce 
\begin{equation}\label{eq:f-xi-h*}
 f_h(\xi)=q_{a,\zeta,\kappa ,h}(\hat w_h)\big|_{\zeta=h^{3/8}\xi}\,,
\end{equation}
where $q_{a,\zeta,\kappa ,h}$
is the quadratic form introduced in \eqref{eq:qf-fo-w}.
 We rewrite \eqref{eq:qf-u-tranc*} as follows
\begin{equation}\label{eq:int:f-xi-h}
\int_{\R}f_h(\xi)d\xi\leq   \big(\beta_a+h^{1/2}M_3(a)\kappa \big)m_h +\mathcal O(h^{3/4}M_h)+\mathcal O(h^{\frac32-\theta})\,.
\end{equation}
Fix a positive constant $\varepsilon<1$. 
Then by Proposition~\ref{prop:1dw},
\begin{equation}\label{eq:lb-f-h-xi}
f_h(\xi)
\geq \int_{-h^{-\delta}}^{h^{-\delta}}\Big( \beta_a+\hat c_0(a)\min\big((h^{3/8}\xi-\zeta_a)^2,\varepsilon\big)+h^{1/2}M_3(a)\kappa - C_\epsilon h \Big)|\hat w_h|^2(1-h^{1/2}\kappa\tau)d\tau\,.
\end{equation}
Inserting this into \eqref{eq:int:f-xi-h} we get
\[
\int_{\R}\int_{-h^{-\delta}}^{h^{-\delta}}\hat c_0(a)\min\big((h^{3/8}\xi-\zeta_a)^2,\varepsilon\big)|\hat w_h|^2(1-h^{1/2}\kappa\tau)d\xi d\tau=\mathcal O(h^{3/4}M_h)+\mathcal O(h^{\frac32-\theta})\,,\]
from which we infer the following two estimates
\begin{equation}\label{eq:tang-est1}
\int_{|h^{3/8}\xi-\zeta_a|^2<\varepsilon}\int_{-h^{-\delta}}^{h^{-\delta}} |h^{3/8}\xi-\zeta_a|^2|\hat w_{h}|^2(1-h^{1/2}\kappa\tau)d\xi d\tau=\mathcal O(h^{3/4}M_h)+\mathcal O(h^{\frac32-\theta})\,,
\end{equation}
and
\begin{equation}\label{eq:tang-est2}
\int_{|h^{3/8}\xi-\zeta_a|^2\geq \varepsilon}\int_{-h^{-\delta}}^{h^{-\delta}} |\hat w_{h}|^2(1-h^{1/2}\kappa\tau)d\xi d\tau=\mathcal O(h^{3/4}M_h)+\mathcal O(h^{\frac32-\theta})\,.
\end{equation}

{\it Step~3.}

Noticing the simple decomposition
\begin{multline}\label{eq:dec-u}
 \int_{\R}\int_{-h^{-\delta}}^{h^{-\delta}}|\hat w_h|^2(1-h^{1/2}\kappa\tau)d\xi d\tau=
\int_{|h^{3/8}\xi-\zeta_a|^2<\varepsilon}\int_{-h^{-\delta}}^{h^{-\delta}}|\hat w_h|^2(1-h^{1/2}\kappa\tau)d\xi d\tau\\
+ \int_{|h^{3/8}\xi-\zeta_a|^2\geq \varepsilon}\int_{-h^{-\delta}}^{h^{-\delta}}|\hat w_h|^2(1-h^{1/2}\kappa\tau)d\xi d\tau\,,
\end{multline} 
we get from \eqref{eq:tang-est2} and \eqref{eq:norm-u-w*},
\begin{equation}\label{eq:norm-u-w**}
\int_{|h^{3/8}\xi-\zeta_a|^2<\varepsilon}\int_{-h^{-\delta}}^{h^{-\delta}}|\hat w_h|^2(1-h^{1/2}\kappa\tau)d\xi d\tau=m_h+\mathcal O(h^{3/4}M_h)+\mathcal O(h^{\frac32-\theta})\,.
\end{equation}
Similarly, we decompose the integral in \eqref{eq:int:f-xi-h} as follows
\begin{equation}\label{eq:dec-f}
\int_{\R}f_h(\xi)d\xi= \int_{|h^{3/8}\xi-\zeta_a|^2<\varepsilon}f_h(\xi)d\xi+ \int_{|h^{3/8}\xi-\zeta_a|^2\geq \varepsilon}f_h(\xi)d\xi\,.
\end{equation}
We write a lower bound of the integral on $\{|h^{3/8}\xi-\zeta_a|^2\geq \varepsilon\}$  by using \eqref{eq:lb-f-h-xi}. Noting that  $\hat c_0(a)>0$, we get by \eqref{eq:norm-u-w**},
\[
 \int_{|h^{3/8}\xi-\zeta_a|^2<\varepsilon}f_h(\xi)d\xi\geq \big(\beta_a + h^{1/2} M_3(a) \kappa+\mathcal O(h)\big) m_h +\mathcal O(h^{3/4}M_h)+\mathcal O(h^{\frac32-\theta})\,.
\]
Inserting this into \eqref{eq:dec-f} and using  \eqref{eq:int:f-xi-h}, we get
\begin{equation}\label{eq:tang-est3}
\int_{|h^{3/8}\xi-\zeta_a|^2\geq \varepsilon}f_h(\xi)d\xi=\mathcal O(h^{3/4}M_h)+\mathcal O(h^{\frac32-\theta})\,.
\end{equation}
{\it Step~4.}

We write a lower bound for $f_h(\xi)$ by gathering \eqref{eq:f-xi-h*} and \eqref{eq:min-max-1dw} thereby obtaining 
\[ \int_{|h^{3/8}\xi-\zeta_a|^2\geq \varepsilon}f_h(\xi)d\xi \geq (1-Ch^{\frac12-2\delta})
\int_{|h^{3/8}\xi-\zeta_a|^2\geq \varepsilon}\int_{\R}\Big(|\partial_\tau\hat w_h|^2+|(b_a(\tau) \tau +h^{3/8}\xi)\hat w_h|^2\Big)d\xi d\tau\,.\]
Using \eqref{eq:tang-est3} and the inequality (note that $|b_a|\leq 1$ since $|a|<1$)
\[ (b_a(\tau) \tau +h^{3/8}\xi)^2\geq \frac12(h^{3/8}\xi)^2-2\tau^2\,,\]
we get
\begin{equation}\label{eq:est-freq}
  \frac12\int_{|h^{3/8}\xi-\zeta_a|^2\geq \varepsilon}\int_{\R} |h^{3/8}\xi\hat w_{h}|^2 d\xi d\tau 
  \leq 2\int_{|h^{3/8}\xi-\zeta_a|^2\geq \varepsilon}\int_{\R}\tau^2|\hat w_{h}|^2d\xi d\tau +\mathcal O(h^{3/4}M_h)+\mathcal O(h^{\frac32-\theta})\,.
 \end{equation}
Let $p=\frac1\theta$ and $q=\frac{1}{1-\theta}$.  By the H\"older inequality, \eqref{eq:dec-g-bnd1} and \eqref{eq:tang-est2}, we write
\begin{align*}
\int_{|h^{3/8}\xi-\zeta_a|^2\geq \varepsilon}&\int_{\R} \underset{=\tau^2|\hat w_h|^{2\theta}|\hat w_h|^{2-2\theta}}{\underbrace{\tau^2|\hat w_h|^2}}d\xi d\tau\\
&\leq\left(\int_{|\xi_h-\zeta_a|^2\geq \varepsilon}\int_{\R} \tau^{2p}|\hat w_h|^{2p\theta} d\xi d\tau\right)^{1/p}\left(\int_{|\xi_h-\zeta_a|^2\geq \varepsilon}\int_{\R} |\hat w_h|^{q(2-2\theta)} d\xi d\tau\right)^{1/q}\\
&\leq \left(\int_{\R^2} \tau^{2p} |w_h|^2d\tau d s \right)^{1/p} \left(\int_{|\xi_h-\zeta_a|^2\geq \varepsilon}\int_{\R} |\hat w_h|^2 d\xi d\tau\right)^{1/q}\\
&=\mathcal O(h^{\frac34(1-\theta)}M_h)+\mathcal O(M_h^\theta\,  h^{(1-\theta)(\frac32-\theta)})\\
&=\mathcal O(h^{\frac34(1-\theta)}M_h)+\mathcal O(h^{\frac32-\frac{5\theta}{2}})\,,
\end{align*}
where, in the last step, we used Young's inequality, 
\begin{align*} M_h^\theta\, h^{(1-\theta)(\frac32-\theta)}&=M_h\theta h^{\theta(\frac34-\theta)}h^{(1-\theta)(\frac32-\theta)-\theta(\frac34-\theta)}\\
& \leq \theta M_h h^{\frac34-\theta}+(1-\theta)h^{\frac32-\theta}h^{-\frac\theta{1-\theta}(\frac34-\theta)}\\
&\leq\theta M_h h^{\frac34-\theta}+(1-\theta)h^{\frac32-\frac{5\theta}2}~{\rm for~}0<\theta<\frac38\,.
\end{align*}
Inserting this estimate into \eqref{eq:est-freq}, we get
\[
\int_{|h^{3/8}\xi-\zeta_a|^2\geq \varepsilon}\int_{\R} |h^{3/8}\xi\hat w_{h}|^2 d\xi d\tau = \mathcal O(h^{\frac34-\theta}M_h)+\mathcal O(h^{\frac32-\frac{5\theta}2})\,.
\]
Collecting the foregoing estimate and those in \eqref{eq:tang-est1} and \eqref{eq:tang-est2}, we deduce that
\begin{multline*}
\int_{\R^2} |(h^{3/8}\partial_\sigma -i\zeta_a)w_h|^2d\sigma d\tau= \int_\R \int_{-h^{-\delta}}^{h^{-\delta}} |h^{3/8}\xi-\zeta_a|^2|\hat w_h|^2d\xi d\tau\\
= \mathcal O(h^{\frac34-\theta}M_h) +\mathcal O( h^{\frac32-\frac{5\theta}2})\,. 
\end{multline*}
With \eqref{eq:norm-u-w} and \eqref{eq:def-Mh} in mind, 
this implies \eqref{eq:5.9a} as stated in the proposition.
\end{proof}

\section{Localization of bound states}\label{sec:local}

In this section, we fix  a labeling  $n\geq 1$ and denote by $\psi_{h,n}$ a normalized eigenfunction of the operator $\mathcal P_{h}$ with eigenvalue $\lambda_n(h)$.  
  By Theorem~\ref{thm:up}, it holds
\begin{equation}\label{eq:Qh-psi-hn}
\mathcal Q_h(\psi_{h,n})\leq \big(h\beta_a+h^{3/2}M_3(a)k_{\max}+C_1 h^{7/4}\big)\|\psi_{h,n}\|_{L^2(\Omega)}^2\,, 
\end{equation}
where $\mathcal Q_h$ is the quadratic form introduced in \eqref{eq:qf-MLD}. 

The decay estimates in Subsections~\ref{sec:n-est} and \ref{sec:t-est} follow by standard semiclassical Agmon estimates. We refer to \cite{HM, fournais2006accurate} for details in the case of the Laplacian with a smooth magnetic field, and to \cite{assaad2020band} for adaptations in the piecewise constant field discussed here.

 Using the aforementioned decay estimates, the bound state $\psi_{h,n}$  satisfies the hypotheses in Section~\ref{sec:loc-fct}.  Namely the estimates in \eqref{eq:hyp-g-qf}$_\theta$, \eqref{eq:hyp-g-dec} and \eqref{eq:hyp-g-dec*} hold with $g_h=\psi_{h,n}$, $r_h=1$ and for any $\theta\in(0,\frac38)$. Consequently, we will be able to estimate its  tangential derivative (see Proposition~\ref{prop:tang-freq}). Estimating the second  order tangential derivative of $\psi_{h,n}$ (as in Proposition~\ref{rem:tang-freq}) requires the analysis  of the decay of its first order tangential derivative  in order to verify the hypotheses of Section~\ref{sec:loc-fct}.

\subsection{Decay away from the edge}\label{sec:n-est}~

 The derivation of an Agmon decay estimate relies on the following  useful lower bound of the quadratic form \cite[Sec.~4.3]{assaad2020band}. For every $R_0>1$, there exists a positive constant $C_0$  and $h_0>0$ such that,  for $h\in (0,h_0]$,
 \begin{equation}\label{eq:lb-qf-well}
 \mathcal Q_h(u)\geq \int_\Omega \big(U_{h,a}(x)-C_0R_0^{-2}h\big)|u(x)|^2\,dx\quad(u\in H^1_0(\Omega)),\end{equation}
where $\mathcal Q_h$ is  introduced in \eqref{eq:qf-MLD} and
\[U_{h,a}(x)=\begin{cases}
|a|h &{\rm if~}{\rm dist}(x,\Gamma)>R_0 h^{1/2} \\
\beta_a h&{\rm if~}{\rm dist}(x,\Gamma)<R_0 h^{1/2}
\end{cases}\,.
\]
Note that the decay property is a consequence of $\beta_a < |a|$.
Following  \cite[Thm.~8.2.4]{fournais2010spectral}, it results from the foregoing lower bound that the eigenfunction $\psi_{h,n}$ decays  roughly like $ \exp\left(- \alpha_0 h^{-1/2}d(x,\Gamma) \right)$, for some constant $\alpha_0>0$.  More precisely, the following holds
\begin{equation}\label{eq:exp-dec-psi}
\int_\Omega\big( |\psi_{h,n}|^2+h^{-1}|(h\nabla-i\Fb)\psi_{h,n}|^2\big)\exp \left(2\alpha_0 h^{-1/2}d(x,\Gamma)\right)dx\leq C\,.
\end{equation}

\subsection{Decay along the edge}\label{sec:t-est}
Here we discuss tangential estimates along the edge $\Gamma$. Recall that $s=0$ corresponds to  the (unique) point of maximal curvature. 

The starting point is the following refined lower bound of the quadratic form \cite[Sec.~4.3]{assaad2020band}
\begin{equation}\label{eq:qh-u-lb}
\mathcal Q_h(u)\geq \int_\Omega\big(U_{h,a}^\Gamma(x)-C_0  h^{\frac{7}4}\big)\big)|u|^2\,dx\quad(u\in H^1_0(\Omega)),
\end{equation}
where, with  $x=\Phi(s,t)$, 
\[ U^\Gamma_{h,a}(x) =\begin{cases}
|a|h&{\rm if~}{\rm dist}(x,\Gamma)\geq 2h^{\frac16}\\
\beta_ah+M_3(a)\kappa(s)h^{\frac32}&{\rm if}~{\rm dist}(x,\Gamma)<2h^{\frac16}
\end{cases}.\]
Here we recall that $M_3(a)$ is negative so the potential in the second zone is minimal at the point of maximal curvature.
The lower bound \eqref{eq:qh-u-lb}  can be derived along the same arguments in \cite[Prop.~8.3.3, Rem.~8.3.6]{fournais2010spectral} and by using Proposition~\ref{prop:1dw}.

The eigenfunction 
$\psi_{n,h}$ decays exponentially roughly like $ \exp \left(- \alpha_1 h^{-1/8}s(x)\right)$, for some constant $\alpha_1>0$. More precisely, picking $t_0$ sufficiently  small so  that the Frenet  coordinates recalled in Appendix~\ref{sec:frenet} are valid in $\{d(x,\Gamma)<t_0\}$, we have 
\begin{equation}\label{eq:exp-dec-psi*}
 \int_{d(x,\Gamma)\leq t_0} \big(|\psi_{h,n}(x)|^2+h^{-1}|(h\nabla-i\Fb)\psi_{h,n}|^2\big)\exp\left(2\alpha_1 h^{-1/8}|s(x)|\right)dx\leq C\,. 
\end{equation}
\begin{rem}\label{rem:hyp-sec-tf}
 We observe, by collecting \eqref{eq:Qh-psi-hn}, \eqref{eq:exp-dec-psi}  and \eqref{eq:exp-dec-psi*}, that the eigenfunction $g_h=\psi_{h,n}$ satisfies the hypotheses of Proposition~\ref{prop:g-tang-freq}, namely 
\begin{itemize}
\item \eqref{eq:hyp-g-qf}  holds for any $\theta\in(0,\frac38)$;
\item \eqref{eq:hyp-g-dec} and \eqref{eq:hyp-g-dec*} hold with $0<\alpha\leq \min(2\alpha_1,2\alpha_2)$ and $r_h=1$\,.
\end{itemize}
\end{rem}

\subsection{Estimating tangential frequency}

The localization of the eigenfunction $\psi_{h,n}$ is to be measured by two parameters $\rho\in(0,\frac12)$  and  $\eta\in(0,\frac18)$. We will choose $\rho=\frac12-\delta$ with  $\delta\in(0,\frac1{12})$, i.e. we are assuming 
\[  \frac{5}{12}<\rho<\frac12\,. \]
We introduce the following function
\begin{equation}\label{eq:ef-tranc}
 u_{h,n}(\sigma,\tau)=h^{5/16}\chi(h^{\eta}\sigma)\chi(h^{\delta}\tau)\tilde \psi_{h,n}\big(h^{1/8}\sigma,h^{1/2}\tau\big)\,,
\end{equation}
 where $\tilde \psi_{h,n}$ is the function assigned to $\psi_{h,n}$ by the Frenet coordinates as in \eqref{eq:u-(s,t)},  $\chi\in C_c^\infty(\R)$, ${\supp}\,\chi\subset[-1,1]$,  $0\leq \chi\leq 1$  and $\chi=1$ on $[-1/2,1/2]$. Note that $ u_{h,n}$  can be seen as a function on $\R^2$,  and by \eqref{eq:g-norm} (applied with $g_h= \psi_{h,n}$), its $L^2$-norm satisfies
\begin{equation}\label{eq:norm-u}
\| u_{h,n}\|^2_{L^2(\R^2)}=\|\psi_{h,n}\|^2_{L^2(\Om)}(1+\mathcal O(h^{1/2}))=1+\mathcal O(h^{1/2})\,,
\end{equation}
since $\psi_{h,n}$ is normalized in $L^2(\Omega)$.

Using Proposition~\ref{prop:g-tang-freq}, we can estimate the tangential derivative of $ u_{h,n}$. More precisely, we apply this proposition with  $g_h=\psi_{h,n}$, $r_h=1$ and \emph{any} $0<\theta<\frac38$ (see Remark~\ref{rem:hyp-sec-tf}).  In this  case, the function 
  introduced in \eqref{eq:g-tranc} is given by $w_h= u_{h,n}$. 
\begin{proposition}\label{prop:tang-freq}
For all $\theta\in(0,\frac38)$, there exist constants $C_\theta,h_\theta>0$ such that, for all $h\in(0,h_\theta]$, 
\[ \|(h^{3/8}\partial_\sigma-i\zeta_a) u_{h,n} \|_{L^2(\R^2)}\leq C_\theta \, h^{\frac38-\theta}\,.  \]
\end{proposition}

We can estimate higher order tangential derivatives of $u_{h,n}$. 

\begin{proposition}\label{rem:tang-freq}
For all $\vartheta\in(0,\frac34)$, there exist constants $C_\vartheta,h_\vartheta>0$ such that, for all $h\in(0,h_\vartheta]$, 
\begin{equation}\label{eq:tang-freq''}
\|(h^{3/8}\partial_\sigma-i\zeta_a)^2 u_{h,n}\|_{L^2(\R^2)}\leq  C_{\vartheta} \, h^{\frac{3}4-\vartheta}\,,
\end{equation}
where $ u_{h,n}$ is introduced in \eqref{eq:ef-tranc}.
\end{proposition}
 
Before proceeding with the proof of Proposition~\ref{rem:tang-freq}, we introduce  the notation, \break $r_h=\tO(h^\gamma)$ for a positive number $\gamma$, to mean the following
\begin{equation}\label{eq:not-tO}
\forall\,\vartheta\in(0,\gamma),\,\exists\,C_\vartheta,h_\vartheta>0,\,\forall h\in(0,h_\vartheta),~|r_h|\leq C_\vartheta h^{\gamma-\vartheta}\,.
\end{equation}
\begin{proof}[Proof of Proposition~\ref{rem:tang-freq}]
We will apply Proposition~\ref{prop:g-tang-freq} with an adequate choice of the function $g_h$ defining the function $w_h$ in \eqref{eq:g-tranc}.

We  introduce the function $\varphi_h$ on $\Omega$ as follows
\begin{equation}\label{eq:defvarphi}
\varphi_h(x)=f(x)\psi_{h,n}(x)\,, 
\end{equation}
where $f(x)= (1-\chi\big({\rm dist}(x,\partial\Omega)/t_1\big))\, \chi\big({\rm dist}(x,\Gamma)/t_0\big)$, $t_1$ and $t_0$  are constants so that the set\break$\{x\in\Omega~:~{\rm dist}(x,\partial\Omega)>t_1\}$   contains the point of maximum curvature and  the transformation in \eqref{Frenet} is a diffeomorphism, $\chi\in C_c^\infty(\R)$, ${\supp}\,\chi\subset[-1,1]$,  $0\leq \chi\leq 1$  and $\chi=1$ on $[-1/2,1/2]$. 
Then we
define 
\begin{equation}\label{eq:def-gh-td}
\tilde g_h(s,t)=(h^{1/2}\partial_s-i\zeta_a)\tilde \varphi_h(s,t)\,,
\end{equation}
where $\tilde \varphi_h$ is the function assigned to $\varphi_h$ by  \eqref{eq:u-(s,t)}.
Notice that, using the notation in~\eqref{eq:not-tO},  the conclusion of Proposition~\ref{prop:tang-freq} can be written as
\begin{equation}\label{eq:norm-gh}
\|g_h\|_{L^2(\Omega)}=\tO(h^{3/8})\,.
\end{equation}
We will show that $g_h$  satisfies \eqref{eq:hyp-g-qf}$_\theta$  for \emph{any} $\theta\in(0,\frac38)$, and that \eqref{eq:hyp-g-dec} and \eqref{eq:hyp-g-dec*} hold with
\begin{equation}\label{eq:def-rh-dec} 
r_h=\|g_h\|_{L^2(\Omega)}^2+h^{3/4}\,. 
\end{equation} 
This will be done in  several steps  outlined below.
\begin{itemize}
\item In Step~1, we establish rough decay estimates for $g_h$  in the normal and tangential directions (see \eqref{eq:exp-dec-g}). These estimates  are nevertheless weaker than the estimates in \eqref{eq:hyp-g-dec} and \eqref{eq:hyp-g-dec*} that  we wish to prove.
\item In Step~2, we show that $g_h$ is in the domain of  the operator $\mathcal P_h$ introduced in  \eqref{eq:P}.
\item In  Step~3, using the rough estimates obtained in  Steps~1 and ~2, we can verify that \eqref{eq:hyp-g-qf} holds for any $\theta\in(0,\frac38)$.
\item  In Step~4,  using the estimates obtained in Steps~1 and~3,  and the Agmon method, we derive the decay estimates for $g_h$ as in \eqref{eq:hyp-g-dec} and  \eqref{eq:hyp-g-dec*} with $r_h$ given in \eqref{eq:def-rh-dec}.
\item In Step~5, we can apply the conclusion of Proposition~\ref{prop:g-tang-freq} and conclude the proof of Proposition~\ref{rem:tang-freq}.
\end{itemize}

{\it Step~1.}

We show that the function $g_h$ decays exponentially in the normal and tangential directions. We select  the constant $t_0$  so that the two functions
\[x\mapsto {\rm dist}(x,\Gamma)\quad{\rm and}\quad x\mapsto s(x) \]
are smooth in the neighborhood,  $\Gamma_{2t_0}$,  of the edge $\Gamma$. Consequently,   the transformation  in \eqref{Frenet} is valid in $\Gamma_{2t_0}$. 
 Since we 
encounter integrals of  the function $g_h$, which  is supported in $\Gamma_{t_0}\cap\Omega$,  we select the gauge given in Lemma~\ref{lem:Anew2}.  In particular, by \eqref{eq:A_tild1}, we have 
\begin{equation}\label{eq:6.17a}
 |\Fb(x)|=\mathcal O({\rm dist}(x,\Gamma)) ~{\rm on~}\Omega\cap\Gamma_{t_0}\,.
 \end{equation}
 Let $\alpha_2 \in (0,\frac12\min(\alpha_0,\alpha_1))$, where $\alpha_0,\alpha_1$ are the positive constants in \eqref{eq:exp-dec-psi} and \eqref{eq:exp-dec-psi*}. We introduce on $\Omega$ the weight functions 
\begin{equation}\label{eq:weights}
\Phi_{\rm norm}(x)=\exp\left(\frac{\alpha_2\,{\rm dist}(x,\Gamma)}{h^{1/2}} \right)~{\rm and}~\Phi_{\rm tan}(x)=\exp\left(\frac{\alpha_2\, s(x)}{h^{1/8}} \right)\,.
\end{equation}
By Remark~\ref{rem:hyp-sec-tf}, we can use \eqref{eq:dec-g-bnd1} for $\psi_{h,n}$.  It results  from   \eqref{eq:exp-dec-psi*}, \eqref{eq:6.17a}, the H\"older inequality, and our choice of $\alpha_2$, that, for $j\in \{1,2\}$, 
\begin{align}\label{eq:norm:F2psi}
~\int_{\Omega} |\Fb|^{2j}|\psi_{h,n}|^2\Phi_{\rm tan}^2dx&= 
\int_{\Omega\cap\Gamma_{t_0}} |\Fb|^{2j}|\psi_{h,n}|^2\Phi_{\rm tan}^2dx+\mathcal O(h^\infty)\nonumber\\
& \leq A_{4j}( \psi_{h,n})^{1/2}  \big\| \Phi_{\rm tan}^2\psi_{h,n}\big\|_{L^2(\Omega)}+\mathcal O(h^\infty)=\mathcal O(h^j)\,,
\end{align}
where $A_{4j}(\cdot)$ is defined in~\eqref{eq:dec-g-bnd1} and
\begin{align*}
\int_{\Omega} |\Fb\cdot (h\nabla-i\Fb)\psi_{h,n}|^2\Phi_{\rm tan}^2dx&= 
\int_{\Omega\cap\Gamma_{t_0}} |\Fb\cdot (h\nabla-i\Fb)\psi_{h,n}|^2\Phi_{\rm tan}^2dx+\mathcal O(h^\infty)\\
&\leq  A_4(\psi_{h,n})^{1/2} \big\| \Phi_{\rm tan}^2(h\nabla-i\Fb)\psi_{h,n}\big\|_{L^2(\Omega)}+\mathcal O(h^\infty)\\
&=\mathcal O(h^2)\,.
\end{align*} 
In a similar fashion, we estimate the $L^2(\Omega)$-norms of $\Fb\psi_{h,n}\Phi_{\rm norm}$, $(\Fb\cdot\Fb) \psi_{h,n}\Phi_{\rm norm}$ and $\Phi_{\rm norm}\Fb\cdot(h\nabla-i\Fb)\psi_{h,n}$ using \eqref{eq:exp-dec-psi}. Eventually, we get the following estimates
\begin{equation}\label{eq:exp-dec-F-psi}
\begin{aligned}
&\Big \|\Fb \psi_{h,n} \Phi_{\rm norm}\Big\|_{L^2(\Omega\cap\Gamma_{2t_0};\R^2)}+\Big\|\Fb \psi_{h,n} \Phi_{\rm tan}\Big\|_{L^2(\Omega\cap\Gamma_{2t_0};\R^2)}\leq Ch^{1/2}\\
&\Big \|\Fb\cdot\nabla\Big( \psi_{h,n}\Phi_{\rm norm}\Big)\Big\|_{L^2(\Omega\cap\Gamma_{2t_0};\R^2)}+\Big\|\Fb \cdot\nabla\Big(\psi_{h,n}\Phi_{\rm tan}\Big)\Big\|_{L^2(\Omega\cap\Gamma_{2t_0})}\leq C
\end{aligned}\,.
 \end{equation}
Furthermore, the following two estimates hold
\begin{equation}\label{eq:exp-dec-H1}
\begin{aligned}
&\big\|\psi_{h,n}\Phi_{\rm norm}\Big\|_{L^2(\Omega\cap\Gamma_{2t_0})}+\Big\|\psi_{h,n}\Phi_{\rm tan}\big\|_{L^2(\Omega\cap\Gamma_{2t_0})}\leq C\\
&\big\|\psi_{h,n}\Phi_{\rm norm}\big\|_{H^1(\Omega\cap\Gamma_{2t_0})}+\big\|\psi_{h,n}\Phi_{\rm tan}\big\|_{H^1(\Omega\cap\Gamma_{2t_0})}\leq Ch^{-1/2}
\end{aligned}
\,.
\end{equation}
Notice that for  $w_\#:=\psi_{h,n}\Phi_\#$, ($\#\in\{\rm norm,tan\}$), we have, with $\mathcal P_h$ the operator introduced in \eqref{eq:P}
\[\mathcal P_hw_\#=\lambda_n(h) w_\#-2h\nabla\Phi_{\#}\cdot(h\nabla-i\Fb)\psi_{h,n}-h^2\Delta\Phi_{\#}\,\psi_{h,n}.\]
Hence, noting that $\mathcal P_h = -h^2\Delta +2ih\Fb\cdot\nabla +ih\,{\rm div}\Fb +|\Fb|^2$,  we find by \eqref{eq:0},~\eqref{eq:norm:F2psi}  and \eqref{eq:exp-dec-F-psi},
\begin{multline*}
h^2\|\Delta w_\#\|_{L^2(\Omega\cap\Gamma_{2t_0})} \leq \Big(\| \mathcal P_hw_\#\|_{L^2(\Omega)} +\|(h\nabla-i\Fb)w_\#\|_{L^2(\Omega\cap\Gamma_{2t_0})}+ h\|{\rm div}\,\Fb \, w_\#\|_{L^2(\Omega\cap\Gamma_{2t_0})}\\+2h\|\Fb\cdot\nabla w_\#\|_{L^2(\Omega\cap\Gamma_{t_0})}+\big\| |\Fb|^2 w_\#\big\|_{L^2(\Omega\cap\Gamma_{2t_0})}\Big)=\mathcal O(h)\,.
\end{multline*}
By the  $L^2$-elliptic estimates for the Dirichlet problem in $\Gamma_{2t_0}\cap\Omega$,  and noting that $w_\#$ satisfies the Dirichlet condition, 
\[ \| w_\#\|_{H^2(\Omega\cap\Gamma_{t_0})}\leq C(t_0,\Omega)\big(\|\Delta w_\#\|_{L^2(\Omega\cap\Gamma_{2t_0})}+\| w_\#\|_{L^2(\Omega\cap\Gamma_{2t_0})}\big)\,.\]
Consequently, we get the following estimate
\begin{equation}\label{eq:exp-dec-H2}
\big\|\psi_{h,n} \Phi_{\rm norm}\big\|_{H^2(\Omega\cap\Gamma_{t_0})}+\big\|\psi_{h,n} \Phi_{\rm tan}\big\|_{H^2(\Omega\cap\Gamma_{t_0})}\leq Ch^{-1}\,.
\end{equation}
 Now we can derive decay estimates of the function $g_h$ introduced in  \eqref{eq:def-gh-td}.  Controlling the decay of the magnetic  gradient of $g_h$  requires a decay estimate of  $\psi_{h,n}$ in the $H^2$ norm. Actually, collecting \eqref{eq:exp-dec-H1} and \eqref{eq:exp-dec-H2}, 
we observe that 
\begin{equation}\label{eq:exp-dec-g}
\begin{aligned}
&\big \|g_h \Phi_{\rm norm}\big\|_{L^2(\Gamma_{t_0})}+h^{-1/2}\big\|\big((h\nabla-i\Fb)g_h\big) \Phi_{\rm norm} \big)\big\|_{L^2(\Gamma_{t_0};\R^2)}\leq C\\
&\big \|g_h \Phi_{\rm tan} \big\|_{L^2(\Gamma_{t_0})}+h^{-1/2}\big\|\big((h\nabla-i\Fb)g_h\big) \Phi_{\rm tan} \big\|_{L^2(\Gamma_{t_0};\R^2)}\leq C
\end{aligned}\,.
 \end{equation}

{\it Step~2.} 
 By the definition of  $g_h$ in~\eqref{eq:def-gh-td}, this function is compactly  supported in $\Om\cap\Gamma_{t_0}$. Hence, there exists a regular  open set $\omega$ such that,   
 for $h\in (0,h_0]$,  $\supp g_h \subset \omega \subset \bar \omega \subset \Om\cap\Gamma_{2t_0}$. Consequently $g_h$ satisfies the Dirichlet boundary condition on $\partial\omega$. To prove that $g_h$ is in the domain of the operator $\mathcal P_h$, it suffices to establish that
\begin{equation}\label{eq:R1}
\partial_s\tilde\psi_{h,n}\in H^2\Big(\Phi^{-1}\big(\omega \big)\Big)\,.
\end{equation}
To that end, we consider the spectral equation satisfied by the eigenfunction $\psi_{h,n}$
\begin{equation}\label{eq:R2}
-(h\nabla-i\Fb)^2\psi_{h,n}=\lambda_n(h)\psi_{h,n}\,.
\end{equation}
Using~\eqref{eq:op-FC} with the potential $\tilde\Fb$ in~\eqref{eq:gauge},~\eqref{eq:R2} reads in the $(s,t)$-coordinates as follows
\begin{equation}\label{eq:R3}
-\left(\mathfrak a^{-1}(h\partial_s-i\tilde{F}_{1})\mathfrak a^{-1}(h\partial_s-i\tilde{F}_{1})+h^2\mathfrak a^{-1}\partial_t\mathfrak a\partial_t\right)\tilde\psi_{h,n}=\lambda_n(h)\tilde\psi_{h,n}\,,
\end{equation}
that is
\begin{equation}\label{eq:R4}
h^2(\mathfrak a^{-2}\partial^2_s\tilde\psi_{h,n}+\partial^2_t\tilde\psi_{h,n})=f_1(s,t)\partial_s\tilde\psi_{h,n}
+f_2(s,t)\partial_t\tilde\psi_{h,n}
+f_3(s,t)\tilde\psi_{h,n}\,,
\end{equation}
where
\begin{align*}
f_1(s,t)&=-h^2\mathfrak a^{-3}tk'(s)-2i \mathfrak a^{-2}b_a(t)\Big(t-\frac{t^2}{2}k(s)\Big),\\
f_2(s,t)&=h^2\mathfrak a^{-1}k(s),\\
f_3(s,t)&=-ih\mathfrak a^{-3}tk'(s)b_a(t)\Big(t-\frac{t^2}{2}k(s)\Big)+h \mathfrak a^{-2}\frac{t^2}{2}k'(s)+\mathfrak a^{-2} b_a^2(t)\Big(t-\frac{t^2}{2}k(s)\Big)^2  -\lambda_n(h)\,.
\end{align*} 
 We differentiate with respect to $s$ in~\eqref{eq:R4}, and get
\begin{multline}\label{eq:R5}
h^2(\mathfrak a^{-2}\partial^2_s+\partial^2_t) (\partial_s\tilde\psi_{h,n})=(f_1-h^2\partial_s\mathfrak a^{-2})\partial_s^2\tilde\psi_{h,n}+f_2\, \partial_s\partial_t\tilde\psi_{h,n}+(\partial_sf_1+f_3)\, \partial_s\tilde\psi_{h,n}\\+\partial_sf_2\, \partial_t\tilde\psi_{h,n}+\partial_sf_3\, \tilde\psi_{h,n}.
\end{multline}
Having $s\mapsto k(s)$ smooth, $\mathfrak a=1-tk(s)$ for $t\in(-2t_0,2t_0)$, and $\psi_{n,h}\in\dom \mathcal P_h$ ensure that the function in the RHS of~\eqref{eq:R5} is in $L^2\big(\Phi^{-1}(\Om\cap\Gamma_{2t_0})\big)$. Hence $\partial_s \tilde\psi_{h,n} \in H^1 (\Om\cap\Gamma_{2t_0})$ and satisfies
\begin{equation}\label{eq:R6}
(\mathfrak a^{-2}\partial^2_s+\partial^2_t)\partial_s \tilde\psi_{h,n} \in L^2\big(\Phi^{-1}(\Om\cap\Gamma_{2t_0})\big).
\end{equation}
Hence \eqref{eq:R1} follows from~\eqref{eq:R6} using  the interior elliptic estimates associated with the differential operator $L:=(\mathfrak a^{-2}\partial^2_s+\partial^2_t)$.

{\it Step~3.}  We prove that
\begin{equation}\label{eq:qf-gh}
\mathcal Q_h(g_h)=\lambda_n(h)\|g_h\|_{L^2(\Omega)}^2+\tO (h^{\frac{5}{2}})\,,
\end{equation}
where $\mathcal Q_h$ is the the quadratic form introduced in \eqref{eq:qf-MLD}\,.  \\
  With the notation introduced   in \eqref{eq:not-tO},  the estimates in \eqref{eq:0} and \eqref{eq:qf-gh} yield  \eqref{eq:hyp-g-qf} for \emph{any} 
$\theta\in(0,\frac38)$.

We start by noticing that
\begin{equation}\label{eq:t1}\langle \mathcal P_h \varphi_h, G_h\rangle _{L^2(\Omega)}=\lambda_n(h) \langle \varphi_h,G_h\rangle_{L^2(\Omega)}+
\langle (\mathcal  P_h-\lambda_n(h))\varphi_h,G_h\rangle_{L^2(\Omega)}\,, \end{equation}
where $\varphi_h$ is defined in \eqref{eq:defvarphi} and 
\[\tilde G_h(s,t)=-(h^{1/2}\partial_s-i\zeta_a)g_h\,.\]
 Recall that $\varphi_h$ and $G_h$ are compactly supported in $\Omega\cap\Gamma_{t_0}$ so that we can use the Frenet coordinates valid near the edge $\Gamma$.   
By \eqref{eq:exp-dec-H2} we have
\begin{equation}\label{6.21aa}
 \|(\mathcal  P_h-\lambda_n(h))\varphi_h\|_{L^2(\Omega)}=\mathcal O(h^\infty)
 \end{equation}
and by  \eqref{eq:exp-dec-g}
\begin{equation} \label{6.21ab} \|G_h\|_{L^2(\Omega)}=\mathcal O(1).
\end{equation}
By H\"older's inequality, we infer from \eqref{6.21aa} and \eqref{6.21ab}
\begin{equation} \label{6.21ab*}
\langle (\mathcal  P_h-\lambda_n(h))\varphi_h,G_h\rangle_{L^2(\Omega)}=\mathcal O(h^\infty)\,.
\end{equation}

Furthermore, computing the integrals in the Frenet  coordinates  and integrating by parts, we find 
\begin{equation}\label{6.21a}  \langle \varphi_h,G_h\rangle_{L^2(\Omega)}=\langle \mathfrak a(h^{1/2}\partial_s-i\zeta_a)\tilde \varphi_h+h^{1/2}(\partial_s\mathfrak a)\tilde\varphi_h,\tilde g_h \rangle_{L^2(\R^2)}=\|g_h\|_{L^2(\Omega)}^2+\mathcal O(h^{9/8})\|g_h\|_{L^2(\Omega)}\,.
\end{equation}
Here we get the $\mathcal O(h^{9/8})$ remainder by  using that $\partial_s\mathfrak a=\mathcal  O(ts)$, the H\"older inequality and Remark~\ref{rem:hyp-sec-tf} on the decay estimates in \eqref{eq:dec-g-bnd1} and \eqref{eq:dec-g-bnd1*} for $\psi_{h,n}$    as follows
\[ |\langle \mathfrak a (\partial_s\mathfrak a)\tilde\varphi_h,\tilde g_h \rangle_{L^2(\R^2)} |\leq C \big(A_4(\psi_{h,n})B_4(\psi_{h,n}) \big)^{1/4}\|g_h\|_{L^2(\R^2)}=\mathcal O(h^{5/8})\|g_h\|_{L^2(\R^2)}\,.\]
 By \eqref{eq:0} and \eqref{eq:norm-gh}, we infer from \eqref{6.21a}
\begin{equation}\label{6.21a*}
\lambda_n(h)\langle \varphi_h,G_h\rangle_{L^2(\Omega)}=\lambda_n(h)\|g_h\|_{L^2(\Omega)}^2+\tO  (h^{5/2})\,.
\end{equation} 
Therefore,   inserting the estimates in \eqref{6.21a*}  and \eqref{6.21ab*} into \eqref{eq:t1},  we find
\begin{equation}\label{eq:p-qf-gh}
\langle \mathcal P_h \varphi_h, G_h\rangle _{L^2(\Omega)}=\lambda_n(h) \|g_h\|_{L^2(\Omega)}^2+\tO(h^{5/2})\,.
\end{equation}
Now,  by Lemma~\ref{lem:Adecom-form} (used with $\phi=0$), we get 
\begin{equation}\label{eq:Rh}
{\rm Re}\langle \mathcal P_h \varphi_h, G_h\rangle =\mathcal Q_h(g_h)-h^{1/2}{\rm Re}\langle  R_h,g_h\rangle_{L^2(\Omega)}\,,
\end{equation}
where the function $R_h$ is defined  via \eqref{eq:u-(s,t)} as follows,
\begin{equation}\label{eq:fct-Rh}
\tilde R_h(s,t)= (h\partial_s-i\tilde{F}_{1})\Big( \big(\partial_s \mathfrak a^{-1}-i\mathfrak a^{-1}\partial_s\tilde F_1)\big)(h\partial_s-i\tilde{F}_{1})\tilde\varphi_h -i\mathfrak a^{-1}(\partial_s\tilde F_1)\tilde\varphi_h\Big) +h^2\partial_t\big(\partial_s\mathfrak a \big)\partial_t\tilde\varphi_h\,.
\end{equation}
 Our choice of gauge in Lemma~\ref{lem:Anew2} ensures that $\tilde F_2=0$ and $\tilde F_1=\mathcal O(t)$. By Remark~\ref{rem:hyp-sec-tf} and \eqref{eq:A_tild2}, we have
\[ \int_{\R}\int_{-t_0}^{t_0}  |t|^N\big(|\tilde\varphi_h|^2+\mathfrak a^{-1}h^{-1}|(h\partial_s-i\tilde F_1)\tilde\varphi_h|^2+h|\partial_t\tilde\varphi_h|^2\big) \mathfrak a  dsdt =\mathcal O(h^{N/2}) \]
and
\[ \int_{\R}\int_{-t_0}^{t_0} |s|^N\big(|\tilde\varphi_h|^2+\mathfrak a^{-1}h^{-1}|(h\partial_s-i\tilde F_1)\tilde\varphi_h|^2+h|\partial_t\tilde\varphi_h|^2\big) \mathfrak a  dsdt =\mathcal O(h^{N/8}) \,.\]
Furthermore, by \eqref{eq:exp-dec-H2}, 
\begin{equation*} \int_{\R}\int_{-t_0}^{t_0}  |t|^N\big(|\partial_s^2\tilde\varphi_h|^2+|\partial_t^2\tilde\varphi_h|^2\big)  dsdt =\mathcal O(h^{\frac{N}2-2})
\end{equation*}
and 
\begin{equation*}
\int_{\R}\int_{-t_0}^{t_0}  |s|^N\big(|\partial_s^2\tilde\varphi_h|^2+|\partial_t^2\tilde\varphi_h|^2\big)  dsdt =\mathcal O(h^{\frac{N}8-2})\,.  
\end{equation*}
Now we  can estimate  $\tilde R_h$ in \eqref{eq:fct-Rh}, by expressing it as follows
\[
\tilde R_h=m_1(h\partial_s-i\tilde F_1)^2\tilde\varphi_h+(m_2+h\partial_sm_1)(h\partial_s-i\tilde F_1)\tilde\varphi_h+
h(\partial_sm_2)\tilde\varphi_h+h^2m_3\partial_t^2\tilde\varphi_h+h^2(\partial_tm_3)\partial_t\tilde\varphi_h\,,
\]
where
\begin{align*}
&m_1=\partial_s\mathfrak a^{-1}-i\mathfrak a^{-1}\partial_s\tilde F_1=\mathcal O(ts),\quad \partial_sm_1=\mathcal O(t),\\
&m_2=-i\mathfrak a^{-1}\partial_s\tilde F_1=\mathcal O(t^2s),\quad \partial_sm_2=\mathcal O(t^3s^2),\\
&m_3=\partial_s\mathfrak a=\mathcal O(ts),\quad \partial_tm_3=\mathcal O(s)\,.
\end{align*}
We get then that the norm of $R_h$ satisfies,
\begin{equation}\label{eq:norm-Rh}
\|R_h\|_{L^2(\Omega)}=\mathcal O(h^{13/8})\,.
\end{equation}
By  H\"older's inequality, we infer from \eqref{eq:norm-Rh} and \eqref{eq:norm-gh} the following estimate
\[h^{1/2}| {\rm Re}\langle  R_h,g_h\rangle_{L^2(\Omega)}|  \leq h^{1/2}\|R_h\|_{L^2(\Omega)}\|g_h\|_{L^2(\Omega)}=\tO(h^{5/2})\,.\] Consequently, \eqref{eq:p-qf-gh} and \eqref{eq:Rh} yield \eqref{eq:qf-gh}.

{\it Step~4.} 

We refine the exponential decay of $g_h$. To that end,  consider a fixed constant $0<\alpha<\frac14\alpha_2$, where $\alpha_2$ is the constant in \eqref{eq:weights}, and a real-valued Lipshitz function $\phi_{h,\alpha}\geq 0$, which  will be either
\[ \phi_{h,\alpha}(x)=\phi_{h,\alpha}^{\rm norm}(x):=\alpha h^{-1/2}{\rm dist}(x,\Gamma)~{\rm or~}\phi_{h,\alpha}(x)=\phi_{h,\alpha}^{\rm tan}(x):=\alpha h^{-1/8}s(x)\,.\]
We introduce the function $G_{h,\alpha}$ defined via \eqref{eq:u-(s,t)} as follows
\[ \tilde G_{h,\alpha}(s,t)=-(h^{1/2}\partial_s-i\zeta_a)\big(e^{2\phi_{h,\alpha}}\tilde g_h(s,t)\big)\,.\]
Since $\alpha<\frac14\alpha_2$,  we infer from \eqref{eq:exp-dec-H1} and \eqref{eq:exp-dec-g} 
\begin{multline*}
 \int_{\Omega} \big({\rm dist}(x,\Gamma)\big)^2 \big|e^{\phi_{h,\alpha}}\varphi_h(x)\big|^2 dx=\mathcal O(h),\\
  \int_{\Omega} \big(s(x)\big)^2 \big|e^{\phi_{h,\alpha}}\varphi_h(x)\big|^2 dx=\mathcal O(h^{1/4}),\\
  \|G_{h,\alpha}\|_{L^2(\Omega)}=\mathcal O(1)\,, 
\end{multline*}
and also
\[\langle \mathcal P_h\varphi_h,G_{h,\alpha}\rangle_{L^2(\Omega)}=\lambda_n(h)\|e^{\phi_{h,\alpha}}g_h\|_{L^2(\Omega)}^2+\tO (h^{\frac{19}{8}})\]
which results in a similar fashion to \eqref{eq:p-qf-gh}.

Now, we write by Lemma~\ref{lem:Adecom-form},
\[ {\rm Re}\langle \mathcal P_h \varphi_h, G_{h,\alpha}\rangle =\mathcal Q_h(e^{\phi_{h,\alpha}}g_h)
-h^2\big\||\nabla \phi_{h,\alpha}|e^{\phi_{h,\alpha}}g_h\big\|_{L^2(\Omega)}^2 -h^{1/2}{\rm Re}\langle  R_h,e^{2\phi_{h,\alpha}}g_h\rangle_{L^2(\Omega)}\,, \]
where $R_h$ is introduced in \eqref{eq:fct-Rh}.  Since $\alpha<\frac14\alpha_2$, we get from \eqref{eq:exp-dec-H1} and \eqref{eq:exp-dec-H2},
\[\big\|e^{\phi_{h,\alpha}} R_h\big\|_{L^2(\Omega)}=\mathcal O(h^{9/8})~{\rm and~}\langle  R_h,e^{2\phi_{h,\alpha}}g_h\rangle_{L^2(\Omega)}=\mathcal O(h^{9/8})\|g_h\|_{L^2(\Omega)} \,. \]
 Collecting the foregoing estimates, we get
\begin{equation}\label{eq:agmon-qf-gh}
\mathcal Q_h(e^{\phi_{h,\alpha}}g_h)=\lambda_n(h)\|e^{\phi_{h,\alpha}}g_h\|_{L^2(\Omega)}^2+\tO (h^{5/2})\,.
\end{equation}
Now we can select $\alpha>0$ small enough so that the following two estimates 
hold. The first estimate is
\begin{equation}\label{eq:dec-est-ghN}
\int_\Omega\big( |g_h|^2+h^{-1}|(h\nabla-i\Fb)g_h|^2\big)\exp \left(\alpha h^{-1/2}d(x,\Gamma)\right)dx\leq C\|g_h\|_{L^2(\Omega)}^2+\tO(h^{3/2})\,,
\end{equation}
and it follows after choosing $\phi_{h,\alpha}=\alpha h^{-1/2}{\rm dist}(x,\Gamma)$ and using \eqref{eq:lb-qf-well}. The second estimate follows by choosing $\phi_{h,\alpha}=\alpha h^{-1/8}s(x)$ and using \eqref{eq:hyp-g-dec*}; it reads as follows 
\begin{equation}\label{eq:dec-est-ghN*}
\int_\Omega\big( |g_h|^2+h^{-1}|(h\nabla-i\Fb)g_h|^2\big)\exp \left(\alpha h^{-1/8}s(x)\right)dx\leq C\|g_h\|_{L^2(\Omega)}^2+\tO(h)\,.
\end{equation}
~
{\it Step~5.} 

 Let $\theta\in(0,\frac38)$. Collecting the estimates in \eqref{eq:qf-gh}, \eqref{eq:dec-est-ghN} and \eqref{eq:dec-est-ghN*}, we observe that the function $g_h$ satisfies \eqref{eq:hyp-g-qf}$_\theta$, \eqref{eq:hyp-g-dec} and \eqref{eq:hyp-g-dec*} with $r_h=\mathcal O(h^{\frac34-\theta})$. We can then apply Proposition~\ref{prop:g-tang-freq} and get  (recall that $\|w_h\|_{L^2(\Omega)}\sim \|g_h\|_{L^2(\Omega)}\leq \sqrt{r_h}$ by \eqref{eq:g-norm})
\[ \|(h^{3/8}\partial_\sigma-i\zeta_a)w_h\|_{L^2(\Omega)}\leq C_\theta h^{\frac38-\frac\theta2}\big(\|g_h\|_{L^2(\Omega)}+\sqrt{r_h}+ h^{\frac38-\frac{3\theta}4}\big)=\mathcal O(h^{\frac34-\frac{5\theta}4})\,. \]
Since this  holds for any $\theta\in(0,\frac38)$, we get that $\|(h^{3/8}\partial_\sigma-i\zeta_a)w_h\|_{L^2(\Omega)}=\tO(h^{3/4})$, thereby finishing the proof of Proposition~\ref{rem:tang-freq}.
\end{proof}

\section{Lower bound}\label{sec:lb}
We fix a labeling $n\geq 1$ corresponding to the  eigenvalue $\lambda_n(h)$ of the operator $\mathcal P_h$ introduced in \eqref{eq:P}.   The purpose of this section is to obtain  an accurate  lower bound for $\lambda_n(h)$.  This will be done by doing a spectral reduction via various auxiliary operators.

\subsection{Useful operators}
 We introduce operators, on the real line and in the plane, which will be useful to carry out a spectral reduction for the operator $\mathcal P_{h}$ and deduce the eigenvalue  lower bounds that match  with the established eigenvalue asymptotics in Theorem~\ref{thm:main}.

These new operators are defined via the spectral characteristics of the model operator introduced  in Subsection~\ref{sec:p-bnd-fc},   namely, the spectral constants $\beta_a>0$ and  $\zeta_a<0$ introduced in \eqref{eq:mub}  and \eqref{eq:main-ct}, and the positive normalized eigenfunction $\phi_a\in L^2(\R)$ corresponding to $\beta_a$.
We introduce the following  two operators
\begin{equation}\label{eq:proj:R-}
R_0^-:  \psi \in L^2(\R^2)\mapsto \int_{\R}\phi_a(\tau)\psi(\cdot,\tau)d\tau\in L^2(\R)\,,
\end{equation}
and
\begin{equation}\label{eq:proj:R+}
R_0^+:f\in L^2(\R)\mapsto f\otimes \phi_a\in L^2(\R^2)\,,
\end{equation}
 where $(f\otimes \phi_a)(\sigma,\tau):=f(\sigma)\phi_a(\tau)\,.$\\
 Note that $R_0^+R_0^-$ is an  orthogonal  projector on $L^2(\mathbb R^2)$ whose image is $L^2(\mathbb R)\otimes {\rm span} (\phi_a) $. 
It is easy to check that the operator norms of $R_0^\pm$  are equal to $1$,  hence, for any $f\in L^2(\R)$ and $\psi\in L^2(\R^2)$, we have
\begin{equation}\label{eq:proj-R-norm}
\| R_0^+f\|_{L^2(\R)}\leq \|f\|_{L^2(\R)},\quad \|R_0^-\psi\|_{L^2(\R)}\leq \|\psi\|_{L^2(\R^2)},\quad
\|R_0^+R_0^-\psi\|_{L^2(\R^2)}\leq \|\psi\|_{L^2(\R^2)}\,.
\end{equation}  
If we denote by $\pi_a$ the projector in $L^2(\mathbb R_\tau)$ on the vector space generated by $\phi_a$, we notice that
\begin{equation}\label{eq:Pia}
\Pi_0:=R_0^+R_0^- = I \otimes \pi_a \,.
\end{equation}
\subsection{Structure of bound states}
 Our aim is to  determine a rough approximation of the bound state $\psi_{h,n}$ of $\mathcal P_h$, satisfying  
  \begin{equation}\label{eq:7.4a}
  \mathcal P_h\psi_{h,n}=\lambda_n(h)\psi_{h,n}\,,
  \end{equation}
  this approximation being  valid near the point of maximum curvature and reading as follows in the Frenet coordinates 
\[\tilde \psi_{h,n}(s,t) \approx h^{ -5/16}e^{i\zeta_a s/h^{1/2}}\phi_a\left( h^{-1/2}t\right)\,.\] 
Associated with   $\psi_{h,n}$, we introduced  in \eqref{eq:ef-tranc}  the function  $ u_{h,n}$  which can be seen as a function on $\R^2$ with $L^2$-norm satisfying \eqref{eq:norm-u}. We recall that
\begin{equation*}
 u_{h,n}(\sigma,\tau)=h^{5/16}\chi(h^{\eta}\sigma)\chi(h^{\delta}\tau)\tilde \psi_{h,n}\big(h^{1/8}\sigma,h^{1/2}\tau\big)\,,
\end{equation*}
 where $\tilde \psi_{h,n}$ is the function assigned to $\psi_{h,n}$   by \eqref{eq:u-(s,t)},    $\chi\in C_c^\infty(\R)$, ${\supp}\,\chi\subset[-1,1]$,  $0\leq \chi\leq 1$  and $\chi=1$ on $[-1/2,1/2]$.\\ 
 We consider the function defined as follows 
\begin{equation}\label{eq:ef-tranc-v}
v_{h,n}(\sigma,\tau)=e^{-i\zeta_a\sigma/h^{3/8}} u_{h,n}(\sigma,\tau)\,.
\end{equation}
Approximating the function $v_{h,n}\sim \chi(h^\eta\sigma)\chi(h^\delta\tau) \phi_a(\tau)$  is the aim of the next proposition, which also yields an approximation of the bound state $\psi_{h,n}$ by the previous considerations.

\begin{proposition}\label{prop:app-ef} Let $\mathcal P_{h}^{\rm new}$ be the operator in \eqref{eq:P1}. It holds the following.
\begin{enumerate}
\item $\big\|\mathcal P_{h}^{\rm new}v_{h,n}-\big(h^{-1}\lambda_n(h)-\beta_a\big)v_{h,n} \|_{L^2(\R^2)}=\mathcal O(h^\infty)$\,;
\item $ \|v_{h,n}\|_{L^2(\R^2)}=1+\mathcal O(h^{1/2})$\,; 
\item $\|v_{h,n}-\Pi_0 v_{h,n}\|_{L^2(\R^2)}=\mathcal O(h^{1/4})$\,;
\item $\|\partial_\tau v_{h,n}-\partial_\tau \Pi_0 v_{h,n}\|_{L^2(\R^2)}+\|\tau( v_{h,n}- \Pi_0 v_{h,n})\|_{L^2(\R^2)} =\mathcal O(h^{1/4})$\,.
\end{enumerate}
\end{proposition}
~
\begin{proof}~\medskip\\
{\it Proof of item (1).\;}\medskip 
Let $z_h$ be the function supported near $\Gamma$ and defined in the Frenet coordinates by means of \eqref{eq:u-(s,t)} as follows
\begin{equation}\label{eq:fct-zh}
\tilde z_h(s,t)=\chi(h^{-\frac {1}8+\eta}s)\chi(h^{-\frac {1}{2}+\delta}t)\,.
\end{equation} 
We introduce  the function involving the commutator of $\mathcal P_h$ and $z_h$  acting on $\psi_{h,n}$, 
\begin{equation}\label{eq:comm-fh} 
 f_h=[ \mathcal P_h,z_h ]\psi_{h,n}=\big(\mathcal P_hz_h-z_h\mathcal P_h\big)\psi_{h,n}\,.
 \end{equation}
By Remark~\ref{rem:hyp-sec-tf}, we may use the localization estimates in \eqref{eq:dec-g-bnd2} and \eqref{eq:dec-g-bnd2*}  with $g_h=\psi_{h,n}$ and $r_h=1$. Consequently, 
\[ \int_{\R^2}|\tilde f_h(s,t)|^2dsdt\leq C\int_\Omega|f_h(x)|^2dx=\mathcal O(h^\infty)\,,\]
where $\tilde f_h$  which is  assigned to the function $f_h$ in \eqref{eq:comm-fh}  is supported in the set \break $ \{\{|s|\geq \frac12h^{\eta-\frac18}\}\cup\{|t|\geq \frac12h^{\delta-\frac12}\}\}\cap\{ \{|s|\leq h^{\eta-\frac18}\}\cap\{|t|\leq h^{\delta-\frac12}\}\}   $.\\

We infer from \eqref{eq:7.4a},  \eqref{eq:op-Ph-st}, \eqref{eq:op-sigma-tau} and 
\eqref{eq:ef-tranc},
\[ \check{\mathcal P}_{h} u_{h,n}-\lambda_n(h) u_{h,n}=h^{5/16}\check f_h\,,\]
 where
\[\check f_h(\sigma,\tau)=\tilde f_h(h^{1/8}\sigma,h^{1/2}\tau)\,.\] 
 Consequently, 
after performing the change of variable $(\sigma=h^{-1/8}s,\tau=h^{-1/2}t)$,
\begin{equation}\label{eq:Pu}
\|\check{\mathcal P}_{h} u_{h,n}-\lambda_n(h) u_{h,n}\|^2_{ L^2(\R^2)}
=\|\tilde f_h\|^2_{L^2(\R^2)}=\mathcal O(h^\infty)\,.
\end{equation}
By~\eqref{eq:P1} and \eqref{eq:ef-tranc-v}, we observe that
\[\check{\mathcal P}_{h}u_{h,n}=he^{i\zeta_a\sigma/h^{3/8}}(\mathcal P_{h}^{\rm new}+\beta_a)v_{h,n}, \]
which after being inserted into \eqref{eq:Pu}, yields the estimate in item (1).\medskip\\
\begin{rem}\label{rem:item(1)H1}
By \eqref{eq:R1}, $\partial_\sigma v_{h,n}\in H^2(\R^2)$. Furthermore,  by \eqref{eq:exp-dec-H2}, the function $f_h$ in \eqref{eq:comm-fh}  satisfies
$\|\partial_\sigma \check f_h\|_{L^2(\R^2)}=\mathcal O(h^\infty)$. A slight adjustment of the proof of item~(1) then yields
\[\big\|\partial_\sigma\mathcal P_{h}^{\rm new}v_{h,n}-\big(h^{-1}\lambda_n(h)-\beta_a\big)\partial_\sigma v_{h,n} \|_{L^2(\R^2)}=\mathcal O(h^\infty)\,.\]
\end{rem}
{\it Proof of item (2).}\medskip\\
By the normalization of $\psi_{h,n}$ and Remark~\ref{rem:hyp-sec-tf}, we have
\[1= \int_\Omega|\psi_{h,n}|^2dx=\int_{\{|s(x)|< h^{-\eta+\frac18},|t(x)|< h^{-\delta+\frac12}\}}|\psi_{h,n}|^2dx+\mathcal O(h^\infty)\,,\]
\[\int_\Omega (1-z_h^2)|\psi_h|^2dx=\mathcal O(h^\infty) \]
and
\[\int_{\Omega} {\rm dist}(x,\Gamma)|\psi_{h,n}|^2dx=\mathcal O(h^{1/2})\,. \]
We notice that  the function  $z_h$ introduced above  in \eqref{eq:fct-zh}  equals  $1$\break in ${\{|s(x)|< \frac12h^{-\eta+\frac18},|t(x)|< \frac12h^{-\delta+\frac12}\}}$.
 
Now we infer from \eqref{eq:A_tild2}
\[
\int_{\{| s|< h^{-\eta+\frac18}\,,\,|t|< h^{-\delta+\frac18}\}}|\tilde \psi_{h,n}(s,t)|^2 |t|dsdt\leq C\int_{\Omega} {\rm dist}(x,\Gamma)|\psi_{h,n}|^2dx=\mathcal O(h^{1/2})\]
and
\begin{align*}
\int_{\{| s|< h^{-\eta+\frac18},\,|t|< h^{-\delta+\frac18}\}}|\tilde \psi_{h,n}(s,t)|^2dsdt&=
\int_{\{| s|< h^{-\eta+\frac18},\, |t|< h^{-\delta+\frac18}\}}|\tilde \psi_{h,n}(s,t)|^2(1-tk(s))dsdt\\
& \qquad+\int_{\{| s|< h^{-\eta+\frac18},\, |t|< h^{-\delta+\frac18}\}}|\tilde \psi_{h,n}(s,t)|^2\,tk(s)dsdt\\
&=1+\mathcal O(h^{1/2})\,.
\end{align*}
Similarly we get
\[ \int_{\{| s|< \frac12h^{-\eta+\frac18},\,|t|< \frac12h^{-\delta+\frac18}\}}(1-\tilde z_h^2)\, |\tilde \psi_{h,n}(s,t)|^2dsdt=\mathcal O(h^{1/2}).\]
Consequently, returning to  \eqref{eq:ef-tranc-v}, doing a change of variables and noticing that $\tilde z_h$ is supported in ${\{| s|< h^{-\eta+\frac18},|t|< h^{-\delta+\frac18}\}}$, we get
 \begin{align*}
 \|v_{h,n}\|^2_{L^2(\R^2)}&=\int_{\{| s|< h^{-\eta+\frac18},\,|t|< h^{-\delta+\frac18}\}}|\tilde \psi_{h,n}|^2dsdt 
 -\int_{\{| s|< h^{-\eta+\frac18},\,|t|<  h^{-\delta+\frac18}\}}(1-\tilde z_h^2)|\tilde\psi_h|^2dsdt\\
 &= 1+\mathcal O(h^{1/2}).
\end{align*}
{\it Proof of items (3) and (4).}\medskip\\
{\it Step~1.}
We recall that the    $\tO$ notation was introduced in \eqref{eq:not-tO}. Note that Proposition~\ref{prop:tang-freq} yields
\begin{equation}\label{eq:ds-v}
\|h^{3/8}\partial_\sigma v_{h,n}\|_{L^2(\R^2)}=\tO (h^{3/8})\,.
\end{equation}
By Remark~\ref{rem:hyp-sec-tf}, we can use  \eqref{eq:Qh-gh} and \eqref{eq:qf-u-tranc} with  $g_h=\psi_{h,n}$, $r_h=1$ (and $w_h=\check u_{h,n}$).  In the same vein, we can use  \eqref{eq:dec-g-bnd1} and \eqref{eq:dec-g-bnd1*} too.  Since $u_{h,n}=e^{i\zeta_a\sigma/h^{3/8}}v_{h,n}$, we get 
\begin{equation}\label{eq:qf-v-tranc}
\int_{\R^2}\Big(|\partial_\tau v_{h,n}|^2+ |h^{3/8}\partial_\sigma v_{h,n} +i(b_a(\tau) \tau +\zeta_a)v_{h,n}|^2\Big)d\tau\,d\sigma\leq \Big(\beta_a+\mathcal O(h^{1/2})\Big)\|v_{h,n}\|_{L^2(\R^2)}^2\,.
\end{equation} 
By Cauchy's inequality and \eqref{eq:ds-v}, we obtain for any $\epsilon >0$, 
\begin{align*}
\int_{\R^2}& |h^{3/8}\partial_\sigma v_{h,n} +i(b_a(\tau) \tau +\zeta_a)v_{h,n}|^2d\sigma d\tau
\\&\geq \int_{\R^2}\Big((1-\epsilon) |(b_a(\tau) \tau +\zeta_a)v_{h,n}|^2
-\epsilon^{-1}|h^{3/8}\partial_\sigma v_{h,n}|^2\Big)d\sigma d\tau\\
&\geq  (1-\epsilon) \int_{\R^2} |(b_a(\tau) \tau +\zeta_a)v_{h,n}|^2d\sigma d\tau-\tO(\epsilon^{-1} h^{3/4})\,.\end{align*}
We choose $\epsilon=h^{3/8}$ and insert the resulting inequality into \eqref{eq:qf-v-tranc} to get:
\begin{equation}\label{eq:qf-v-tranc*}
\int_{\R^2} \Big( |\partial_\tau v_{h,n}|^2+ |(b_a(\tau) \tau +\zeta_a)v_{h,n}|^2\Big)d\tau\,d\sigma\leq 
\beta_a+\tO(h^{3/8})\,.
\end{equation}
{\it Step~2.}   In light of \eqref{eq:Pia}, let us introduce 
\begin{equation} \label{eq:decomp-v-r,r}
r:= \Pi_0 v_{h,n} \mbox{ and }  r_\bot := (I-\Pi_0) v_{h,n} = (I \otimes (I-\pi_a) ) v_{h,n}\,.
\end{equation}
Using the last relation,  and since the orthogonal projection $\pi_a$ commutes with the operator $\mathfrak h_a[\zeta_a]$, we have the following two identities,  for almost every $\sigma\in\R$, 
\[ \int_{\R} |v_{h,n}(\sigma,\tau)|^2 d\tau
= \int_\R|r(\sigma,\tau)|^2d\tau+ \int_\R|r_\bot(\sigma,\tau)|^2d\tau\]
and
\begin{equation}\label{eq:decomp-q-orth}
\begin{aligned}
q_{\zeta_a}\big(v_{h,n}(\sigma,\cdot)\big)&:= \int_{\R} \Big( |\partial_\tau v_{h,n}(\sigma,\tau)|^2+ |(b_a(\tau) \tau +\zeta_a)v_{h,n}(\sigma,\tau)|^2\Big)d\tau\\
&=q_{\zeta_a}\big(r(\sigma,\cdot)\big)+q_{\zeta_a}(r_\bot\big(\sigma,\cdot)\big)\\
&\geq \beta_a\int_\R|r(\sigma,\tau)|^2d\tau+\mu_2(\zeta_a)\int_\R|r_\bot(\sigma,\tau)|^2d\tau\,,
\end{aligned}
\end{equation}
by the min-max principle,  where $\mu_2(\zeta_a)$ is the second eigenvalue of the operator $\mathfrak h_a[\zeta_a]$, satisfying $\mu_2(\zeta_a)>\beta_a$ (see Section~\ref{sec:step1}). Integrating with respect to $\sigma$, we get
\begin{multline}\label{eq:qf-v-tranc**}
\int_{\R^2} \Big( |\partial_\tau v_{h,n}(\sigma,\tau)|^2+ |(b_a(\tau) \tau +\zeta_a)v_{h,n}(\sigma,\tau)|^2\Big)d\sigma d\tau\\
\geq \beta_a\int_{\R^2}|r(\sigma,\tau)|^2 d\sigma d\tau +\mu_2(\zeta_a)\int_{\R^2}|r_\bot(\sigma,\tau)|^2d\sigma d\tau\,.
\end{multline} 
 We deduce from \eqref{eq:qf-v-tranc*} and the first  item in  Proposition~\ref{prop:app-ef}
\begin{equation}\label{eq:norm-v-tranc}
(\mu_2(\zeta_a)-\beta_a)\int_{\R^2}|r_\bot(\sigma,\tau)|^2d\sigma d\tau \leq \tO(h^{3/8})\int_{\R^2}|r(\sigma,\tau)|^2d\sigma d\tau\,,
\end{equation}
\begin{equation}\label{eq:norm-v-tranc=1}
\int_{\R^2}|r(\sigma,\tau)|^2d\sigma d\tau =1+ \tO(h^{3/8})\,,
\end{equation}
and 
\begin{equation}\label{eq:qf-v-tranc***}
 \int_{\R^2} \Big( |\partial_\tau r_\bot(\sigma,\tau)|^2+ |(b_a(\tau) \tau +\zeta_a)r_\bot(\sigma,\tau)|^2\Big)d\sigma d\tau \leq \tO(h^{3/8})\int_{\R^2}|r(\sigma,\tau)|^2d\sigma d\tau\,. 
 \end{equation} 
{\it Step~3.}
Coming back to the definition of $r_\bot $ in \eqref{eq:decomp-v-r,r}, we still have to improve the error term in  \eqref{eq:norm-v-tranc} to get   the estimate  of the third item  in Proposition~\ref{prop:app-ef}.

 To that end, we will estimate the terms involving $\partial_\sigma v_{h,n}$ in \eqref{eq:qf-v-tranc}.    By    \eqref{eq:Pia} and  dominated convergence, it is clear that $\Pi_0$ commutes with $\partial_\sigma$  when acting on compactly supported functions of $H^1(\R^2)$, 
 \begin{equation}\label{eq:7.20}   \Pi_0 \partial_\sigma =\partial_\sigma \Pi_0\,.
  \end{equation}
By \eqref{eq:FH-orth}, $\phi_a$ is orthogonal to $(b_a(\tau) \tau +\zeta_a)\phi_a$ in $L^2(\mathbb R)$, so
\[
\pi_a (b_a(\tau) \tau + \zeta_a) \pi_a =0\,,
\]
which implies, by taking the tensor product,
\begin{equation}\label{eq:7.21}
\Pi_0 (b_a(\tau) \tau + \zeta_a) \Pi_0 =0\,.
\end{equation}
By \eqref{eq:decomp-v-r,r}, \eqref{eq:7.20} and \eqref{eq:7.21}, we get
\begin{equation*}
\big\langle  r(\sigma,\tau), i(b_a(\tau) \tau +\zeta_a)\partial_\sigma r(\sigma,\tau)\big\rangle_{L^2(\R^2)} =0 \,.
\end{equation*}
Now, we inspect the term
 \begin{align}\label{eq:v}
\langle \partial_\sigma v_{h,n}, & i(b_a(\tau) \tau +\zeta_a)r\rangle_{L^2(\R^2)} =-\langle  v_{h,n}, i(b_a(\tau) \tau +\zeta_a)\partial_\sigma r\rangle_{L^2(\R^2)} \nonumber\\
 &=-\underset{=0}{\underbrace{\langle  r, i(b_a(\tau) \tau +\zeta_a)\partial_\sigma r\rangle_{L^2(\R^2)}} }-\langle  r_\bot , i(b_a(\tau) \tau +\zeta_a)\partial_\sigma r\rangle_{L^2(\R^2)} \nonumber\\
&  =  -\langle  r_\bot, i(b_a(\tau) \tau +\zeta_a)\partial_\sigma r\rangle_{L^2(\R^2)}=   -\langle  (b_a(\tau) \tau +\zeta_a)r_\bot, i\partial_\sigma r\rangle_{L^2(\R^2)}\,.
\end{align} 
 Since 
\begin{align*}
 \|h^{3/8}\partial_\sigma r\|_{L^2(\R^2)} &=h^{3/8}\|\Pi_0\partial_\sigma v_{h,n}\|_{L^2(\R^2)}&\qquad
 [\text{by }\eqref{eq:7.20}]\\
 &\leq h^{3/8}\|\partial_\sigma v_{h,n}\|_{L^2(\R^2)} &\qquad[\text{by }\eqref{eq:proj-R-norm}]\\
 &=\tO(h^{3/8})&\qquad[\text{by } \eqref{eq:ds-v}]
 \end{align*}
we get by the Cauchy-Schwarz inequality,
~\eqref{eq:v} and \eqref{eq:qf-v-tranc***}
\begin{equation}\label{eq:comm-orth}
h^{3/8}| \langle \partial_\sigma v_{h,n},  i(b_a(\tau) \tau +\zeta_a)r\rangle_{L^2(\R^2)}|\leq \|(b_a(\tau)\tau+\zeta_a)r_\bot\|_{L^2(\R^2)}\|h^{3/8}\partial_\sigma r\|_{L^2(\R^2)}=\tO(h^{9/16})\,.
\end{equation} 
Now, we can  estimate the  following  inner product  term by using \eqref{eq:decomp-v-r,r} and  \eqref{eq:comm-orth},
\begin{align}\label{eq:com-term1}
\langle h^{3/8}&\partial_\sigma v_{h,n},  i(b_a(\tau) \tau +\zeta_a)v_{h,n}\rangle_{L^2(\R^2)}\nonumber\\
&=\langle h^{3/8} \partial_\sigma v_{h,n}, i(b_a(\tau) \tau +\zeta_a)r_\bot\rangle_{L^2(\R^2)} +\langle h^{3/8}\partial_\sigma v_{h,n}, i(b_a(\tau) \tau +\zeta_a)r\rangle_{L^2(\R^2)}
\nonumber\\
&=\langle h^{3/8} \partial_\sigma v_{h,n}, i(b_a(\tau) \tau +\zeta_a)r_\bot\rangle_{L^2(\R^2)} +\tO(h^{9/16})\,.
\end{align}
 By the Cauchy-Schwarz inequality,  \eqref{eq:ds-v},~\eqref{eq:qf-v-tranc***} and~\eqref{eq:com-term1}, we get
\begin{multline}\label{eq:dv-bv}
 \big|\langle h^{3/8}\partial_\sigma v_{h,n},  i(b_a(\tau) \tau +\zeta_a)v_{h,n}\rangle_{L^2(\R^2)}\big|\leq \|  h^{3/8}\partial_\sigma v_{h,n}\|\,\|(b_a(\tau) \tau +\zeta_a)r_\bot \| +\tO(h^{9/16})\\
 =\tO(h^{9/16})=o(h^{1/2})\,.\end{multline}
  Consequently,
 \begin{align*}
 \|h^{3/8}\partial_\sigma v_{h,n}+  i(b_a(\tau) \tau +\zeta_a)v_{h,n}\|^2_{L^2(\R^2)}&=
  \|h^{3/8}\partial_\sigma v_{h,n}\|^2_{L^2(\R^2)}+ \|(b_a(\tau) \tau +\zeta_a)v_{h,n}\|^2_{L^2(\R^2)}\\
  &\quad+2{\rm Re}\langle h^{3/8}\partial_\sigma v_{h,n},  i(b_a(\tau) \tau +\zeta_a)v_{h,n}\rangle_{L^2(\R^2)}\\
  & \geq \|(b_a(\tau) \tau +\zeta_a)v_{h,n}\|^2_{L^2(\R^2)}+o(h^{1/2})\,.
 \end{align*} 
Inserting the previous inequality into \eqref{eq:qf-v-tranc} we get the following improvement of \eqref{eq:qf-v-tranc*}
\begin{equation}\label{eq:eq:qf-v-tranc-imp}
\int_{\R^2} \Big( |\partial_\tau v_{h,n}|^2+ |(b_a(\tau) \tau +\zeta_a)v_{h,n}|^2\Big)d\tau\,d\sigma\leq 
\beta_a+\mathcal O(h^{1/2})\,. 
\end{equation}
{\it Step 4.}  Now we are ready to finish the proof of items (3) and (4). By \eqref{eq:qf-v-tranc**} and \eqref{eq:decomp-q-orth}, we infer from \eqref{eq:eq:qf-v-tranc-imp} and~\eqref{eq:decomp-v-r,r},
\[ (\mu_2(\zeta_a)-\beta_a)\int_{\R^2}|r_\bot(\sigma,\tau)|^2d\sigma d\tau \leq \mathcal O(h^{1/2})\int_{\R^2}|r(\sigma,\tau)|^2d\sigma d\tau \]
and
\[  \int_{\R^2} \Big( |\partial_\tau r_\bot(\sigma,\tau)|^2+ |(b_a(\tau) \tau +\zeta_a)r_\bot(\sigma,\tau)|^2\Big)d\sigma d\tau \leq \mathcal O(h^{1/2})\int_{\R^2}|r(\sigma,\tau)|^2d\sigma d\tau\,. \]
With  \eqref{eq:norm-v-tranc=1} in hand, we get the estimates of items (3) and (4) of Proposition~\ref{prop:app-ef}.
\end{proof}
%
\subsection{Projection on a refined quasi-mode}
We would like to improve  the approximation $\clb v_{h,n}\sim \chi(h^\eta\sigma)\chi(h^\delta\tau) \phi_a(\tau)$ obtained in Proposition~\ref{prop:app-ef}  by two  ways which eventually are correlated: (i)  displaying the curvature effects in $v_{h,n}$ and (ii) getting better estimates of the errors.  Along the proof of Proposition~\ref{prop:app-ef},  curvature effects were neglected and absorbed in the error terms.  Not  neglecting the curvature,  we  get the approximation  $v_{h,n}\sim \chi(h^\eta\sigma)\chi(h^\delta\tau) \phi_{a,h}(\tau)$ where   $\phi_{a,h}(\tau)$ corrects $\phi_a(\tau)$ via curvature dependent terms (see \eqref{eq:phi-a,h}).  This  is precisely stated in Proposition~\ref{prop:app-ef*} after introducing the necessary preliminaries. 
\subsubsection{Preliminaries}
In this subsection,  we write $\kappa =k(0)=k_{\max}$ and $k_2=k''(0)$.
  We consider the weighted $L^2$ space
\begin{equation}\label{eq:space-Xh}
  X_{h,\delta}=L^2\big((-h^{-\delta},h^{-\delta}); (1- h^{1/2} \kappa \tau) d\tau\big)
  \end{equation}
endowed with the Hilbertian norm
\[ \|f\|_{X_{h,\delta}}=\left(\int_{-h^{-\delta}}^{h^{-\delta}} |f(\tau)|^2(1-h^{1/2}\kappa\tau)d\tau\right)^{1/2}\,.\]
This norm is equivalent to the usual norm of $L^2(-h^{-\delta},h^{-\delta})$ provided $h$ is sufficiently small.

With domain  $H^2(-h^{-\delta},h^{-\delta})\cap H^1_0(-h^{-\delta},h^{-\delta})$, consider  
the  operator in \eqref{eq:H-beta} for $\xi=\zeta_a$, 
\begin{multline}\label{eq:H-beta*}
\mathcal H_{a,\kappa ,h}=-\frac {d^2}{d\tau^2}+(b_a(\tau) \tau +\zeta_a)^2\\+\kappa  h^\frac 12(1- \kappa h^\frac 12\tau)^{-1}\partial_\tau+2\kappa   h^\frac 12 \tau\left(b_a(\tau) \tau +\zeta_a-\kappa  h^\frac 12 b_a(\tau) \frac {\tau^2}2\right)^2\\-\kappa  h^\frac 12 b_a(\tau) \tau ^2 (b_a(\tau) \tau +\zeta_a)+\kappa ^2 hb_a(\tau)^2\frac {\tau^4}4\,,
\end{multline}
   which is  self-adjoint  on the space $X_{h,\delta}$. This operator can  be decomposed as follows
\begin{equation}\label{eq:exp-H-beta}
\mathcal H_{a,\kappa ,h}=\mathfrak h[\zeta_a] + \kappa h^{1/2}\mathfrak h^{(1)}[\zeta_a]+hL_h\,,
\end{equation}
where $\mathfrak h[\zeta_a]$ is introduced in \eqref{eq:ha} and
\begin{equation}\label{eq:h1-Lha}
\mathfrak h^{(1)}[\zeta_a]=  \partial_\tau+2\tau(b_a(\tau) \tau +\zeta_a)^2-b_a(\tau) \tau ^2(b_a(\tau) \tau +\zeta_a)\,,
\end{equation}
and
\begin{equation}\label{eq:h1-Lhb}
 L_h=q_{1,h}(\tau)\partial_\tau+q_{2,h}(\tau)\mbox{ with }  |q_{1,h}(\tau)|\leq C_1|\tau|,\quad |q_{2,h}(\tau)|\leq C_2(1+|\tau|^5)\,,\end{equation}
where $C_1,C_2$ are positive constants independent of $h,\tau$.

We introduce the following quasi-mode in the space $X_{h,\delta}$,
\begin{equation}\label{eq:phi-a,h}
\phi_{a,h}(\tau) =\chi (h^\delta \tau) \left(\phi_a(\tau) +h^{1/2}\kappa\, \phi_{a}^{\rm cor}(\tau)\right)\,,
\end{equation}
where $\chi\in C_c^\infty(\R;[0,1])$, ${\supp}\,\chi\subset[-1,1]$,   $\chi=1$ on $[-1/2,1/2]$.  The function $\phi_a$ is the  positive ground state of $\mathfrak h[\zeta_a]$ with corresponding ground state energy $\beta_a$:
 \[\big(\mathfrak h[\zeta_a]-\beta_a\big)\phi_a=0 \,.\]
We now explain the construction of $\phi_a^{\rm cor}$. 
By \eqref{eq:exp-H-beta},  starting from some $\phi_a^{\rm cor}$ to be determined, 
 \begin{multline}\label{eq:decom-H-phi-a-cor}
 \big( \mathcal H_{a,\kappa ,h}-\beta_a-h^{1/2}\kappa M_3(a)\big) \left(\phi_a +h^{1/2}\kappa\, \phi_{a}^{\rm cor}\right)\\=
\kappa h^{1/2}\left(\big(\mathfrak h[\zeta_a]-\beta_a)\phi_a^{\rm cor}+\big(\mathfrak h^{(1)}[\zeta_a]-M_3(a)\big)\phi_a\right)+h\mathcal R_{a,h}\,,
 \end{multline}
where
 \[\mathcal R_{a,h}=L_h\left(\phi_a+h^{1/2}\kappa\, \phi_{a}^{\rm cor}\right)
 +\kappa^2\big( \mathfrak h^{(1)}[\zeta_a] -M_3(a)\big)\phi_a^{\rm cor}\,.
 \]
 Note that, by Remark~\ref{prop:mom},
$\mathfrak h^{(1)}[\zeta_a]\phi_a-M_3(a)\phi_a$ is orthogonal to $ \phi_a$ in $L^2(\R)$\,. 
Hence we can choose 
\begin{equation}\label{eq:phi-a-cor}
\phi_{a}^{\rm cor} =-{\mathfrak R}_a\big(
\mathfrak h^{(1)}[\zeta_a]\phi_a-M_3(a)\phi_a\big)\,,
\end{equation}
 so that the coefficient of $h^{1/2}$ in \eqref{eq:decom-H-phi-a-cor} vanishes.  In this  way, we  infer from \eqref{eq:decom-H-phi-a-cor},
\[
 \big( \mathcal H_{a,\kappa ,h}-\beta_a-h^{1/2}\kappa M_3(a)\big) \left(\phi_a +h^{1/2}\kappa\, \phi_{a}^{\rm cor}\right)=h\mathcal R_{a,h}\,.
 \]
 Notice that $\phi_{a,h}$ is constructed   so that it  has compact support in $(-h^{-\delta}, h^{-\delta})$ hence satisfies the Dirichlet conditions at $\tau=\pm h^{-\delta}$.  
 Since,  $\phi_a$ and $\phi_a^{\rm cor}$ decay exponentially at infinity by Lemma~\ref{lem:res}, we deduce 
\begin{equation}\label{eq:7.30a}
 \| \mathcal H_{a,\kappa ,h}\phi_{a,h}-(\beta_a+h^{1/2}\kappa M_3(a))\phi_{a,h}\|_{X_{h,\delta}}=\mathcal O(h)\,.
 \end{equation}
We denote by $\phi^{\rm gs}_{a,h}$ the normalized ground state of the Dirichlet realization of $\mathcal H_{a,\kappa ,h}$ in the weighted space 
$X_{h,\delta}$ (i.e. in $L^2((-h^{-\delta},h^{-\delta});(1-h^{1/2}\kappa \tau)d\tau)$).  By \eqref{eq:min-max-1dw}, the min-max principle  and  Proposition~\ref{prop:1dw}, we have
\begin{equation}\label{eq:w-op-gap-LB}
 \lambda_1( \mathcal H_{a,\kappa ,h})=\beta_a+h^{1/2}\kappa M_3(a)+\mathcal O(h)\mbox{ and } \lambda_2( \mathcal H_{a,\kappa ,h})\geq \mu_2(\zeta_a)+o(1) \,,
 \end{equation}
 so we infer from \eqref{eq:7.30a} and the H$\ddot{\rm o}$lder inequality
\[ \big\langle \big(\mathcal H_{a,\kappa ,h}\phi_{a,h}- \lambda_1( \mathcal H_{a,\kappa ,h}) \big)\big(\phi_{a,h}^{\rm gs} -\phi_{a,h}\big),\phi_{a,h}^{\rm gs} -\phi_{a,h}\big\rangle_{X_{h,\delta}}=\mathcal O(h)\|\phi_{a,h}^{\rm gs} -\phi_{a,h}\|_{X_{h,\delta}}\,.\]
Thus, by the spectral theorem,
\begin{equation}\label{eq:gs-phi-a,h}
\big\| \phi_{a,h}^{\rm gs} -\phi_{a,h}\big\|_{X_{h,\delta}}+\big\| \tau\big(\phi_{a,h}^{\rm gs} -\phi_{a,h}\big)\big\|_{X_{h,\delta}}+
\big\| \partial_\tau\big(\phi_{a,h}^{\rm gs} -\phi_{a,h}\big)\big\|_{X_{h,\delta}}=\mathcal  O(h)\,.
\end{equation}
\subsubsection{ New projections}
  We fix $h_0>0$ so that $1-h^{\frac12-\delta}_0\kappa>\frac12$. In the sequel, the parameter $h$ varies in the interval $(0,h_0)$.
 Consider the space 
\begin{equation}\label{eq:space-Xh-R2}
   X_{h,\delta}^2=L^2\big(\R\times(-h^{-\delta},h^{\delta}); (1- h^{1/2} \kappa \tau) d\sigma d\tau\big)\end{equation}
 endowed with the weighted norm
 \[\|v\|_{X_{h,\delta}^2}=\left(\int_\R\int_{-h^{-\delta}}^{h^{-\delta}} |v(\sigma,\tau)|^2(1-h^{1/2}\kappa\tau)d\sigma d\tau\right)^{1/2} \] 
which is equivalent to the usual norm of $L^2\big(\R\times(-h^{-\delta},h^{\delta})\big)$.

 We introduce the following  two operators
\begin{equation}\label{eq:proj:R-h}
R_h^-: v\in { X_{h,\delta}^2}\mapsto \int_{\R}\phi_{a,h}(\tau)v(\cdot,\tau)(1-h^{1/2}\kappa \tau)d\tau\in  L^2(\mathbb R)\,,
\end{equation}
and
\begin{equation}\label{eq:proj:R+h}
R_h^+:f\in L^2(\R)\mapsto f\otimes \phi_{a,h}\in X_{h,\delta}^2\quad{\rm where~}f\otimes \phi_{a,h}(\sigma,\tau)=f(\sigma)\phi_{a,h}(\tau)\,.
\end{equation}
 The image of  $R_h^+R_h^-$ is  $ L^2(\R)\otimes {\rm span} (\phi_{a,h})$.
Furthermore, for all $ v\in X_{h,\delta}^2$,  the functions $R_h^+R_h^-v$ and $v-R_h^+R_h^-v$ are orthogonal in $ X_{h,\delta}^2$, since the operator $R_h^+R_h^-$ can be expressed as follows
\begin{equation}\label{eq:Piah}
	\Pi_h:=R_h^+R_h^- = I \otimes \pi_{a,h}\,,
	\end{equation}
	where $\pi_{a,h}$ is the orthogonal   projection, in the weighted Hilbert space $X_{h,\delta}$,  on the space ${\rm span\,}\phi_{a,h}$\,.
With this projection in hand, we can approximate the truncated bound state $v_{h,n}$, introduced in \eqref{eq:ef-tranc-v}, with better error terms, thereby improving  Proposition~\ref{prop:app-ef}.

\begin{proposition}\label{prop:app-ef*}
The following holds 
\[ \|v_{h,n}-  \Pi_h v_{h,n}\| _{X_{h,\delta}^2}+\|\partial_\tau(v_{h,n}- \Pi_h v_{h,n})\| _{X_{h,\delta}^2}
+\|\tau (v_{h,n}- \Pi_h v_{h,n})\| _{X_{h,\delta}^2}=\tO(h^{5/16})\,,\]
where $\Pi_h$ is the projection in \eqref{eq:Piah}.
\end{proposition}
~

\begin{rem}\label{rem:app-ef*L2}
By \eqref{eq:phi-a,h} and \eqref{eq:decom-H-phi-a-cor}, we observe that, 
\[\|(\Pi_h-\Pi_0)v_{h,n}\|_{L^2(\R^2)}+\|(\partial_\tau\Pi_h-\partial_\tau \Pi_0)v_{h,n}\|_{L^2(\R^2)}+
\|\tau(\Pi_h-\Pi_0)v_{h,n}\|_{L^2(\R^2)}=\mathcal O(h^{1/2})\,,\]
where $\Pi_0$ is the projection introduced in \eqref{eq:Pia}. 
Since the norm of $X_{h,\delta}^2$ is equivalent to the usual norm of $L^2$, Proposition~\ref{prop:app-ef*} yields the following improvement  of Proposition~\ref{prop:app-ef}, 
\begin{equation}\label{eq:app-ef*}
\|v_{h,n}- \Pi_0 v_{h,n}\| _{L^2(\R^2)}+\|\partial_\tau(v_{h,n}- \Pi_0 v_{h,n})\| _{L^2(\R^2)}
+\|\tau (v_{h,n}- \Pi_0 v_{h,n})\| _{L^2(\R^2)}=\tO(h^{5/16})\,,
\end{equation}
where $\Pi_0$ is the projection in \eqref{eq:Pia}. This remark  will be useful in the next subsection.
\end{rem}
~\\

\begin{proof}[Proof of Proposition~\ref{prop:app-ef*}]~

{\it Step~1.}

We give here preliminary estimates that we will use in Step~3 below. Firstly, by Remark~\ref{rem:hyp-sec-tf},
\begin{equation}\label{eq:t4-vh}
\int_{\R^2} \tau^4|v_{h,n}(\sigma,\tau)|^2d\sigma d\tau =\mathcal O(1)\,.
\end{equation}
Secondly, we will prove that
\begin{equation}\label{eq:qf-IPtermh}
\langle h^{3/8}\partial_\sigma v_{h,n},(b_a(\tau)\tau+\zeta_a)v_{h,n}\rangle_{L^2(\R^2)}=\tO(h^{5/8})\,,
\end{equation}
By  \eqref{eq:ds-v} and  Proposition~\ref{prop:app-ef},
\begin{multline*}
 |\langle h^{3/8}\partial_\sigma v_{h,n},(b_a(\tau)\tau+\zeta_a)(v_{h,n}-\Pi_0v_{h,n})\rangle_{L^2(\R^2)}|\\ \leq \|h^{3/8}\partial_\sigma v_{h,n}\|_{L^2(\R^2)}\|(b_a(\tau)\tau+\zeta_a)(v_{h,n}-\Pi_0 v_{h,n} )\|_{L^2(\R^2)}=\tO(h^{5/8})\,.\end{multline*}
Similarly, using  \eqref{eq:7.20} and H$\ddot{\rm o}$lder's inequality, we write
\begin{multline*}
 |\langle  (b_a(\tau)\tau+\zeta_a)h^{3/8}\partial_\sigma\Pi_0v_{h,n}, v_{h,n}-\Pi_0 v_{h,n}  \rangle_{L^2(\R^2)}|\\
 \leq \|h^{3/8}\Pi_0\partial_\sigma v_{h,n}\|_{L^2(\R^2)}\|(b_a(\tau)\tau+\zeta_a)(v_{h,n}-\Pi_0 v_{h,n} )\|_{L^2(\R^2)}=\tO(h^{5/8})\,. 
 \end{multline*}
Now, writing 
$v_{h,n}=\Pi_0v_{h,n}+(v_{h,n}-\Pi_0v_{h,n})$ and collecting the foregoing estimates, we get 
\begin{align*}
\langle h^{3/8}&\partial_\sigma v_{h,n},(b_a(\tau)\tau+\zeta_a)v_{h,n}\rangle_{L^2(\R^2)}\\
&=
\langle h^{3/8}\partial_\sigma v_{h,n},(b_a(\tau)\tau+\zeta_a)\Pi_0v_{h,n}\rangle_{L^2(\R^2)}
+\tO(h^{5/8})\\
&=-\langle  (b_a(\tau)\tau+\zeta_a)v_{h,n},h^{3/8}\partial_\sigma\Pi_0v_{h,n}\rangle_{L^2(\R^2)}+\tO(h^{5/8})\quad\rm  {[by~integration~by~parts]}\,.
\end{align*}
Again, decomposing $v_{h,n}$ by the projection $\Pi_0$ and observing  that \eqref{eq:7.21} yields
\[\langle  (b_a(\tau)\tau+\zeta_a)\Pi_0v_{h,n},h^{3/8}\partial_\sigma\Pi_0v_{h,n}\rangle_{L^2(\R^2)}=0\,,\]
we get 
\begin{align*}
\langle h^{3/8}&\partial_\sigma v_{h,n},(b_a(\tau)\tau+\zeta_a)v_{h,n}\rangle_{L^2(\R^2)}\\
&=-\langle  (b_a(\tau)\tau+\zeta_a)h^{3/8}\partial_\sigma\Pi_0v_{h,n}, v_{h,n}-\Pi_0 v_{h,n}  \rangle_{L^2(\R^2)}+\tO(h^{5/8})=\tO(h^{5/8})\,,
\end{align*}
thereby obtaining \eqref{eq:qf-IPtermh}.\\

{\it Step~2:}

We introduce operators  involving the ground state $\phi_{a,h}^{\rm gs}$ as follows. First we introduce the operators,
\begin{equation}\label{eq:proj:R-h*}
\tilde R_h^-: v\in  X^2_{h,\delta}\mapsto \int_{\R}\phi_{a,h}^{\rm gs}(\tau)v(\cdot,\tau)(1-h^{1/2}\kappa \tau)d\tau  \in L^2(\R) \,,
\end{equation}
and
\begin{equation}\label{eq:proj:R+h*}
\tilde R_h^+:f\in L^2(\R)\mapsto f\otimes \phi_{a,h}^{\rm gs}\in X_{h,\delta}^2 \quad{\rm where~}(f\otimes \phi_{a,h}^{\rm gs})(\sigma,\tau)=f(\sigma)\phi_{a,h}^{\rm gs}(\tau)\,.
\end{equation}
Denoting by $\tilde \pi_{a,h}$ the orthogonal projection, in $X_{h,\delta}$,  on the space ${\rm span\,}\phi_{a,h}^{\rm gs}$, we introduce
	\begin{equation}\label{eq:t-Piah}
	\tilde\Pi_h:=\tilde R_h^+\tilde R_h^- = I \otimes \tilde\pi_{a,h} \,.
	\end{equation}
By \eqref{eq:gs-phi-a,h} and \eqref{eq:Piah}, we observe that, for all $g\in X_{h,\delta}$ and $f\in X_{h,,\delta}^2$, we have
\[
\|(\tilde R_h^--R_h^-)g\|_{X_{h,\delta}}=\mathcal O(h)\|g\|_{X_{h,\delta}},~\|(\tilde \Pi_h-\Pi_h )f\|_{X_{h,\delta}^2}=\mathcal O(h)\|f\|_{X_{h,\delta}^2}\,.
\]
So if  we prove that
\begin{equation}\label{eq:prop-app-t-ef*}
 \|v_{h,n}-  \tilde\Pi_h v_{h,n}\| _{X_{h,\delta}^2}+\|\partial_\tau(v_{h,n}- \tilde\Pi_h v_{h,n})\| _{X_{h,\delta}^2}
+\|\tau (v_{h,n}- \tilde\Pi_h v_{h,n})\| _{X_{h,\delta}^2}=\tO(h^{5/16})\,,\end{equation}
then we deduce the estimate in Proposition~\ref{prop:app-ef*}.\\

{\it Step~3.}

Adapting the proof of Proposition~\ref{prop:app-ef}, we  prove  now \eqref{eq:prop-app-t-ef*}.
  By Remark~\ref{rem:hyp-sec-tf}, we can use   \eqref{eq:qf-u-tranc} with $w_h= u_{h,n}$,  $r_h=1$, $m_h=\| u_{h,n}\|_{X_{h,\delta}^2}^2=1+\mathcal O(h^{1/2})$ and $\theta=\frac14$. Thus
\begin{multline}\label{eq:qf-u-tranc-new}
 \int_{\R}\int_{-h^{-\delta}}^{h^{-\delta}}\left(|\partial_\tau  u_{h,n}|^2+(1+2\kappa  h^\frac 12\tau)\Big|\Big(h^{3/8}\partial_\sigma+i \Big(b_a\tau  -\kappa  h^\frac 12 b_a \frac {\tau^2}2\Big)\Big) u_{h,n}\Big|^2\right)(1-\kappa  h^\frac 12\tau)\,d\sigma d\tau\\
\leq \Big(\beta_a+h^{1/2}M_3(a)\kappa +\mathcal O(h^{3/4})\Big)\| u_{h,n}\|_{X_{h,\delta}^2}^2\,.
\end{multline}
Since $u_{h,n}=e^{i\zeta_a\sigma/h^{3/8}}v_{h,n}$ (by~\eqref{eq:ef-tranc-v}), we get 
\begin{multline}\label{eq:qf-v-tranc-new}
 \int_{\R}\int_{-h^{-\delta}}^{h^{-\delta}}|\partial_\tau  v_{h,n}|^2(1-\kappa  h^\frac 12\tau)\,d\sigma d\tau\\
 + \int_{\R}\int_{-h^{-\delta}}^{h^{-\delta}}(1+2\kappa  h^\frac 12\tau)\Big|\Big(h^{3/8}\partial_\sigma+i \Big(b_a \tau +\zeta_a -\kappa  h^\frac 12 b_a \frac {\tau^2}2\Big)\Big) v_{h,n}\Big|^2(1-\kappa  h^\frac 12\tau)\,d\sigma d\tau\\
\leq \Big(\beta_a+h^{1/2}M_3(a)\kappa +\mathcal O(h^{3/4})\Big)\| v_{h,n}\|_{X_{h,\delta}^2}^2\,.
\end{multline}
Using \eqref{eq:ds-v}, \eqref{eq:qf-IPtermh} and \eqref{eq:t4-vh}, we deduce the following estimate from \eqref{eq:qf-v-tranc-new},
\begin{multline}\label{eq:qf-v-tranc-new*}
\int_{\R}\int_{-h^{-\delta}}^{h^{-\delta}}\left(|\partial_\tau  v_{h,n}|^2+(1+2\kappa  h^\frac 12\tau)\Big|\Big(b_a\tau +\zeta_a -\kappa  h^\frac 12 b_a\frac {\tau^2}2\Big) v_{h,n}\Big|^2\right)(1-\kappa  h^\frac 12\tau)\,d\sigma d\tau\\
\leq \Big(\beta_a+h^{1/2}M_3(a)\kappa +\tO(h^{5/8})\Big)\| v_{h,n}\|_{X_{h,\delta}^2}^2\,,
\end{multline}
where we used also that $\| v_{h,n}\|_{X_{h,\delta}^2}^2=1+\mathcal O(h^{1/2})$, by  \eqref{eq:norm-u} and \eqref{eq:ef-tranc-v}.

Now we get \eqref{eq:prop-app-t-ef*}  by decomposing $v_{h,n}$ in $X_{h,\delta}^2$ in the form
\[ v_{h,n}=\tilde r_h+\tilde r_{h,\bot}, \quad \tilde r_h:=\tilde\Pi_h v_{h,n}, \quad \tilde r_{h,\bot}=(I-\tilde\Pi_h)v_{h,n}\,, \]
and by using the spectral asymptotics for the operator $\mathcal H_{h,a,\kappa}$,  recalled in \eqref{eq:w-op-gap-LB}. 
\end{proof}
~
\subsection{Quasi-modes   for the effective operator}~

Let us start with some heuristic considerations.
 The derivation of the eigenvalue upper bound of  Theorem~\ref{thm:up} suggested  in the tangent variable  the following one dimensional effective operator (see  \eqref{eq:harm}) 
\begin{equation}\label{eq:eff-op-lb}
 H^{\rm harm}_a=-c_2(a)\partial_\sigma^2-\frac{M_3(a)k''(0)}{2}\sigma^2\,,
 \end{equation}
 where $c_2(a)>0$ is introduced in \eqref{eq:main-ct}.\\
 Moreover, 
 by Remark~\ref{rem:eff-qm}, it is natural to consider 
 the following quasi-mode   
\[ v_{h,n}^{\rm app}=\Big(\phi_a(\tau)+2\mathfrak R_a\big((\zeta_a+b_a(\tau)\tau)\phi_a\big) ih^{3/8}\partial_\sigma+k_{\max}h^{1/2}\phi_a^{\rm cor} (\tau)\Big)f_n(\sigma)\]
 where $\mathfrak R_a$ is the regularized resolvent introduced in \eqref{eq:R},  $\phi_a^{\rm cor}$ is the function in \eqref{eq:phi-a-cor}, and $f_n$ is the normalized $n$th eigenfunction of the operator $H^{\rm harm}_a$.
Denoting by $\Pi_{h,n}^{\rm app}$  the orthogonal projection, in $L^2(\R^2)$,  on the space generated by $v_{h,n}^{\rm app}$, 
 we  observe formally, by neglecting the terms with  coefficients having order lower than $h^{3/4}$,
 \[ c_2(a)\Pi_{h,n}^{\rm app}\mathcal P_h^{\rm new}\approx h^{1/2}\big(M_3(a)k_{\max}+h^{1/4}H^{\rm harm}_a\big) \Pi^{\rm new}_n\,,\]
 where  $\Pi^{\rm new}_n$ is  the projection, in $L^2(\R^2)$, on the  space generated by the function $\varphi_a(\tau)f_n(\sigma)$, and 
 \begin{equation}\label{eq:varphi-a-new}
 \varphi_a(\tau):= \phi_a(\tau)-4(b_a(\tau) \tau +\zeta_a)\mathfrak R_a\big((b_a(\tau) \tau +\zeta_a)\phi_a(\tau)\big)\,. 
 \end{equation}
 Guided by these heuristic observations, we will use
  the truncated bound state  $v_{h,n}$ in \eqref{eq:ef-tranc-v} to construct quasi-modes of the operator $H^{\rm harm}_a$, by projecting  $v_{h,n}$  on the vector space generated by the  function $\varphi_a$  introduced in \eqref{eq:varphi-a-new}.
To that end, we introduce the following operator
\begin{equation}\label{eq:proj:R-h**}
 R_0^{{\rm new}}: v\in L^2 (\R^2)\mapsto \int_{\R}\varphi_a(\tau) v(\cdot,\tau) d\tau\in L^2(\R)\,.
\end{equation}

We will prove the following proposition.
\begin{proposition}\label{prop:lb-Rnew-qm} Let $n\in\mathbb N$ be fixed. The following holds:
\begin{enumerate}
\item $\Big\| R^{\rm new}_0v_{h,n}-(1-4I_2(a))R_0^-v_{h,n}\Big\|_{L^2(\R)}=\mathcal O(h^{1/4})$ 
where  $R_0^-$ is the operator in \eqref{eq:proj:R-} and $I_2(a)$ is introduced in \eqref{eq:I2}.\medskip
\item 
$\| R^{\rm new}_0v_{h,n}\|_{L^2(\R)}=1-4I_2(a)+\mathcal O(h^{1/4})$\,.\medskip
\item For every $n\in\mathbb N$, there exists  $h_n>0$ such that, for all $h\in(0,h_n)$,
\begin{equation}
 \langle R_{0}^{\rm new}v_{h,k},R_{0}^{\rm new}v_{h,k'} \rangle_{L^2(\R)}=(1-4I_2(a))^2\delta_{k,k'}+o(1)\qquad (1\leq k,k'\leq n)\,,
 \end{equation}
and
\begin{equation}\label{eq:sp-M}
M_n={\rm span}(R_{0}^{\rm new}v_{h,k},~1\leq k\leq n) ~{\rm satisfies}~{\rm dim}(M_n)=n\,.
\end{equation}
\item 
We have as $h\to0_+$
\[\big\langle \big(H^{\rm harm}_a-h^{-3/4}\Lambda_n(h)\big)R^{\rm new}_0v_{h,n},  R^{\rm new}_0v_{h,n}\big\rangle_{L^2(\R)} = o(1) \|R^{\rm new}_0v_{h,n}\|_{L^2(\R)}^2 \]
where 
\begin{equation*}\label{eq:l}
\Lambda_n(h)=h^{-1}\lambda_n(h)-\beta_a-M_3(a)k_{\max} h^{1/2}\,,\end{equation*}
and $ H^{\rm harm}_a$ is the operator introduced in \eqref{eq:eff-op-lb}.
\end{enumerate}
\end{proposition}
\begin{proof}~\\
{\it Proof of item~(1).}
Consider $\Pi_0=R_0^+R_0^-$  the projection introduced in \eqref{eq:Pia}. By  \eqref{eq:I2}, $R^{\rm new}_0R_0^+ = (1-4I_2(a)) {\rm} Id$, hence,  composing by $R_0^-$ on the right gives
\[R^{\rm new}_0\Pi_0=(1-4I_2(a))R_0^{-}\,. \]
Writing $v_{h,n}=\Pi_0v_{h,n} +(v_{h,n}-\Pi_0v_{h,n})$, we get
\begin{align*}
R^{\rm new}_0v_{h,n}&=R^{\rm new}_0\Pi_0v_{h,n} +R^{\rm new}_0(v_{h,n}-\Pi_0v_{h,n})\\
&=(1-4I_2(a))R_0^{-}v_{h,n} +R^{\rm new}_0(v_{h,n}-\Pi_0v_{h,n}).
\end{align*}
Then we observe that
\[\big\|R^{\rm new}_0( v_{h,n}-\Pi_0v_{h,n})\big\|_{L^2(\R)}\leq \|\varphi_a\|_{L^2(\R)}\|v_{h,n}-\Pi_0v_{h,n}\|_{ L^2(\R^2)}=\mathcal O(h^{1/4})\]
by H$\ddot{\rm o}$lder's inequality and Proposition~\ref{prop:app-ef}. This yields the conclusion of item~(1).
\medskip

{\it Proof of item~(2).}

By \eqref{eq:I2a}, $1-4I_2(a)>0$.
By \eqref{eq:proj:R-} and Proposition~\ref{prop:app-ef}, we have
\[ \|R_0^-v_{h,n}\|_{L^2(\R)}=\|\Pi_0v_{h,n}\|_{L^2(\R^2)}=1+\mathcal O(h^{1/4})\,.\]
Now item~(2) follows from item~(1).\medskip

{\it Proof of item~(3).} 
 If $1\leq k,k'\leq n$  and $k\not=k'$, we have as $h\to0_+$,
\[ \langle v_{h,k},v_{h,k'}\rangle_{L^2(\R^2)}=o(1) +  \delta_{k,k'}\,.\]
By Proposition~\ref{prop:app-ef}, we get further
\[ \langle R_0^-v_{h,k},R_0^-v_{h,k'}\rangle_{L^2(\R)}=\langle \Pi_0v_{h,k},\Pi_0v_{h,k'}\rangle_{L^2(\R^2)}=o(1)+\delta_{k,k'}\,.\]
Thus, by item~(1),
\[  \langle R_0^{\rm new}v_{h,k},R_0^{\rm new}v_{h,k'}\rangle_{L^2(\R)}=o(1)+\delta_{k,k'}\,.\]
With item~(2) in hand, we get the conclusion of item~(3).\medskip

{\it Proof of item~(4).}

{\it Step~1.}
 We introduce the following operator 
\begin{equation}\label{eq:proj:tR-h**}
\tilde R^{{\rm new}}_h: v\in H^1 (\R^2) \mapsto \int_{\R}\phi_{a,h}^{\rm new}(\tau,i\partial_\sigma) v(\cdot,\tau) d\tau\in L^2(\R)\,,
\end{equation}
where $\phi_{a,h}^{\rm new}(\tau,i\partial_\sigma)$ is the  first order differential operator,
\begin{equation}\label{eq:phi-a,h*}
\phi^{\rm new}_{a,h}(\tau,i\partial_\sigma):=\phi_a(\tau) +2h^{3/8}{\mathfrak R}_a\big((b_a(\tau) \tau +\zeta_a)\phi_a(\tau)\big)i\partial_\sigma+\kappa h^{1/2}\phi_a^{\rm cor}(\tau)\,,
\end{equation}
$\kappa=k_{max}$ and $\phi_a^{\rm cor}$ is the function introduced in \eqref{eq:phi-a-cor}.

By  H$\ddot{\rm o}$lder's inequality, there exists a constant $C_1$  such that, for all $v\in H^1(\R^2)$,
\begin{equation}\label{eq:7.57a}
 \|\tilde R^{\rm new}_{h}v\|_{L^2(\R)}\leq C_1\big(\|v\|_{L^2(\R^2)}+\|\partial_\sigma v\|_{L^2(\R^2)}\big)\,.
 \end{equation}
Thus, by Proposition~\ref{prop:app-ef} and Remark~\ref{rem:item(1)H1}, 
\begin{equation}\label{eq:RP=evRP}
\big\| \tilde R^{\rm new}_h\mathcal P_{h}^{\rm new}v_{h,n}-(h^{-1}\lambda_{n} (h)-\beta_a)\tilde R^{\rm new}_hv_{h,n}\big\|_{L^2(\R)}=\mathcal O(h^\infty)\,,\end{equation}
where $\mathcal P_{h}^{\rm new}$ is the operator in \eqref{eq:P1}. \medskip

{\it Step~2.} We prove the following estimate
\begin{equation}\label{eq:Rnew-Ph}
\left\langle \big(c_2(a)\tilde R^{\rm new}_h\mathcal P_{h}^{\rm new} -M_3(a)k_{\max}h^{1/2} R_0^{\rm new} -h^{3/4}H^{\rm harm}_aR_0^{\rm new}\big)v_{h,n},R_0^{\rm new}v_{h,n}\right\rangle_{L^2(\R)}=o(h^{3/4})\,.
\end{equation}
We first observe that it  results from \eqref{eq:proj:R-}, \eqref{eq:ds-v},  \eqref{eq:proj:tR-h**}, and \eqref{eq:phi-a,h*},
\begin{equation}\label{eq:tRnew=R0}
\|\tilde R_h^{\rm new}v_{h,n}-R_0^-v_{h,n}\|_{L^2(\R)}=\mathcal O(h^{1/2})\,.
\end{equation}
 For the sake of simplicity, we write $\kappa=k(0)=k_{\max}$.  We introduce the following functions in $L^2(\R)$,
\begin{equation}\label{eq:f1}
f_1=2{\mathfrak R}_a\big((b_a(\tau) \tau +\zeta_a)\phi_a\big)
\end{equation}
and (see \eqref{eq:h1-Lha} and \eqref{eq:phi-a-cor}) 
\begin{equation}\label{eq:f2} f_2=\phi_a^{\rm cor}={\mathfrak R}_a\Big(
M_3(a)\phi_a-\phi'_a-2\tau(b_a(\tau) \tau +\zeta_a)^2\phi_a+b_a(\tau) \tau ^2 (b_a(\tau) \tau +\zeta_a)\phi_a\Big)\,.
\end{equation}

Recall the operators  $P_0,P_1,P_2,P_3,Q_h$ introduced in \eqref{eq:LOT-Ph} and \eqref{eq:Q(h)}. Noticing the decomposition in \eqref{eq:P2}, we write,  for any function $v$ with compact support in $\R^2$,
\begin{equation}\label{eq:decomp-RPv}
\begin{aligned}
\tilde R^{\rm new}_h&\mathcal P_{h}^{\rm new}v\\
&=\int_\R \phi_a(\tau)P_0v(\sigma,\tau)\,d\tau\\ &\quad +
h^{3/8}\int_{\R} \Big(if_1(\tau)\partial_\sigma P_0+\phi_a(\tau) P_1 \Big)v(\sigma,\tau)d\tau\\
&\quad 
+h^{1/2}\int_{\R}\Big(\phi_a(\tau)P_2+\kappa  f_2 (\tau)P_0 \Big)v(\sigma,\tau)d\tau\\
&\quad +
h^{3/4}\int_{\R}\Big( \phi_a(\tau)P_3+if_1(\tau)\partial_\sigma P_1\Big)v(\sigma,\tau)d\tau\\&\quad  + \mathrm R_{h,n}v,
\end{aligned}
\end{equation}
where
\begin{multline}\label{eq:decom-R}
\mathrm  R_{h,n}v=h^{7/8}\tilde R^{\rm new}_hQ_hv\\ +h^{7/8}\int_{\R}\Big( if_1(\tau)\partial_\sigma P_2 
 +\kappa f_2(\tau)P_0\Big)
v(\sigma,\tau)d\tau \\ 
 +h\kappa \int_\R f_2(\tau)P_2v(\sigma,\tau)d\tau \\ \qquad \qquad  +h^{5/4}\kappa \int_\R f_2(\tau)P_3v (\sigma,\tau)d\tau \\ \qquad \quad  +h^{9/8}\kappa \int_\R if_1(\tau)\partial_\sigma P_3v(\sigma,\tau)d\tau\,.
\end{multline}

We now  compute the first three terms on the right side of \eqref{eq:decomp-RPv}.\\
\item[$(*)$]  For the first term, since $P_0$ is self-adjoint in $L^2(\R)$, we have
\[
\int_\R \phi_a(\tau)P_0v(\sigma,\tau)\,d\tau=\int_\R P_0\phi_a(\tau) v(\sigma,\tau)d\tau=0\,.\]
\item[$(**)$] For the second term, we have 
\begin{align*} \int_{\R} if_1(\tau)\partial_\sigma P_0 v(\sigma,\tau)d\tau&=\int_{\R} iP_0f_1(\tau)\partial_\sigma v(\sigma,\tau)d\tau\\
&=\int_{\R} 2i \phi_a(\tau)(b_a(\tau) \tau +\zeta_a)  \partial_\sigma v(\sigma,\tau)d\tau\,.
\end{align*}
Hence we find, by \eqref{eq:LOT-Ph}, 
\[
\int_{\R} \Big(if_1(\tau)\partial_\sigma P_0+\phi_a(\tau) P_1 \Big)v(\sigma,\tau)d\tau=0\,.
\]
\item[$(***)$]  For the third term, noticing that $$P_0f_2= M_3(a)\phi_a-\phi'_a-2\tau(b_a(\tau) \tau +\zeta_a)^2\phi_a+b_a(\tau) \tau ^2 (b_a(\tau) \tau +\zeta_a)\phi_a$$ and 
\begin{multline*}
 \int_\R \phi_a(\tau) P_2v(\sigma,\tau)d\tau \\
= \kappa \int_\R\Big(-\phi_a'(\tau)+2\tau(b_a(\tau) \tau +\zeta_a)^2\phi_a(\tau)-b_a(\tau) \tau ^2(b_a(\tau) \tau +\zeta_a)\phi_a(\tau) \Big)v \,d\tau\,,\end{multline*}
we get
\begin{multline*} (W_2v)(\sigma):=\int_{\R}\Big(\phi_a(\tau)P_2+\kappa  f_2 (\tau)P_0 \Big)v(\sigma,\tau)d\tau\\
=\int_{\R}\Big(\phi_a(\tau)P_2+\kappa  \big(P_0 f_2 (\tau)\big) \Big)v(\sigma,\tau)d\tau\\
=\kappa \int_\R\Big(M_3(a)\phi_a(\tau)-2\phi_a'(\tau) \Big)v(\sigma,\tau)d\tau\,.\end{multline*}
 By the forgoing computations, \eqref{eq:decomp-RPv}  becomes 
\begin{equation}\label{eq:decomp-RPv-NF}
\tilde R^{\rm new}_h\mathcal P_{h}^{\rm new}v=
h^{1/2}W_2v +
h^{3/4}W_3v  + \mathrm R_{h,n}v\,,
\end{equation}
with
\begin{equation}\label{eq:defW}
(W_3v)(\sigma):=\int_{\R}\Big( \phi_a(\tau)P_3+if_1(\tau)\partial_\sigma P_1\Big)v(\sigma,\tau)d\tau \,.
\end{equation}
We estimate $W_2v_{h,n}$  by  writing $v_{h,n}=\Pi_0v_{h,n}+(v_{h,n}-\Pi_0v_{h,n})$, with $\Pi_0$ the projection introduced  in \eqref{eq:Pia},  and by using \eqref{eq:app-ef*}. Eventually, since $P_0\Pi_0=0$ and $\langle\phi_a,\phi_a'\rangle_{L^2(\R)}=0$,   we get by Remark~\ref{prop:mom},
\begin{equation}\label{eq:P2term}
\big\|W_2v_{h,n}-M_3(a)\kappa R_0^-v_{h,n}\big\|_{L^2(\R)}=o(h^{1/4})\,.
\end{equation}
We still have to estimate the terms involving $W_3$ and $\mathrm R_{h,n}$ in \eqref{eq:decomp-RPv-NF}  when $v=v_{h,n}$. By choosing $\eta$ small enough, 
the following error term
\begin{equation}\label{eq:decom-r}
\mathrm r_n(\sigma,h) :=\mathrm  R_{h,n}v_{h,n}\,,
\end{equation}
 with $\mathrm  R_{h,n}$ introduced in \eqref{eq:decom-R}, satisfies
\begin{equation}\label{eq:term-r}
\langle \mathrm r_n(\cdot,h),R^{\rm new}_0v_{h,n}\rangle_{L^2(\R)}=o(h^{3/4}). 
\end{equation}

The technical proof of \eqref{eq:term-r} is given in Appendix~\ref{sec:app-r}. So  we are left (see \eqref{eq:defW}) with estimating 
\begin{equation}\label{eq:W3}
W_3v_{h,n} =w_1+w_2 \,,
\end{equation}
where
\begin{align*}
w_1(\sigma)&:=\int_\R\phi_a(\tau)P_3v_{h,n}(\sigma,\tau)d\tau\,,\\
w_2(\sigma)&:=\int_\R if_1(\tau)\partial_\sigma P_1v_{h,n}(\sigma,\tau)d\tau\,.\\
 \end{align*}
In light of the definition of $P_3$ in \eqref{eq:LOT-Ph} and $R_0^{-}$ in \eqref{eq:proj:R-}, we write 
\[
w_1(\sigma)=-\partial^2_\sigma R_0^-v_{h,n}(\sigma)
+ \frac {k''(0)\sigma^2}{2}w(\sigma)\,,
\]
where 
\[w(\sigma)=\int_\R \Big(\partial_\tau+2\tau(b_a(\tau) \tau +\zeta_a)^2-b_a(\tau) \tau (b_a(\tau) \tau +\zeta_a) \Big)\phi_a(\tau)\,v_{h,n}(\sigma,\tau)d\tau\,.\]
Using Proposition~\ref{prop:app-ef} and that $v_{h,n}$ is supported in $\{|\sigma|\leq h^{-\eta}\}$, we get  
\[ \|\sigma^2(w-M_3(a)R_0^-v_{h,n})\|_{L^2(\R)}=\mathcal O(h^{\frac14-2\eta})\,.\]
Hence
\begin{equation}\label{eq:w3}  
\Big\|w_{1}-\Big(-\partial^2_\sigma  +\frac{k''(0)M_3(a)}2\sigma^2\Big)R_0^-v_{h,n} \Big\|_{L^2(\R)}=\mathcal O(h^{\frac14-2\eta})\,.
\end{equation}
Furthermore, by \eqref{eq:LOT-Ph} and \eqref{eq:f1}, the term $w_2$ can be expressed as follows
\begin{equation}\label{eq:P3term*}
\begin{aligned}
w_2(\sigma)&=2\partial_\sigma^2\int_\R f_1(\tau) (\zeta_a+b_a(\tau)\tau)v_{h,n}(\sigma,\tau)d\tau\\
&=4\partial_\sigma^2\int_\R(b_a(\tau) \tau +\zeta_a){\mathfrak R}_a\big((b_a(\tau) \tau +\zeta_a)\phi_a(\tau)\big)v_{h,n}(\sigma,\tau)d\tau\,.
\end{aligned}
\end{equation}
Collecting \eqref{eq:w3} and \eqref{eq:P3term*}, along with the definition of $R_0^{\rm new}$ in \eqref{eq:proj:R-h**}, we infer form \eqref{eq:W3}
\begin{equation}\label{eq:P3term**}
\Big\|W_3v_{h,n}-\Big(-\partial^2_\sigma R^{\rm new}_0   +\frac{k''(0)M_3(a)}2\sigma^2R_0^-\Big)v_{h,n} \Big\|_{L^2(\R)}=\mathcal O(h^{\frac14-2\eta})\,.
\end{equation}
By H\"older's inequality, we infer from \eqref{eq:P2term} and  \eqref{eq:P3term**},
\begin{multline*}
 h^{1/2}\big\langle (W_2 -M_3(a)\kappa R_0^{-})v_{h,n},R_0^{\rm new}v_{h,n} \big\rangle_{L^2(\R)}\\
+h^{3/4}\Big\langle W_3v_{h,n} -\Big(-\partial_\sigma^2R_0^{\rm new} +\frac{k''(0)M_3(a)}{2}\sigma^2R_0^-\Big)v_{h,n}, R_0^{\rm new}v_{h,n} \Big\rangle_{L^2(\R)}\\=o(h^{3/4})\|R_0^{\rm new}v_{h,n}\|_{L^2(\R)}\,. \end{multline*}
By   \eqref{eq:decomp-RPv-NF} and \eqref{eq:term-r}, we get from the above estimate
\[
\left\langle \left(\tilde R_h^{\rm new}\mathcal P_h^{\rm new}-h^{1/2}M_3(a)\kappa R_0^{-}-h^{3/4}\tilde H \right)v_{h,n}, R_0^{\rm new}v_{h,n} \right\rangle_{L^2(\R)}
=o(h^{3/4})\|R_0^{\rm new}v_{h,n}\|_{L^2(\R)}\,,
\] 
where
\[\tilde H:=-\partial_\sigma^2R_0^{\rm new} +\frac{k''(0)M_3(a)}{2}\sigma^2R_0^-\,.\]
Finally, by item~(1) and   Proposition~\ref{prop:I2}, we get \eqref{eq:Rnew-Ph}.\\

{\it Step 3:} 

Using Steps~1 and 2, we are now able to finish the proof of item~(4).  
By \eqref{eq:main-ct} and  \eqref{eq:I2a}, $c_2(a)=1-4I_2(a)$, hence \eqref{eq:tRnew=R0} and item~(1) yield that
\begin{equation}\label{eq:tRnew=Rnew}
\|c_2(a)\tilde R_h^{\rm new}v_{h,n}-R_0^{\rm new}v_{h,n}\|_{L^2(\R)}=\mathcal O(h^{1/4})\,.
\end{equation}
Collecting \eqref{eq:RP=evRP}, \eqref{eq:Rnew-Ph} and \eqref{eq:tRnew=Rnew}, we get,
\[ \left\langle h^{3/4}H^{\rm harm}_aR_0^{\rm new}v_{h,n} -\Lambda_n(h)R_0^{\rm new}v_{h,n}, R_0^{\rm new}v_{h,n}\right\rangle_{L^2(\R)} =  \mathcal O(|\Lambda_n(h)|h^{1/4})+o(h^{3/4})\,, \]
where, by \eqref{eq:qh-u-lb} and Theorem~\ref{thm:up},
\[|\Lambda_n(h)|=|h^{-1}\lambda_n(h)-\beta_a-M_3(a)k_{\max} h^{1/2}|=o(h^{1/2})\,.\]
 Thus, we obtain 
\[ \left\langle h^{3/4}H^{\rm harm}_aR_0^{\rm new}v_{h,n} -\Lambda_n(h)R_0^{\rm new}v_{h,n} , R_0^{\rm new}v_{h,n}\right\rangle_{L^2(\R)} =o(h^{3/4})\,.\]
Dividing by $h^{3/4}$ and using item~(2), we get   item~(4).
\end{proof}

  With Proposition~\ref{prop:lb-Rnew-qm} in hand, we can now finish the proof of Theorem~\ref{thm:main}.
 \begin{proof}[Proof of Theorem~\ref{thm:main}]
 The upper bound of $\lambda_n(h)$ follows from Theorem~\ref{thm:up}. For the lower bound of $\lambda_n(h)$,  consider $u=\sum\limits_{k=1}^na_kR_0^{\rm new}v_{h,k}$  such that $\|u\|_{L^2(\R)}=1$, where $R_0^{\rm new}$ is introduced in \eqref{eq:proj:R-h**}. 
 It results   from  Proposition~\ref{prop:lb-Rnew-qm}, 
 \[\left((1-4I_2(a))^2+o(1)\right)\sum\limits_{k=1}^n|a_k|^2=1\]
 and
 \[ \big\langle \big(H^{\rm harm}_a-h^{-3/4}\Lambda_n(h)\big)u,  u\big\rangle_{L^2(\R)} \leq 
 o(1)\sum\limits_{k=1}^n|a_k|^2\,.\]
 Consequently, 
 \[\max_{\substack{u\in M_n\\ \|u\|=1}}\big\langle \big(H^{\rm harm}_a-h^{-3/4}\Lambda_n(h)\big)u,  u\big\rangle_{L^2(\R)} = o(1) \,,\]
 where $M_n$ is the space defined in~\eqref{eq:sp-M}.
 By  the min-max principle 
 \[ \sqrt{ \frac{M_3(a)k''(0)c_2(a)}{2}}(2n-1)\leq h^{-3/4}\Lambda_n(h)+o(1)\,,\]
thereby leading to
\[ \lambda_n(h)\geq \beta_ah+M_3(a)k_{\max}h^{3/2}+\sqrt{ \frac{M_3(a)k''(0)c_2(a)}{2}}(2n-1)h^{7/4}+o(h^{7/4})\,.\]
 \end{proof}
\subsection*{Acknowledgments} AK is partially supported by the Center for Advanced Mathematics Sciences (CAMS, American University of Beirut).

\appendix 

\section{Frenet coordinates near the magnetic edge}\label{sec:frenet}\

We introduce the Frenet coordinates near $\Gamma$. We  refer the reader to~\cite[Appendix~F]{fournais2010spectral} and~\cite{Assaad2019} for a similar setup. 

Let $s\mapsto M(s)\in\Gamma$ be  the arc length parametrization of $\Gamma$ such that
\begin{itemize}
	\item $\nu(s)$ is the unit normal of $\Gamma$ at the point $M(s)$ pointing towards  $P_1$\,;
	\item   $T(s)$ is the unit tangent vector of $\Gamma$ at the point $M(s)$, such that $(T(s),\nu(s))$ is a direct frame, i.e.
	$\mathrm{det}\big(T(s),\nu(s) \big)=1$.
\end{itemize}
We define the curvature $k$ of $\Gamma$  as follows:
$T'(s)=k(s)\nu(s)$.
 \emph{Working under Assumption~\ref{kmax}}, we assume  w.l.o.g that $s_0=0$, where $s_0$ is the unique maximum of the curvature at $\Gamma$ ($k(0)=k_{max}$).

For $t_0>0$, we define the transformation $\Phi= \Phi_{t_0}$ as follows
\begin{equation}\label{Frenet}
\Phi~:~  \R  \times(-t_0,t_0)\, \ni (s,t)\longmapsto M(s)+t \nu(s) \in \Gamma_{t_0} :=\{x\in\R^2~:~{\rm dist}(x,\Gamma)<t_0\}\,.
\end{equation}
We pick $t_0$ sufficiently small so that $\Phi$ is a diffeomorphism, whose  Jacobian is 
\begin{equation}\label{eq:a1}
\mathfrak a(s,t):=J_\Phi(s,t)=1-t\,k(s).
\end{equation}
We consider the following correspondence between functions $u$  in  $ H_\mathrm{loc}^1\big(\Gamma_{t_0} \big)$ and those $\tilde u$ in $  H^1_\mathrm{loc}(\R\times(-{t_0},{t_0}))$: 
\begin{equation}\label{eq:u-(s,t)}
\tilde{u}(s,t)=u\big(\Phi(s,t)\big),
\end{equation}
and vice versa.

Moreover, we assign to the potential $\Fb$ in~\eqref{eq:F} a vector field $\tilde \Fb \in H^1_\mathrm{loc}\big(\R\times(-{t_0},{t_0})\big)$ as follows
\[\Fb(x)=\big(F_{1}(x),F_{2}(x)\big) \mapsto \tilde{\Fb}(s,t)= \big(\tilde F_{1}(s,t),\tilde F_{2} (s,t)),\]
where 
\begin{equation}\label{eq:A_tild1}
\tilde F_{1}(s,t)=\mathfrak a(s,t)\Fb\big(\Phi(s,t) \big)\cdot T(s)\quad\mathrm{and}\quad \tilde F_{2} (s,t)=\Fb\big(\Phi(s,t) \big)\cdot\nu(s).
\end{equation}
Consequently,
\begin{equation}\label{eq:op-FC}
(h\nabla-i\Fb)^2=\mathfrak a^{-1}(h\partial_s-i\tilde{F}_{1})\mathfrak a^{-1}(h\partial_s-i\tilde{F}_{1})+\mathfrak a^{-1}(h\partial_t-i\tilde{F}_{2})\mathfrak a(h\partial_t-i\tilde{F}_{2})\,.
\end{equation}
 Note that
\begin{equation}\label{eq:curl}
\curl \tilde{\Fb}(s,t)=\big(1-tk(s)\big)\curl\Fb\big(\Phi(s,t)\big)=\big(1-tk(s)\big)\big(\mathbbm 1_{\{t>0\}}+a\mathbbm 1_{\{t<0\}}\big),
\end{equation}
where $\curl \tilde{\Fb}=\partial_s \tilde{F_2}-\partial_t \tilde{F_1}$ and $\curl\Fb=\partial_{x_1}{F_2}-\partial_{x_2}{F_1}$ is as in~\eqref{eq:curl-F}.

Furthermore, we present the  change of variable formulae (for functions compactly supported in $\Gamma_{t_0}$):
\begin{equation}\label{eq:A_tild2}
\begin{aligned}
& \int_{\Gamma_{t_0}}|u |^2\,dx=\int_\R\int_{-{t_0}}^{{t_0}}|\tilde u|^2\, \mathfrak a\,dt\,ds\,,\\
&\int_{\Gamma_{t_0}}\big|\big(h\nabla-i \Fb \big)u \big|^2\,dx= \int_{\R} \int_{-{t_0}}^{{t_0}}\left(\mathfrak a^{-2}\big|(h\partial_s-i\tilde{F}_{1})\tilde{u}\big|^2+\big|(h\partial_t-i\tilde{F}_{2})\tilde{u}\big|^2 \right)\, \mathfrak a\,dt\,ds.\\
\end{aligned}
\end{equation}	
 Now, we make a global change of gauge $\omega$ as follows:

\begin{lemma}\label{lem:Anew2}
 There exists a function $\omega
	\in H^2\big(\Phi^{-1}(\Gamma_{t_0}\cap\Om)\big)$  such that 
	\begin{equation*}\label{eq:gauge1}
	\tilde{\Fb}-\nabla_{s,t}\omega=\begin{pmatrix}
	- b_a(t)\big(t-\frac {t^2}2 k(s)\big)\\0
	\end{pmatrix}\ \mathrm{in}\ \Phi^{-1}(\Gamma_{t_0}\cap\Om),\end{equation*}
	 where $t\mapsto b_a(t)$ is defined by $b_a(t)=\mathbbm 1_{\{t>0\}}+a\mathbbm 1_{\{t<0\}}$.	
\end{lemma}
\begin{proof}
	For $(s,t) \in \Phi^{-1}(\Gamma_{t_0}\cap\Om)$, let $\omega(s,t)=\int_0^t \tilde{F}_2(s,t')\,dt'+\int^s_{0} \tilde{F}_1(s',0)\,ds'$. 
	This choice of $\omega$ and~\eqref{eq:curl} establish the lemma.
\end{proof}
 The gauge of Lemma~\ref{lem:Anew2} is adequate when working with functions localized near the edge $\Gamma$.
With this choice of gauge, we have the following identity which is useful to analyze the decay of functions localized near $\Gamma$.
 
\begin{lemma}\label{lem:Adecom-form}
Assume that $\varphi\in H^2(\Omega)$  with compact support in $\Omega\cap\Gamma_{t_0}$.
Let $g$ and $G$ be the functions defined (by means of \eqref{eq:u-(s,t)}) as follows
\[\tilde g(s,t)=(h^{1/2}\partial_s-i\zeta_a)\tilde \varphi(s,t)~{\rm and}~\tilde G(s,t)=-(h^{1/2}\partial_s-i\zeta_a)\big(e^{2\tilde \phi}\tilde g \big)\,, \]
where $\zeta_a$ is the constant in Subsection \ref{sec:p-bnd-fc} and $\phi$ is a  Lipschitz real-valued function  on $\Omega$. If $g\in H^2(\Omega)$,  then
\[ {\rm Re}\langle \mathcal P_h \varphi, G\rangle_{L^2(\Omega)}=\mathcal Q_h(e^{\phi} g) -h^2\big\||\nabla g|e^{\phi}\varphi\big\|_{L^2(\Omega)}^2-h^{1/2}{\rm Re}(\mathsf T_h)\,.\]
Here $\mathcal Q_h$ is the quadratic form introduced in \eqref{eq:Q(h)} and
\begin{multline*}
\mathsf T_h= \Big\langle(h\partial_s-i\tilde{F}_{1})\Big( \big(\partial_s \mathfrak a^{-1}-i\mathfrak a^{-1}\partial_s\tilde F_1)\big)(h\partial_s-i\tilde{F}_{1})\tilde\varphi -i\mathfrak a^{-1}(\partial_s\tilde F_1)\tilde\varphi\Big) +h^2\partial_t\big(\partial_s\mathfrak a \big)\partial_t\tilde\varphi,  e^{2\tilde\phi}\tilde g\Big\rangle_{L^2(\R)}\,.
\end{multline*}
\end{lemma} 
\begin{proof}
We assume that $\tilde F_2=0$ and get from \eqref{eq:op-FC} and \eqref{eq:a1}
\begin{equation}\label{eq:Ap-loc1}
\langle \mathcal P_h \varphi, G\rangle_{L^2(\Omega)}=\langle (h\partial_s-i\tilde{F}_{1})\mathfrak a^{-1}(h\partial_s-i\tilde{F}_{1}) \varphi+h^2\partial_t \mathfrak a\partial_t\varphi\,,(h^{1/2}\partial_s-i\zeta_a)\big(e^{2\phi} g \big)\rangle_{L^2(\R^2)}\,,
\end{equation}
where we dropped the tilde's from the notation for the sake of simplicity.\\
Notice that 
\begin{align*}
 (h^{1/2}\partial_s-i\zeta_a) \partial_t\mathfrak a\partial_t \varphi&=\partial_t\big((h^{1/2}\partial_s-i\zeta_a)\mathfrak a\partial_t \varphi\big)\\
 &=  \partial_t\big(\mathfrak a\partial_t (h^{1/2}\partial_s-i\zeta_a) \varphi\big)+h^{1/2}\partial_t\big(\partial_s\mathfrak a \big)\partial_t \varphi
 \\
 &=\partial_t \mathfrak a\partial_t g+h^{1/2}\partial_t\big(\partial_s\mathfrak a \big)\partial_t \varphi\,,
 \end{align*}
  and
 \begin{align*}
 (h^{1/2}\partial_s&-i\zeta_a)(h\partial_s-i\tilde{F}_{1})\mathfrak a^{-1}(h\partial_s-i\tilde{F}_{1}) \varphi\\
& =(h\partial_s-i\tilde{F}_{1})\Big((h^{1/2}\partial_s-i\zeta_a)-ih^{1/2}(\partial_s\tilde  F_1)\Big)\mathfrak a^{-1}(h\partial_s-i\tilde{F}_{1}) \varphi\\
 &=(h\partial_s-i\tilde{F}_{1})\Big(\mathfrak a^{-1}(h\partial_s-i\tilde F_1)(h^{1/2}\partial_s-i\zeta_a) \varphi-ih^{1/2}(\partial_s\tilde  F_1)\mathfrak a^{-1}(h\partial_s-i\tilde{F}_{1}) \varphi\Big)\\
 &\qquad+h^{1/2}(h\partial_s-i\tilde{F}_{1})\Big((\partial_s\mathfrak a^{-1})(h\partial_s-i\tilde F_1)\varphi -
 i\mathfrak a^{-1}(\partial_s\tilde F_1)\varphi\Big)\\
 &=(h\partial_s-i\tilde{F}_{1})\mathfrak a^{-1}(h\partial_s-i\tilde F_1)g\\
 &\qquad +h^{1/2}(h\partial_s-i\tilde{F}_{1})\Big( \big(\partial_s \mathfrak a^{-1}-i\mathfrak a^{-1}\partial_s\tilde F_1\big)(h\partial_s-i\tilde{F}_{1})\varphi -i\mathfrak a^{-1}(\partial_s\tilde F_1)\varphi\Big)\,.
 \end{align*}
 By integration by parts, we infer from \eqref{eq:Ap-loc1}
 \begin{equation}\label{eq:Ap:loc2}
 \langle \mathcal P_h \varphi, G\rangle_{L^2(\Omega)}= \langle \mathcal P_h g, e^{2\phi}g\rangle_{L^2(\Omega)}-h^{1/2}\mathsf T_h\,.
 \end{equation}
 Finally, by integration by parts, we get
 \[  {\rm Re}\langle \mathcal P_h g, e^{2\phi}g\rangle_{L^2(\Omega)}=\mathcal Q_h(e^{\phi}g)-h^2\big\||\nabla\phi|e^{\phi}g\big\|_{L^2(\Omega)}^2\,.\]
\end{proof}

\section{Control of a remainder term}\label{sec:app-r}

The  aim of this appendix is to prove the estimate in \eqref{eq:term-r}. We fix a positive integer $n\geq 1$ and  two positive constants $\eta\in(0,\frac18) $ and $\delta\in(0,\frac1{12})$.

For all $h>0$,  let $v_{h,n}$ be the function  introduced in \eqref{eq:ef-tranc-v} which is supported in $\{|\sigma|<h^{-\eta},~|\tau|<h^{-\delta} \}$. Moreover, by  \eqref{eq:ef-tranc-v} and Propositions~\ref{prop:tang-freq} and \ref{rem:tang-freq}, we observe that
\begin{equation}\label{eq:Ap-est-v}
\forall\,\theta\in\Big(0,3/8\Big),~\exists\,C_\theta>0,~\|\partial_\sigma^jv_{h,n}\|_{L^2(\R^2)}\leq C_\theta h^{-j\theta}\quad (0\leq j\leq 2)\,.
\end{equation} 
Consider two functions $f\in L^2(\R)$ and $p\in L^1_{\rm loc}(\R^2)$  so that
\[\forall\,~\alpha\geq 1,~\tau^\alpha f(\tau)\in L^2(\R)\,,\]
and there exist $k\geq 1$ and $C$ such that
\[ |p(\sigma,\tau)|\leq C (|\sigma|^k+|\tau|^k+1)\quad (\sigma,\tau\in\R)\,.\]
For $j\in\{0,1,2\}$, we introduce the  function
\begin{equation}\label{eq:error-r}
w_j(\sigma)
=\int_\R f(\tau)p(\sigma,\tau)\partial_\sigma^j v_{h,n}(\sigma,\tau)d\tau\,,
\end{equation}
whose support is included in $\{|\sigma|< h^{-\eta}\}$,  by the considerations on the support of  $v_{h,n}$.
\begin{lem}\label{lem:Ap-est-wj}
Given $\eta\in(0,\frac18)$, there exist two positive constants  $h_0,C>0$ such that
\[ \|w_j\|_{L^2(\R)}\leq C \, h^{-(k+ j/2)\eta}\,, \]
for all $h\in(0,h_0)$ and $j\in\{0,1,2\}$.
\end{lem}
\begin{proof}
By H\"older's inequality
\begin{equation}\label{eq:Ap-est-w*}
|w_j(\sigma)|^2\leq \left( \int_\R |f(\tau)|^2|p(\sigma,\tau)|^2d\tau\right)\left(\int_\R|\partial_\sigma^jv_{h,n}(\sigma,\tau)|^2d\tau \right)\,.
\end{equation}
For $\sigma$ in the support of $w_j$,  we have
\[\int_\R |f(\tau)|^2|p(\sigma,\tau)|^2d\tau\leq 
C\int_{\R} |f(\tau)|^2(1+|\tau|^k+|\sigma|^k)^2d\tau\leq \tilde C_k(1+ h^{-2k\eta})\,.\]
Inserting this into \eqref{eq:Ap-est-w*} then integrating with respect to $\sigma$, we get
\[\int_\R|w_j(\sigma)|^2d\sigma\leq \tilde C_k(1+h^{-2k\eta})\int_{\R^2} |\partial_\sigma^jv_{h,n}(\sigma,\tau)|^2d\sigma d\tau\,.\]
Finally, we use \eqref{eq:Ap-est-v} with $\theta=\eta$.
\end{proof}

We will encounter functions of the form
\begin{equation}\label{eq:w-erro-r}
\mathfrak w_j(\sigma)=\int_\R g(\tau)q(\sigma)\partial_\tau^j v_{h,n}(\sigma,\tau)d\tau\quad (j\in\{1,2\},~\sigma\in\R)\,,
\end{equation}
where $g\in H^j(\R)$  and $q\in H^1_{\rm loc}(\R)$ satisfy
\[\forall\, \alpha \geq 1,\tau^\alpha g^{(i)} (\tau)\in L^2(\R)\quad(1\leq i\leq j)\,,  \]
and
\[ \exists\,k\geq 1,~\exists\,C_k>0,\quad |q(\sigma)|\leq C_k(1+|\sigma|^k)\quad (\sigma\in\R)\,. \]
\begin{lem}\label{lem:Ap-est-wj*}
Given $\eta\in(0,\frac18)$, there exist two positive constants  $h_0$ and $C$ such that
\[ \|
\mathfrak w_j\|_{L^2(\R)}\leq C  h^{-(k+1)\eta} \]
for all $h\in(0,h_0]$ and $j\in\{1,2\}$.
\end{lem}
\begin{proof}
Using integration by parts and that $v_{h,n}$ is with compact support, we get
\[ \mathfrak w_j(\sigma)=(-1)^j \int_\R g^{(j)}(\tau)q(\sigma)v_{h,n}(\sigma,\tau)d\tau\,.\]
This function has the form of functions in Lemma~\ref{lem:Ap-est-wj}, with $f(\tau)=g^{(j)}(\tau)$ and $p(\sigma,\tau)=q(\sigma)$.
\end{proof}

The inner product of the remainder, $\mathrm r_n(\sigma,h)$ in \eqref{eq:decom-r}, and the function, $R^{\rm new}_0v_{h,n}$ in \eqref{eq:proj:R-h**},   can  be expressed as the inner product of a linear combination of functions having the forms in Lemma~\ref{lem:Ap-est-wj} and \ref{lem:Ap-est-wj*}. The polynomials we encounter are of degree  6 at most. More precisely,
\[\langle \mathrm r_{ n}(\cdot,h),R^{\rm new}_0v_{h,n}\rangle_{L^2(\R)}= h^{7/8}A_1 +h^{7/8}A_2+hA_3+h^{9/8}A_{4}+h^{ 5/4}A_{5}\,, \]
where
\begin{equation*}
\begin{array}{ll}
A_1&=\langle a_{1,1},b_1\rangle_{L^2(\R)}
+h^{ 3/8}\langle a_{1,2},b_2\rangle_{L^2(\R)}+h^{1/2}\langle a_{1,3},b_1\rangle_{L^2(\R)},\\
A_2&=\langle a_{2,1},b_2\rangle_{L^2(\R)}
+\langle a_{2,2},b_1\rangle_{L^2(\R)},\\
A_3&=\langle a_{3},b_1\rangle_{L^2(\R)},\quad A_{ 4}=\langle a_{ 4},b_2\rangle_{L^2(\R)},\quad A_{5}=\langle a_{5},b_1\rangle_{L^2(\R)}\,,
\end{array}
\end{equation*}
and
\begin{align*}
a_{1,1}&= \int_\R g_1(\tau)Q_hv_{h,n}d\tau, &\quad   a_{1,2}&= \int_\R g_2(\tau)Q_hv_{h,n}d\tau,&\quad  a_{1,3}= \int_\R g_3(\tau)Q_hv_{h,n}d\tau \\
a_{2,1}&=\int_\R f_1(\tau) P_2 v_{h,n}d\tau,&\quad 
a_{2,2}&= \kappa\int_\R f_2(\tau) P_0 v_{h,n}d\tau,&\\
a_3&=\kappa  \int_\R f_2(\tau) P_2 v_{h,n}d\tau,&\quad a_{4}&=\kappa \int_\R f_1(\tau) P_3 v_{h,n}d\tau,
&\quad a_{5 }=\kappa  \int_\R f_2(\tau) P_3 v_{h,n}d\tau,\\
b_1&=  \int_\R g(\tau) v_{h,n}d\tau\,, & \quad b_2&=i\int_\R g(\tau)\partial_\sigma v_{h,n}d\tau\,.&
\end{align*}
Here, $Q_h$ is the operator introduced in \eqref{eq:Q(h)}, $P_0,P_1,P_2,P_3$ are the operators introduced in \eqref{eq:LOT-Ph}, $f_1,f_2$ are the functions introduced in \eqref{eq:f1}-\eqref{eq:f2}, the functions $g_1,g_2,g_3 $ and $g$ are defined as follows (see \eqref{eq:phi-a,h*} and \eqref{eq:proj:R-h**})
\begin{align*}
&g_1=\phi_a,\quad g_{ 2}=f_1=2{\mathfrak R}_a\big((b_a(\tau) \tau +\zeta_a)\phi_a\big),\\
&g_{ 3} =\kappa f_2=\kappa {\mathfrak R}_a\Big(
M_3(a)\phi_a-\phi'_a-2\tau(b_a(\tau) \tau +\zeta_a)^2\phi_a+b_a(\tau) \tau ^2 (b_a(\tau) \tau +\zeta_a)\phi_a\Big),\\
&
g 
=
\phi_a -4(b_a(\tau) \tau +\zeta_a)\mathfrak R_a\big((b_a(\tau) \tau +\zeta_a)\phi_a)\,.
\end{align*} 
So, we get
\[ \langle \mathrm r_n (\cdot,h),R^{\rm new}_0v_{h,n}\rangle_{L^2(\R)}=\mathcal O(h^{\frac78-8\eta})\,.\]
By choosing $\eta<\frac1{64}$, we get \eqref{eq:term-r}.

\bibliographystyle{alpha}

\end{document}